\documentclass{aastex62}

\usepackage{amsmath}
\usepackage{longtable}
\usepackage{array}
\usepackage{hyperref}
\usepackage{cleveref}

\submitjournal{ApJ}
\accepted{July 14, 2020}

\shorttitle{Hot Core Infrared Spectral Survey}
\shortauthors{Barr et al.}

\begin{document}

\title{High Resolution Infrared Spectroscopy of Hot Molecular Gas in AFGL 2591 and AFGL 2136: \\ Accretion in the Inner Regions of Disks Around Massive Young Stellar Objects}

\correspondingauthor{Andrew Barr}
\email{barr@strw.leidenuniv.nl}

\author[0000-0003-4909-2770]{Andrew G. Barr}
\affiliation{Leiden University \\
Niels Bohrweg 2, 2333 CA Leiden, \\
The Netherlands}

\author[0000-0001-9344-0096]{Adwin Boogert}
\altaffiliation{Staff Astronomer at the
 Infrared Telescope Facility, which is operated by the University of
 Hawaii under contract NNH14CK55B with the National Aeronautics and
 Space Administration.}
\affiliation{Institute for Astronomy \\
University of Hawaii, 2680 Woodlawn Drive, \\
Honolulu, HI 96822, USA}

\author{Curtis N. DeWitt}
\affiliation{USRA, SOFIA, NASA Ames Research Center, \\
MS 232-11, Moffett Field, \\
CA 94035, USA}

\author{Edward Montiel}
\affiliation{USRA, SOFIA, NASA Ames Research Center, \\
MS 232-11, Moffett Field, \\
CA 94035, USA}

\author[0000-0002-8594-2122]{Matthew J. Richter}
\affiliation{University of California, Davis,\\
Phys 539, Davis, \\
CA 95616, USA}

\author[0000-0001-6783-2328]{John H. Lacy}
\affiliation{Department of Astronomy, \\
The University of Texas at Austin,\\
2515 Speedway, Stop C1400, Austin,\\
TX 78712, USA}

\author{David A. Neufeld}
\affiliation{Johns Hopkins University,\\
Baltimore, \\
MD 21218, USA}

\author{Nick Indriolo}
\affiliation{ALMA Project, \\ 
National Astronomical Observatory of Japan, \\
National Institutes of Natural Sciences, \\
2-21-1 Osawa, Mitaka, Tokyo\\
181-8588, Japan}

\author{Yvonne Pendleton}
\affiliation{NASA Ames Research Center,\\
Moffett Field, \\
CA 94035, USA}

\author{Jean Chiar}
\affiliation{Diablo Valley College,\\
321 Golf Club Rd, Pleasant Hill,\\
CA 94523, USA}

\author{Alexander G. G. M. Tielens}
\affiliation{Leiden University \\
Niels Bohrweg 2, 2333 CA Leiden, \\
The Netherlands}
\nocollaboration

\begin{abstract}

We have performed a high resolution 4-13 ${\mu}m$ spectral survey of the hot molecular gas associated with the massive protostars AFGL 2591 and AFGL 2136, utilising the Echelon-Cross-Echelle-Spectrograph (EXES) on-board the Stratospheric Observatory for Infrared Astronomy (SOFIA), and the iSHELL instrument and Texas Echelon Cross Echelle Spectrograph (TEXES) on the NASA Infrared Telescope Facility (IRTF). Here we present results of this survey with analysis of CO, HCN, C$_2$H$_2$, NH$_3$ and CS, deriving the physical conditions for each species. Also from the IRTF, iSHELL data at 3 ${\mu}m$ for AFGL 2591 are presented that show HCN and C$_2$H$_2$ in emission. In the EXES and TEXES data, all species are detected in absorption, and temperatures and abundances are found to be high (600 K and 10$^{-6}$, respectively). Differences of up to an order of magnitude in the abundances of transitions that trace the same ground state level are measured for HCN and C$_2$H$_2$. The mid-infrared continuum is known to originate in a disk, hence we attribute the infrared absorption to arise in the photosphere of the disk. As absorption lines require an outwardly decreasing temperature gradient, we conclude that the disk is heated in the mid-plane by viscous heating due to accretion. We attribute the near-IR emission lines to scattering by molecules in the upper layers of the disk photosphere. The absorption lines trace the disk properties at 50 AU where a high temperature gas-phase chemistry is taking place. Abundances are consistent with chemical models of the inner disk of Herbig disks.

\end{abstract}


\section{Introduction}

Massive star formation begins with the formation of a dense core, a so-called infrared dark cloud \citep{Egan1998, Carey1998}. This core becomes then gravitationally unstable and collapses, resulting in a central protostar embedded in a dusty cocoon \citep{ZY2007, Tan2017}. Because of excess angular momentum, much of the in-falling material is channeled into an accretion disk, which can feed the central protostar. This disk is heated by viscous dissipation in the mid-plane. For massive stars, hydrogen burning starts already during this accretion phase and the resulting high luminosity heats the gas and dust in the direct environment. This is the so-called Hot Core phase which is rich in complex organic molecules \citep{vanderTak2003review, Beuther2007, HvD2009}. Eventually, when the central star is hot enough to create ionising photons, a hyper compact H{\sc{ii}} region will form in the inflating material. Once the cocoon is thin enough, the ionising photons can escape, leading to the ultracompact and compact H{\sc{ii}} region phases  \citep{Beuther2007}.

From an observational perspective, the understanding of these stages is hindered on several accounts. Distances to the nearest massive stars are larger than for low mass stars, meaning that higher angular resolution is required to study them at comparable physical scales as their low mass counterparts \citep{Henning2002}. Furthermore dust extinction is high ($\sim$ $A_v=200$) as massive stars remain deeply embedded in their natal envelope for a large fraction of their early evolution \citep{Beuther2007}. A 40 M$_{\odot}$ star will live for around a few million years with an embedded lifetime of around 1$\times10^5$ yr. This contrasts to low mass stars that have an embedded lifetime of around $10^6$ yr but a main sequence lifetime of over $10^9$ yr. Therefore it is challenging to disentangle the protostar and its direct environment - circumstellar disk, hot core and surrounding molecular cloud \citep{ZY2007, Cesaroni2007}. 

During the star formation process, infalling material onto the central protostar forms a disk, with a bipolar outflow along the perpendicular axis \citep{ZY2007}. Angular momentum is transported to the outer parts of the disk via viscous stresses and gravitational torques, and the disk expands \citep{LBP1974, CS1983, Dullemond2007}. As a result of this, material moves inward and accretes onto the central protostar. Theoretical models of dust disks come in varying degrees of complexity. One such model combines a hot, flared dust surface layer heated by stellar radiation, and viscous dissipation of gravitational energy due to accretion \citep{dAlessio1998}. In this model one can see that for high accretion rates (10$^{-7}$ - 10$^{-6}$M$_{\odot}$/yr), three regimes arise in the disk. The surface region is irradiation dominated, the midplane of the disk is accretion dominated, and the two are connected by an intermediate zone. Here two temperature gradients can be distinguished, with an outwardly decreasing temperature gradient in the vertical direction in the accretion dominated region, and an inwardly decreasing temperature gradient in the vertical direction in the irradiation dominated zone \citep{Dullemond2007}. 

Whilst disks around forming low mass young stellar objects (YSOs) are common \citep{WC2011}, the number of detected disks around massive stars remains constrained to a handful, especially in the case of O-type stars \citep{Patel2005, Kraus2010, MG2014, Johnston2015, Ilee2016, Moscadelli2019, Zapata2019, Maud2019}. Commonly detected around massive stars are massive rotating toroids likely to host stellar clusters \citep{Cesaroni2005, Sandell2003, Olmi2003, Beltran2004, Sollins2005}. The search for disks around O-B stars is undertaken with sub-millimetre (sub-mm) continuum or specific line tracers and infrared (IR) continuum/line observations, as well as maser studies, in order to penetrate the surrounding obscuring dust in these embedded regions. Sub-mm observations probe preferentially the outer parts of the disk and the envelope, while IR observations are sensitive to the innermost regions of the disk \citep{Cesaroni2007, Dullemond2007}. Inner regions of TTauri and Herbig disks are observed in molecular \textit{emission} such as CO and H$_2$O at IR wavelengths \citep{Mandell2012, Bast2013, Bast2011, Pontoppidan2011, Pontoppidan2019, Setterholm2019, Adams2019}. These disks are illuminated by UV radiation from the central protostar under an irradiation angle, resulting in a PDR-like layered structure \citep{TH1985}, with a temperature gradient that decreases towards the mid-plane. The inner portions of massive disks have been studied mainly through atomic line and CO bandhead \textit{emission} \citep{Cooper2013, Ilee2013, Davies2010, Bik2006, Fedriani2020}. In a survey of massive protostars, \citet{Ilee2013} model CO bandhead emission from 20 massive YSOs and find temperatures and densities consistent with emission from an accretion disk, close to the dust sublimation radius. For a further discussion on the inner regions of protoplanetary disks, see the review by \citet{Dullemond2010}.

Observations of ro-vibrational lines in the mid-infrared (MIR) regime of star forming regions offer a unique view. Towards massive YSOs these lines are typically seen in absorption rather than emission. Since the IR source is very small, a very high angular resolution is achieved, comparable to the highest possible obtainable with emission studies ($\approx$50mas). Furthermore, since the lines lie so close together, it is possible to observe many of these lines in one observational setting. Further details about studies of ro-vibrational lines at IR wavelengths can be found in \citet{Lacy2013} along with the various issues that these kind of observations face.  Finally, we emphasise that IR studies allow observations of CH$_4$, C$_2$H$_2$ and CO$_2$ that have no allowed rotational spectrum in the sub-mm.

We have carried out a high spectral resolution spectral survey of the 4-13 ${\mu}m$ region towards AFGL 2591 and AFGL 2136. This article is an follow-up of a previous study \citep{Barr2018} where we reported IR observations of CS in absorption towards the hot core AFGL 2591 VLA 3. Here we extend the discussion to other species detected in our spectra of AFGL 2591 VLA 3, as well as those in the hot core AFGL 2136 IRS1. In section 2 we present the observational details of the spectral survey, in section 3 we outline the analysis techniques used and section 4 contains the results for each molecular species identified. Section 5 discusses the results in the context of the physical structures of the sources, along with the origin of L-band emission lines and the calculation of the abundances. Finally the chemical implications are discussed as well as extending the discussion to other infrared studies of massive protostars.

\section{Observations and Data Reduction} 

\subsection{Target Sources}

AFGL 2591 is a star forming region in the Cygnus X region, at coordinates $\alpha(J2000) =$ 20:29:24.8, $\delta(J2000) =$ +40:11:19.6. It is home to the massive protostar AFGL 2591 VLA 3 (hereafter AFGL 2591) which has a luminosity of 2$\times10^5$ L$_{\odot}$ at a distance of 3.33$\pm$0.11 kpc \citep{Rygl2012}. It is an O-type star with a mass of $\sim$40 M$_{\odot}$ \citep{Sanna2012} and powers a bipolar outflow seen in CO and H$_2$ \citep{BL1983, Mitchell1991, TY1992}. Radio continuum emission is observed at 3.6 cm and 1.3 cm towards AFGL 2591 which coincides with a cluster of maser emission in a helical shell around the outflow \citep{Trinidad2003}. 

AFGL 2136 IRS1 is an O-type star and is the main IR source, in the star forming region of AFGL 2136, coordinates $\alpha(J2000) =$ 18:22:26.385, $\delta(J2000) =$ -13:30:11.97. The luminosity of AFGL 2136 IRS1 (hereafter AFGL 2136) is 1$\times10^5$ L$_{\odot}$ at a distance of 2.2 kpc \citep{Lumsden2013, Urquhart2012, Urquhart2014}. AFGL 2136 is one of the few sources that has a clear signature of a Keplerian disk around a proto-O-type star \citep{Maud2019} with a bipolar outflow in the perpendicular direction to the disk. There is also evidence for a disk wind from Br$\gamma$ emission and SiO line emission \citep{Murukawa2013, Maud2018}. The surrounding environment of AFGL 2136 consists of a bipolar reflection nebula in a fan shape \citep{Murakawa2008} and has deep ice bands associated with ice located in the envelope \citep{Gibb2004}.

\subsection{EXES}

AFGL 2591 and AFGL 2136 were observed with the EXES spectrometer \citep{Richter2018} onboard the Stratospheric Observatory for Infrared Astronomy (SOFIA) flying observatory \citep{Young2012} as part of SOFIA program 05\_0041. In the full spectral survey of AFGL 2591 and AFGL 2136, EXES covered the range 5.5-8 ${\mu}m$ and required 16 wavelength settings in its HIGH-LOW mode, where the high-resolution echelon grating is cross-dispersed with the R 1/2 LOW resolution echelle grating. For AFGL 2136, observations were taken over three EXES flight series. For the analysis in this paper, 8 spectral settings were used, covering wavelengths 6.7-8.0 ${\mu}m$ (1249cm$^{-1}$ - 1494cm$^{-1}$). These observations are summarised in Table \ref{obs}. The slit width was 3.2$''$ for all settings, providing R=55,000 resolution. The fixed slit lengths used were either 3.1$''$ or 2.2$''$, depending on the wavelength setting. In order to remove background night sky emission and telescope thermal emission, the telescope was nodded to an off-source position 15$''$ away from the target coordinates every 1-2 minutes.

The EXES data were reduced with the SOFIA Redux pipeline \citep{Clarke2015}, which has incorporated routines originally developed for the Texas Echelon Cross Echelle Spectrograph (TEXES) \citep{Lacy2003}. The science frames were de-spiked and sequential nod positions subtracted, to remove telluric emission lines and telescope/system thermal emission. An internal blackbody source was observed for flat fielding and flux calibration and then the data were rectified, aligning the spatial and spectral dimensions. The wavenumber solution was calibrated using sky emission spectra produced for each setting by omitting the nod-subtraction step. We used wavenumber values from HITRAN \citep{Rothman2013} to set the wavelength scale. The resulting wavelength solutions are accurate to 0.3 kms$^{-1}$. Where necessary, lines were divided by a local continuum which was fit as a straight line over the transition. This was done in cases where the continuum was uncertain due to fringing or poor atmospheric correction.

\subsection{TEXES}

Spectra were obtained of HCN, C$_2$H$_2$, and NH$_3$ from spectral surveys made with TEXES \citep{Lacy2002}. AFGL 2136 was observed with the Gemini North telescope on 2017 March 17 and 18 (UT) in the TEXES hi-lo spectral mode. In this mode the high resolution echelon grating is cross dispersed with a first order grating, providing a spectral coverage of $\delta\lambda \approx 0.25 \mu m$ with typically 20-30 echelon spectral orders. To prevent order overlap on the detector array a short slit was required, which necessitated nodding the source off of the slit to observe the sky background. Between the two nights the 782-904 cm$^{-1}$ region was covered with a total telescope time of 3$^h$ 30$^m$ including time for source acquisition, spectral setups, flat fields, nod delays, and sky observations. The on-source time at each spectral setting was typically 130 s, with some settings extended if clouds interfered. On the night of 2017 March 20 the spectral region of 1157-1195 cm$^{-1}$ was observed in the TEXES hi-med spectral mode, in which the echelon is cross dispersed with an echelle, which was used in 6$^{th}$ order. In this mode the spectral coverage in each setting is less, with $\delta \lambda \approx 0.006 \lambda$, but a longer slit can be used so that the source can be nodded along the slit, improving the observing efficiency. The total telescope time was 40$^m$, and the on-source time at each spectral setting was 260 s. The data are publicly available from the Gemini archive.

AFGL 2591 was observed with TEXES on the NASA Infrared Telescope Facility (IRTF) on 2018 July 7 and 8, and 2018 Sep. 29 and 30. The 728-849 cm$^{-1}$ spectral region was covered on July 7, the 849-999 cm$^{-1}$ region was covered on July 8, and the 727-905 cm$^{-1}$ region was covered on Sep. 18. These observations were in the TEXES hi-lo mode, with a total telescope time of 9$^h$ and an on-source time per spectral setting of 130 s. The 1107-1120 cm$^{-1}$ region was covered on July 8, the 1227-1250 cm$^{-1}$ region on July 12, and the 751-841 cm$^{-1}$ region on Sep. 30, all in hi-med mode. The total telescope time for these observations was 9$^h$ 10$^m$. The on-source time per spectral setting was 972 s on July 8 and 12, and 583 s for most Sep. 30 observations, except that additional time was spent when clouds interfered. 

No comparison sources for correction of telluric and instrumental features were available without spectral features and at least comparable in flux to the primary targets. Consequently, the observed spectra were divided by spectra of an ambient temperature black body and a model of the telluric absorption. The telluric model was based on measurements of atmospheric pressure, temperature, and water vapour from a weather balloon launched from Hilo each night and spectral parameters from the HITRAN database.  The balloon data are available from the National Centres for Environmental Information Integrated Global Radiosonde Archive (IGRA) at ncdc.noaa.gov. The HITRAN database is available from cfa.harvard.edu. The atmospheric ozone is measured less frequently, and the measurement closest to the observation dates were used. They are available from the NOAA Earth System Research Laboratories at esrl.noaa.gov. The water vapour and ozone abundances in the telluric model were adjusted to give the best correction and the best fit to the observed telluric emission spectrum. Most spectral lines were corrected well, but instrumental fringing was not always removed as well, so the corrected spectra were divided by a polynomial baseline.

\subsection{iSHELL}

AFGL 2591 and AFGL 2136 were observed with iSHELL \citep{Rayner2016} at the IRTF telescope on Maunakea at a spectral resolving power of 80,000. The observations are summarised in Table \ref{obs}. The observations for AFGL 2591 were described in \citep{Barr2018}. AFGL 2136 was observed on UT 2018-10-10 from 05:45 to 06:52 at an airmass range of 1.55-1.80, during moderate weather conditions as part of program 2018B095. The target was nodded along the 15$''$ long slit to be able to subtract background emission from the sky and hardware. Spectra affected by clouds were discarded in the data reduction. iSHELL's internal lamp was used to obtain flat field images. The spectra were reduced with the Spextool package version 5.0.1 \citep{Cushing2004}. Correction for telluric absorption lines was not done with a standard star but with the
program Xtellcor model developed at the IRTF and makes use of the atmospheric models calculated by the Planetary Spectrum Generator \citep{Villanueva2018}. The Doppler shift of AFGL 2136 at the time of the observations was +37 kms$^{-1}$ (including an assumed VLSR=+23 kms$^{-1}$), and thus telluric and target CO lines are well separated. The blaze shape of the echelle orders was corrected for using the flat fields. The continuum signal-to-noise ratio as measured from the scatter in the data points is $\sim$250 at the native sampling of 3 pixels per resolution element. AFGL 2591 was also observed with iSHELL's L1, L2, and L3 modes at R=80,000 on UT 2018-10-05 at 04:42-08:52 as part of program 2018B095. The airmass range was 1.08-1.20. The data reduction process was the same as for the M-band, although standard stars were used for the telluric absorption correction: BS 8830 (L1), and Vega (L2+L3). The continuum signal-to-noise values between the telluric lines ranges from $\sim$220 in L1 to $\sim$400 in the L2 and L3 modes.  The Doppler shift of AFGL 2591 at the time of the L-band observations was -13 kms$^{-1}$ (including an assumed VLSR=-6 kms$^{-1}$).

\begin{deluxetable*}{ccccccccc}[htb]
\tablecaption{Summary of Observations \label{obs}}
\tablecolumns{9}
\tabletypesize{\scriptsize}
\tablehead{
\colhead{Source} & \colhead{Telescope} & \colhead{Instrument} & \colhead{Date (UT)} & \colhead{Wavelength } & \colhead{Instrument} & \colhead{Slit} & Resolving & \colhead{Integration}
\\
& & & & Range (${\mu}m$) & Mode & Width ($''$) & Power & Time (min)
}
\startdata
AFGL 2591 & IRTF & iSHELL & 2017-07-05 & 4.51-5.24 & M1 \& M2 & 0.375 & 80,000 & 51 \\
 & IRTF & iSHELL & 2018-10-05 & 2.750-3.459 & L1, L2 \& L3 & 0.375 & 80,000 & 93 \\
 & SOFIA & EXES &  2017-03-17 & 7.34-7.68 & High-low & 3.2 & 55,000 & 30 \\
 & SOFIA & EXES & 2017-03-21 & 6.69-7.05 & High-low & 3.2 & 55,000 & 20 \\
 & SOFIA & EXES & 2017-03-22 & 7.19-7.37 & High-low & 3.2 & 55,000 & 14 \\
 & SOFIA & EXES & 2017-03-23 & 7.67-8.01 \& 7.02-7.21 & High-low & 3.2 & 55,000 & 25 \\
 & IRTF & TEXES & 2018-07-08 & 8.93-9.03 & High-med & 1.4 & 85,000 & 65 \\
 & IRTF & TEXES & 2018-07-12 & 8-8.15 & High-med & 1.4 & 85,000 & 32 \\
 & IRTF & TEXES & 2018-09-30 & 11.89-13.32 & High-med & 1.4 & 85,000 & 185 \\    
AFGL 2136 & IRTF & iSHELL & 2018-10-10 & 4.51-5.24 & M1 \& M2 & 0.375 & 80,000 & 42 \\
& SOFIA & EXES & 2017-03-17 & 7.34-7.69 & High-low & 3.2 & 55,000 & 53 \\
& SOFIA & EXES & 2017-03-21 & 6.69-6.88 & High-low & 3.2 & 55,000 & 43 \\
& SOFIA & EXES & 2017-03-23 & 7.67-7.85 \& 8.87-7.05 & High-low & 3.2 & 55,000 & 90 \\
& SOFIA & EXES & 2017-05-25 & 7.19-7.37 & High-low & 3.2 & 55,000 & 30 \\
& SOFIA & EXES & 2019-04-24 & 7.83-8.01 & High-low & 3.2 & 55,000 & 36 \\
& SOFIA & EXES & 2019-04-26 & 7.19-3.37 & High-low & 3.2 & 55,000 & 41 \\
& Gemini & TEXES & 2017-03-17 to 2017-03-18 & 11.06-12.79 & High-low & 1.4 & 85,000 & 15 \\
& Gemini & TEXES & 2017-03-20 & 8.37-8.64 & High-med & 1.4 & 85,000 & 22 \\
\enddata
\label{obs}
\end{deluxetable*}

\section{Analysis}

The detected absorption lines were fitted with one Gaussian line profile for each velocity component detected. The free parameters that were allowed to vary were the line strength, line width, line velocity and continuum level. The fitting was conducted in velocity space, therefore the fit was a function of the local standard of rest (LSR) velocity and full width at half maximum (FWHM) both in units of kms$^{-1}$. The fit was then converted into an optical depth profile, and then into a column density profile via:

\begin{equation}
\label{eqn:1}
\tau(v) = -\ln(flux/cont)
\end{equation}

followed by, 

\begin{equation}
\label{eqn:2}
\dfrac{dN_l}{dv} = \tau(v) \dfrac{g_l}{g_u} \dfrac{8 \pi}{A \lambda^3}
\end{equation}

where $\tau(v)$ is the optical depth profile as a function of velocity, g$_l$ and g$_u$ are the statistical weights of the lower and upper states respectively, A is the spontaneous emission coefficient for the transition and $\lambda$ is the wavelength. The fit was then integrated in velocity space to derive a column density in the lower energy level of the transition. 

Each state-specific column density was then plotted in a rotation diagram by plotting ln(N$_l$/g$_l$) against E$_l$, where N$_l$ is the integrated column density in each transition and E$_l$ is the energy of the lower level in Kelvin. The rotational temperature is given by the Boltzmann equation:

\begin{equation}
\label{eqn:3}
\ln\dfrac{N_l}{g_l} = \ln N - \ln Z(T) - \dfrac{E_l}{kT}
\end{equation}

where the temperature is -1/gradient of the fit to the rotation diagram. Here $Z$ is the partition function at the rotational temperature and N is the total column density of the species. In this way the excitation temperature of each species was calculated. Initially, column densities were calculated for both absorption and emission lines. However a more extensive analysis of the absorption lines was carried out using stellar atmosphere theory to calculate the abundances with respect to H, assuming a dust/H ratio, and this utilised the excitation temperatures derived from the initial rotation diagrams. The analysis process is discussed further in section 5.2. Results from the rotation diagram analysis are presented in Tables \ref{sum2591} \& \ref{sum2136} and parameters of the gaussian fitting are given in Tables \ref{gl2591} and \ref{gl2136} in the Appendix. 

\begin{figure}[t!]
\centering
\begin{tabular}{@{}cccc@{}}
\includegraphics[width=0.5\textwidth]{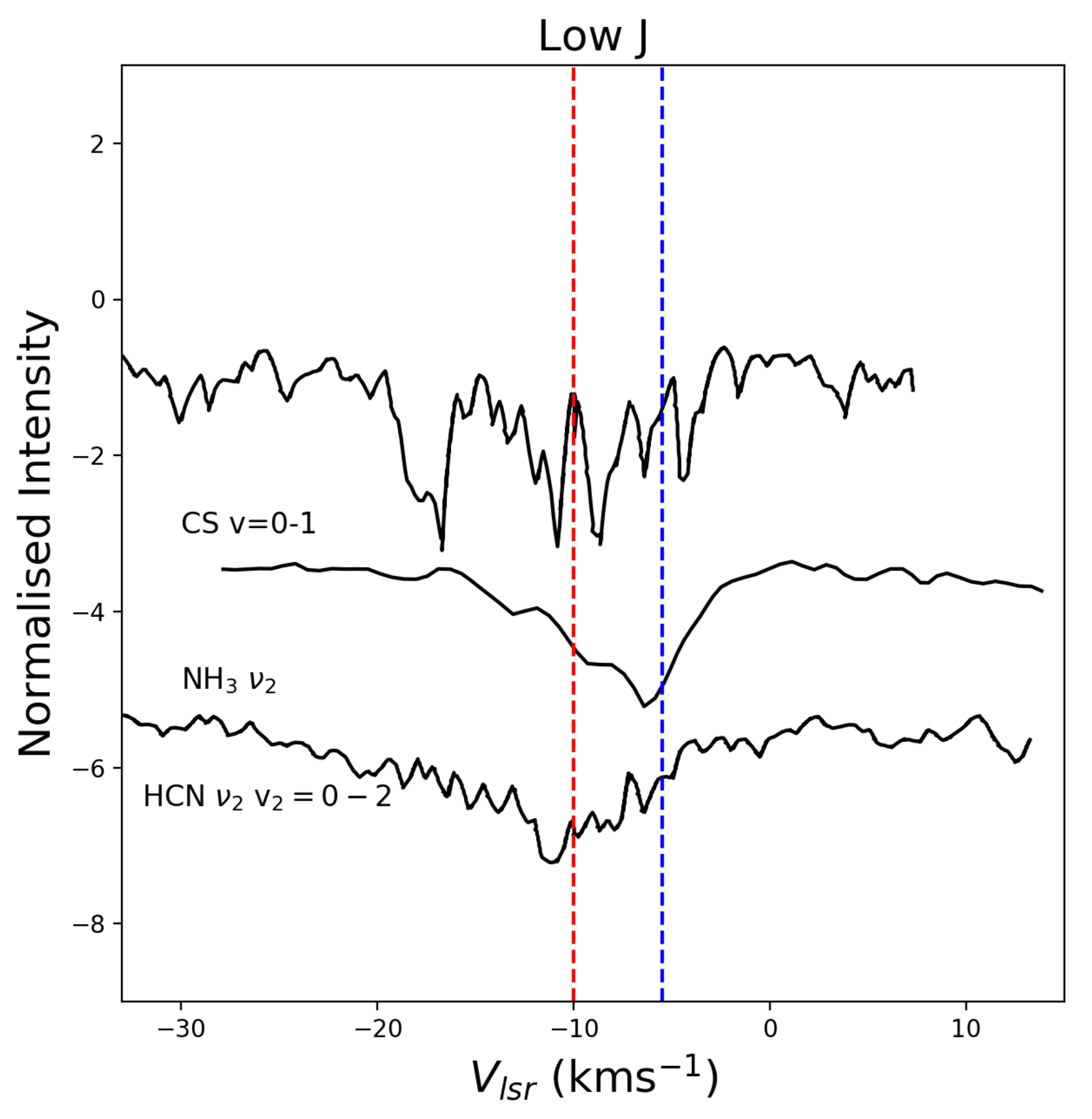}
\includegraphics[width=0.5\textwidth]{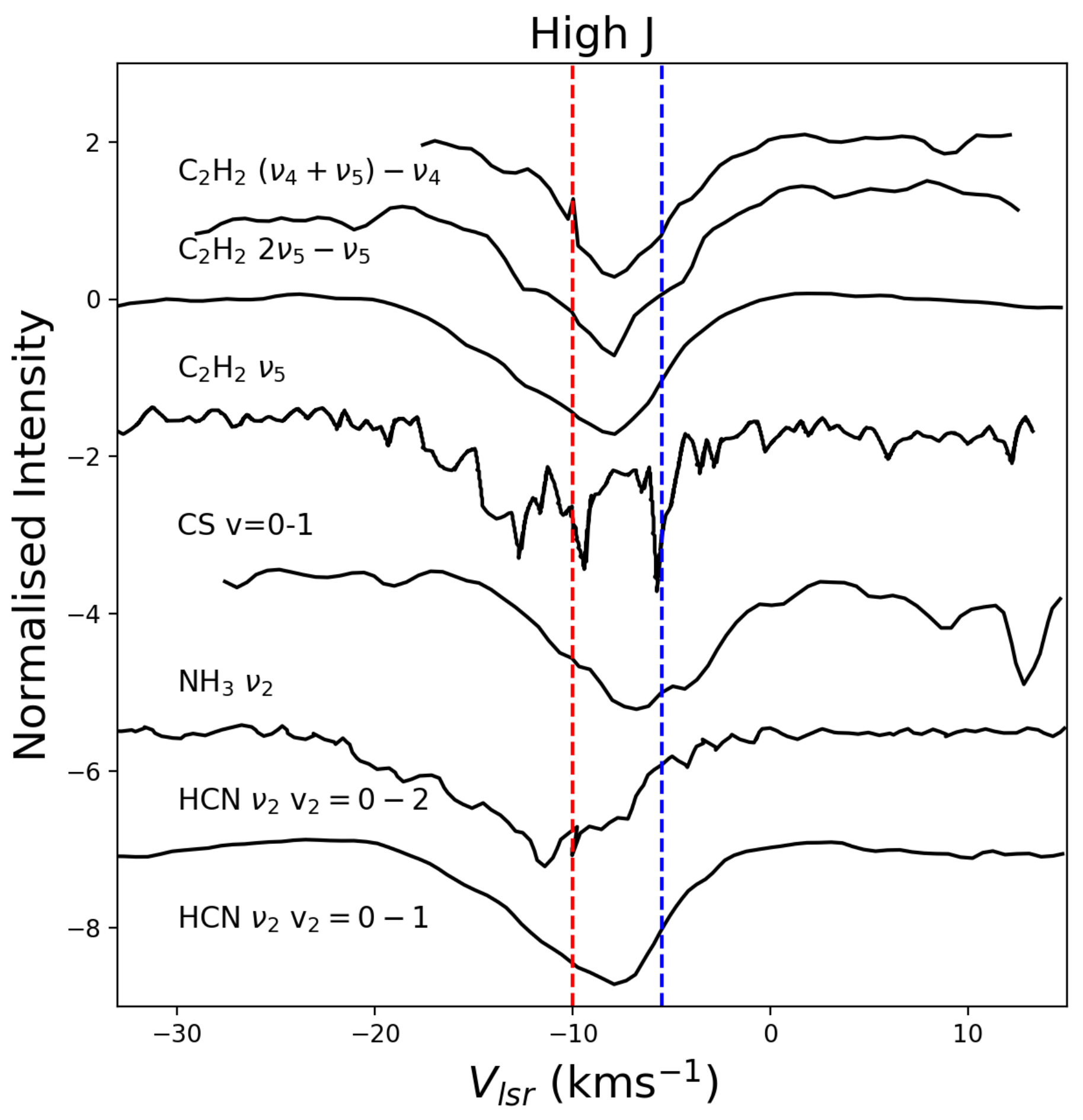} 
\end{tabular}
\caption{Normalised average line profiles for each band detected in AFGL 2591 plotted against the LSR velocity. Transitions are separated into high and low J levels before the average is calculated (high for E$_l$ $>$ 90 K). To guide the eye, we mark $v_{lsr}=-10$ km/s with a red dashed line in both panels. The blue dashed line denotes the systemic velocity of the source which is -5.5 kms$^{-1}$. CS and HCN v$_2=0-2$ transitions are detected with EXES at 7 ${\mu}m$ whilst the other bands/species are detected with TEXES at 13 ${\mu}m$. }
\label{Lineprof2591}
\end{figure}

\begin{figure}[t!]
\centering
\begin{tabular}{@{}cccc@{}}
\includegraphics[width=0.5\textwidth]{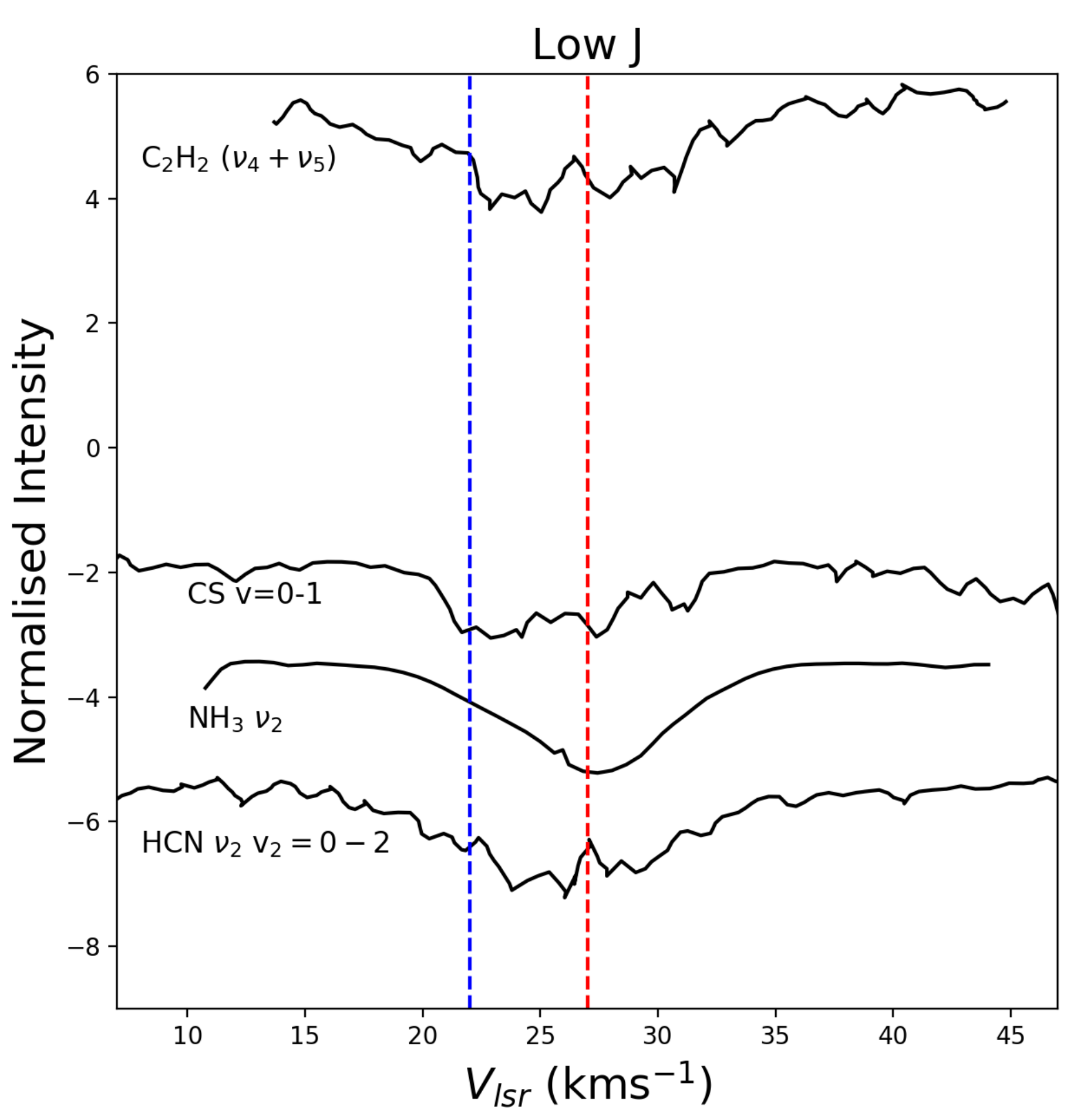}
\includegraphics[width=0.5\textwidth]{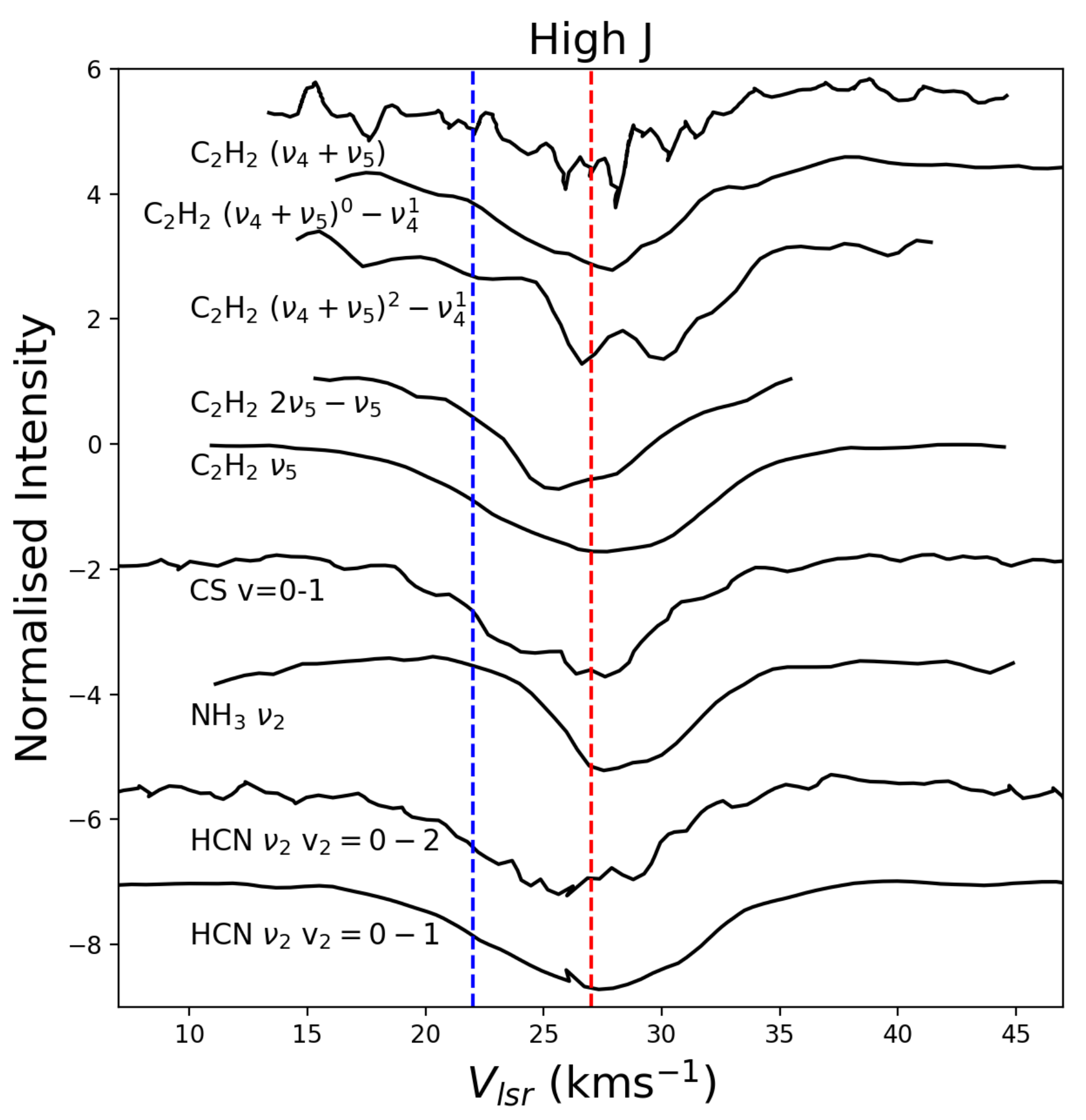}
\end{tabular}
\caption{Normalised average line profiles for each band detected in AFGL 2136 plotted against the LSR velocity. Transitions are separated into high and low J levels before the average is calculated (high for E$_l$ $>$ 90 K). To guide the eye, we mark $v_{lsr}=27$ km/s with a red dashed line in both panels. The blue dashed line denotes the systemic velocity of the source which is 22 kms$^{-1}$ \citep{vanderTak2003}. CS, HCN v$_2=0-2$ and C$_2$H$_2$ ($\nu_4+\nu_5$) transitions are detected with EXES at 7 ${\mu}m$ whilst the other bands/species are detected with TEXES at 13 ${\mu}m$. NH$_3$ transitions are detected from 10 to 13 ${\mu}m$. } 
\label{Lineprof2136}
\end{figure}

The rotational diagrams reveal different behaviour for the low and high J lines. Averaged line profiles were obtained for high and low energy lines by the mathematical averaging of all normalised, unblended lines of a given species. The distinction between high and low energy was taken as low energy lines having E$_l\leq90$ K. The weighted average of the line profiles was taken to give the average line profiles which are presented in Figures \ref{Lineprof2591} and \ref{Lineprof2136}.

\section{Results}

Line widths and centroid velocities quoted in this section are given as weighted averages for the lines detected of a given band. The line widths have been corrected for the resolving power of the different instruments used. In all cases, missing lines of a band are either lost in the wings of deep telluric lines, blended with other hot core lines or additionally, in the case of TEXES data, lost in order gaps. Some lines are affected by systematic error such as proximity to other hot core lines or poor baselines due to fringing or poor atmospheric correction. This makes it harder to put a constraint on the continuum of such lines. The result is that these lines exhibit a greater error in line parameters than that which is portrayed by the fitting routine. 

Also, for the TEXES data, it can be the case that lines lie close to the edge of gaps between orders. Therefore these lines will be slightly underproduced as some of their flux gets lost at the edges of the detector. For the TEXES data of AFGL 2591, additional systematic errors affect the data such as spurious emission wings on the shoulders of absorption lines due to the removal of strong fringing, and also CO$_2$ atmospheric residuals beyond around 13 ${\mu}m$.

\pagebreak

\subsection{AFGL 2591}

\subsubsection{CO}

\begin{figure}[h!]
\centering
\begin{tabular}{@{}cccc@{}}
\includegraphics[width=0.45\textwidth]{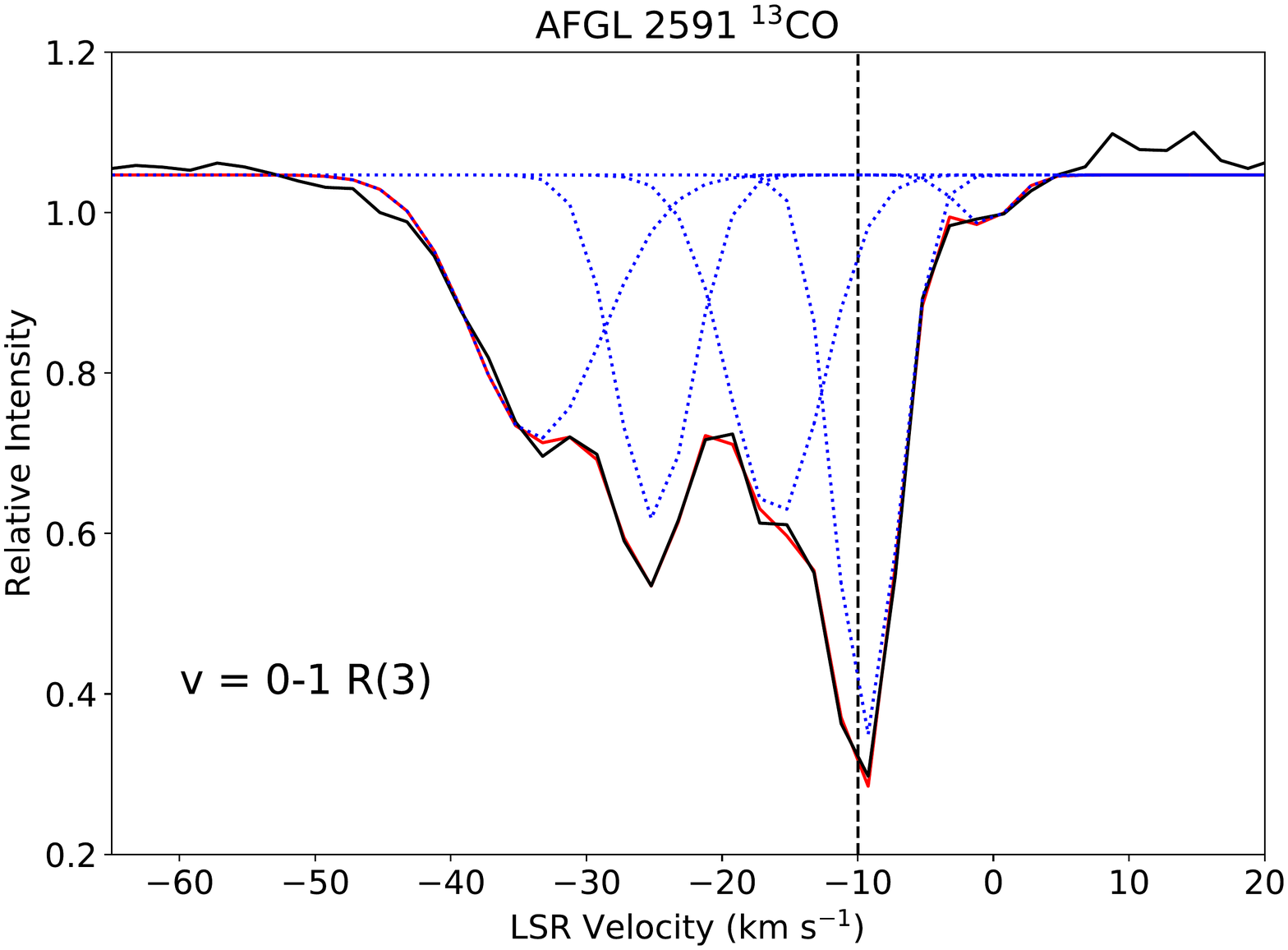}
\includegraphics[width=0.45\textwidth]{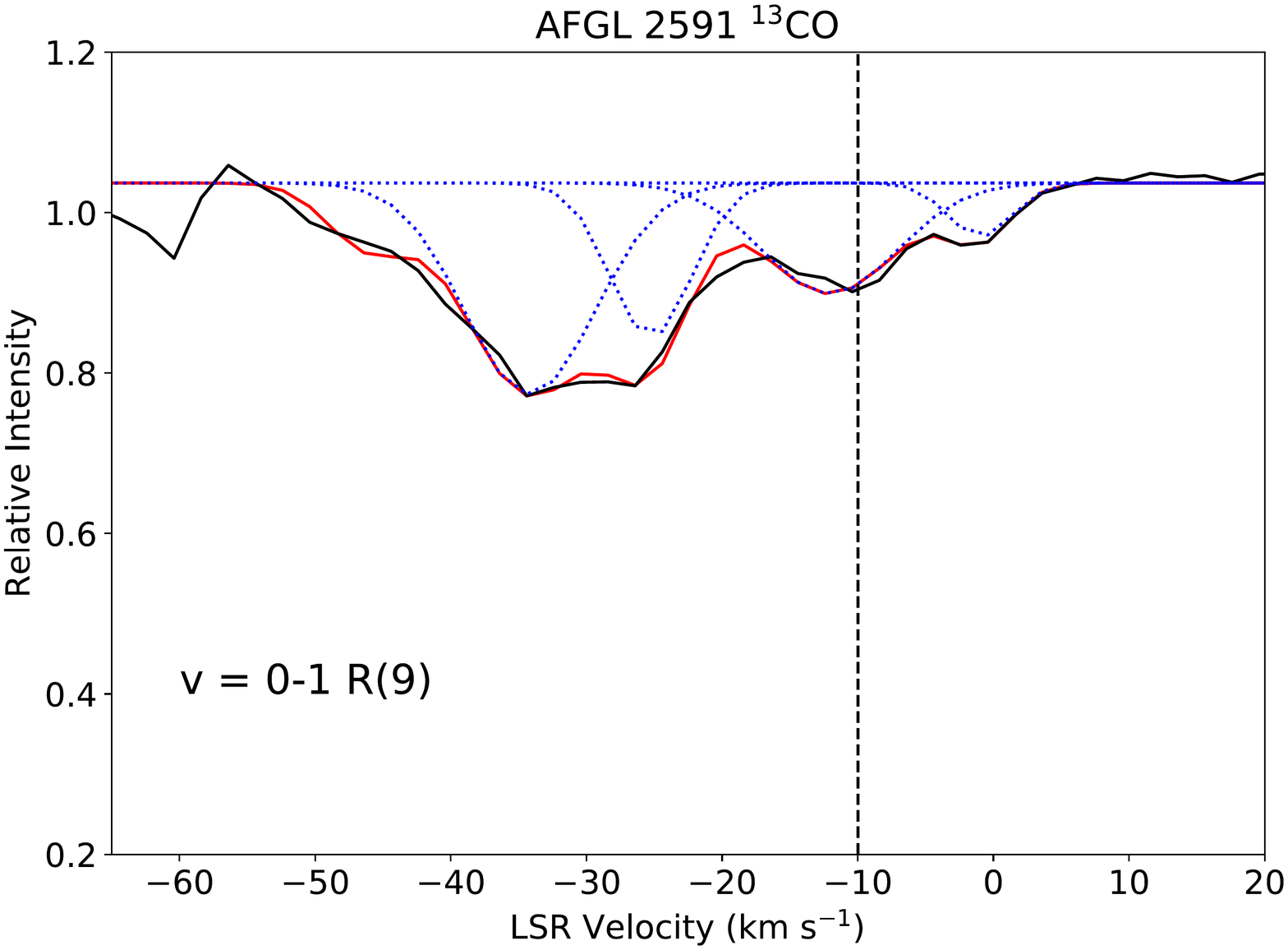} \\
\includegraphics[width=.45\textwidth]{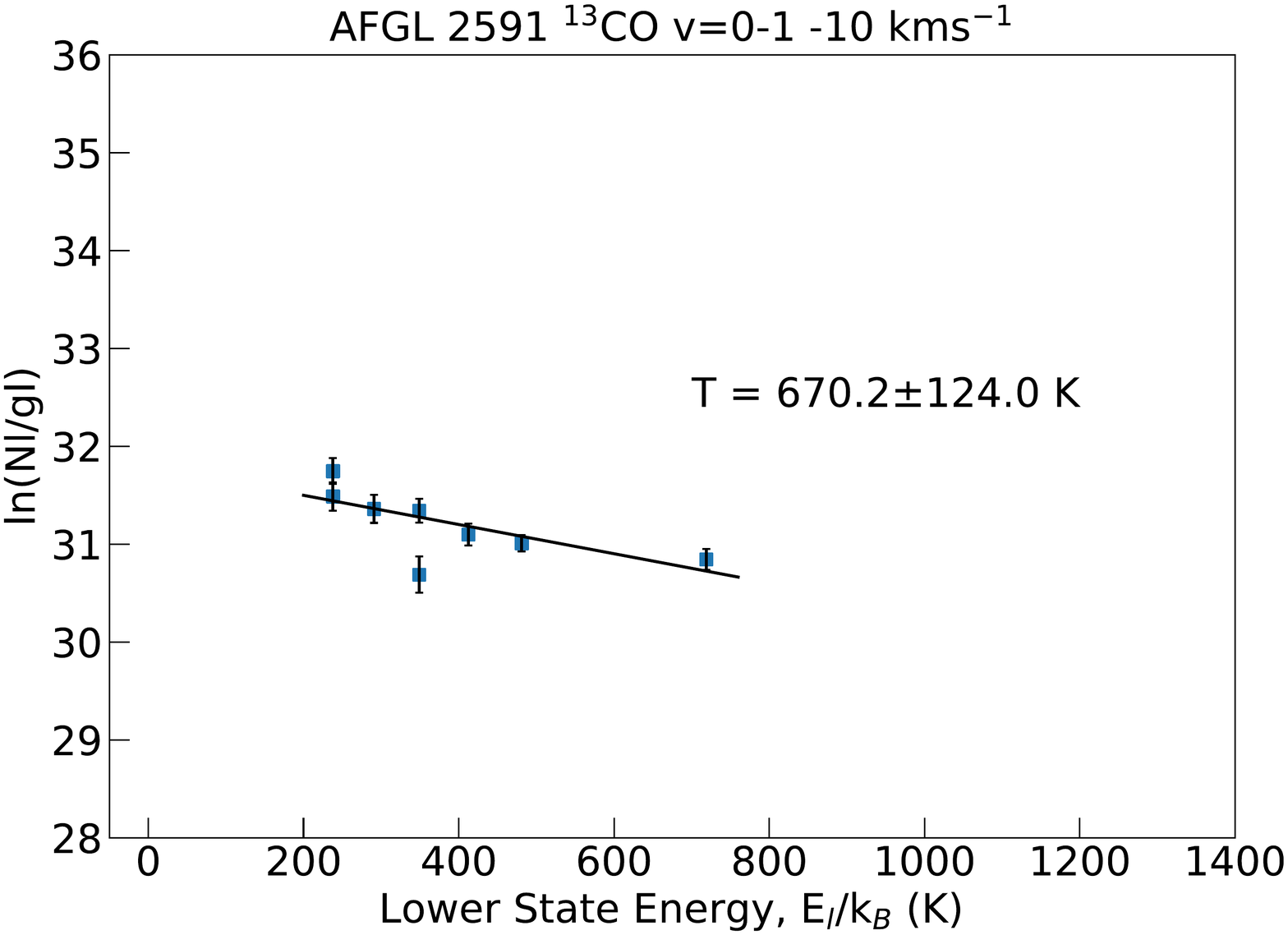} 
\includegraphics[width=.45\textwidth]{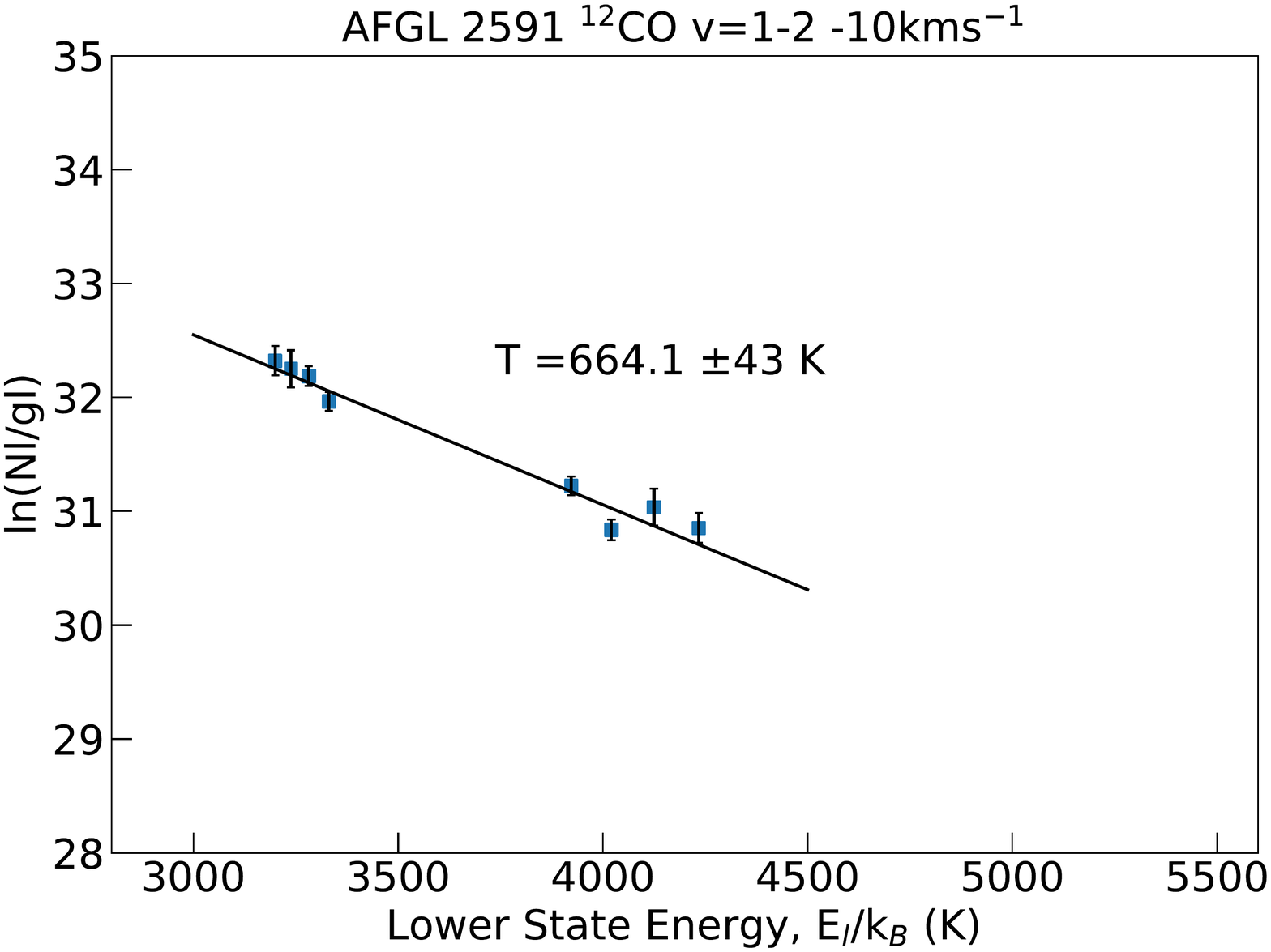} 
\end{tabular}
\caption{Individual line profiles for a high J (right) and low J (left) transition of $^{13}$CO for AFGL 2591 detected with iSHELL at 4.5 ${\mu}m$. Blue dotted lines represent the individual gaussian fits and the red solid line is the overall fit. The black dashed line denoted -10 kms$^{-1}$. Rotation diagrams of $^{13}$CO v=0-1 and $^{12}$CO v=1-2 are also shown for the -10 kms$^{-1}$ velocity component.
}
\label{CO2591}
\end{figure}

\begin{figure}[b!]
\centering
\includegraphics[width=75mm,scale=1.5]{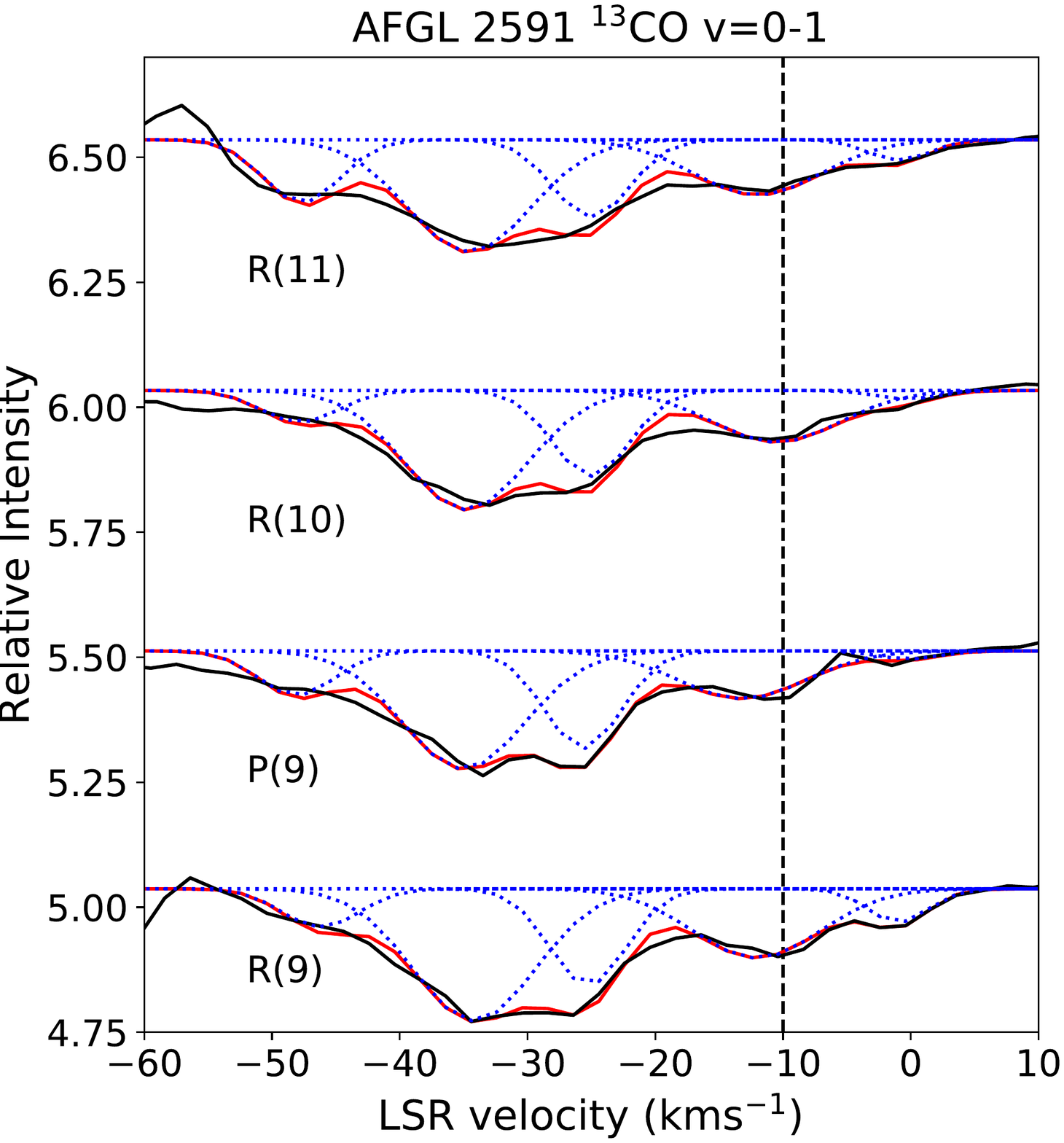}
\includegraphics[width=75mm,scale=1.5]{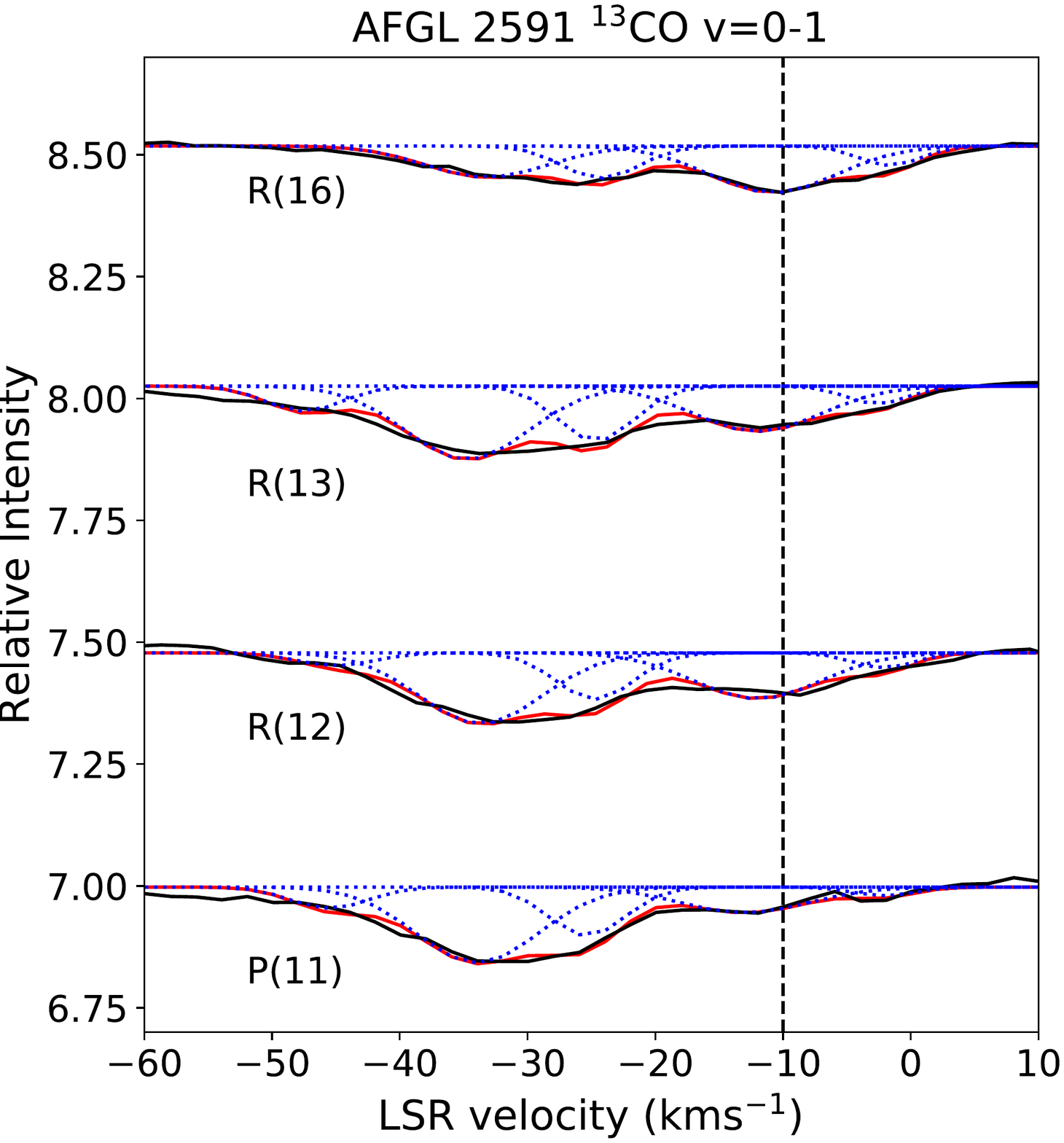}\\
\includegraphics[width=75mm,scale=1.5]{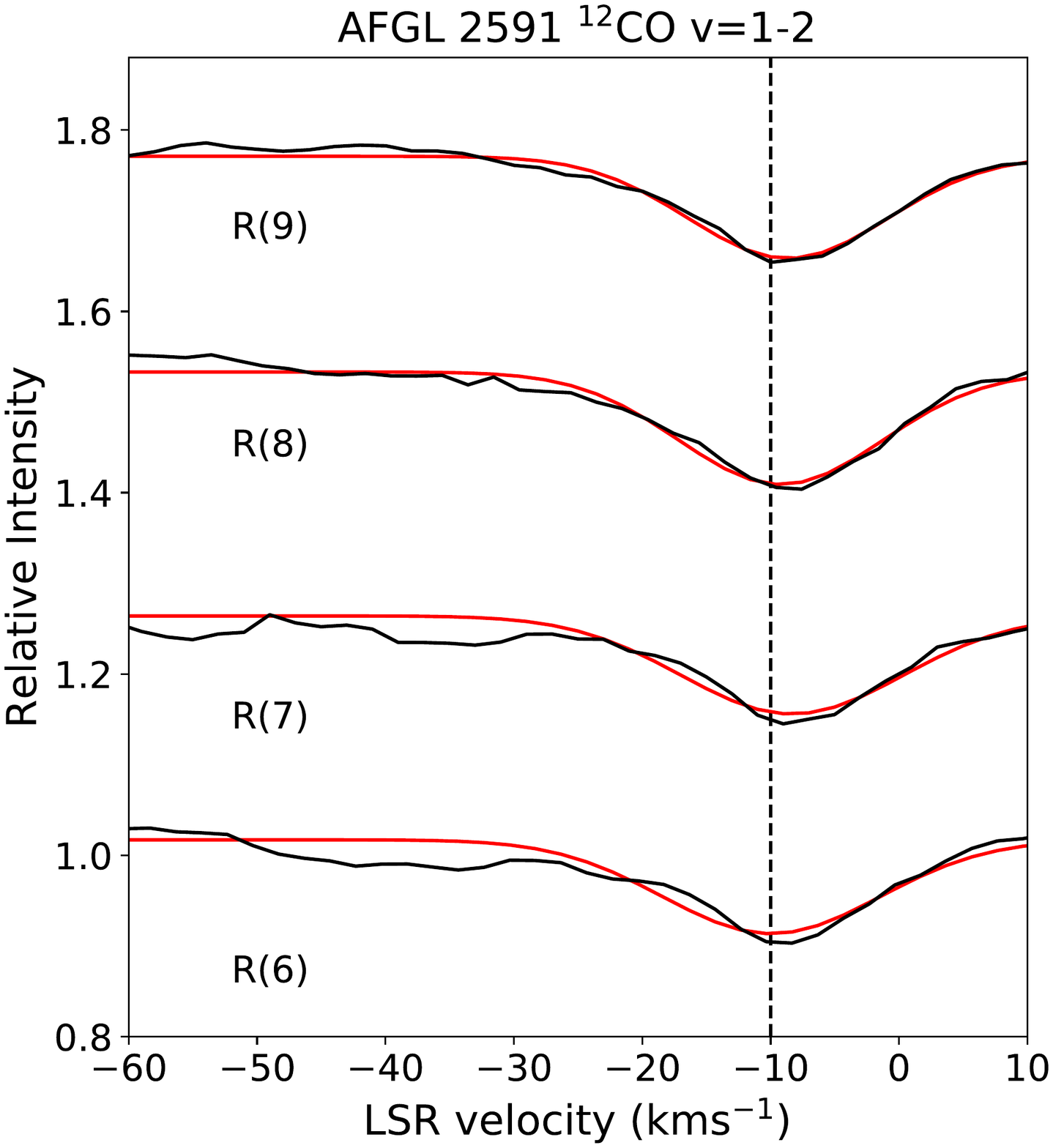}
\includegraphics[width=75mm,scale=1.5]{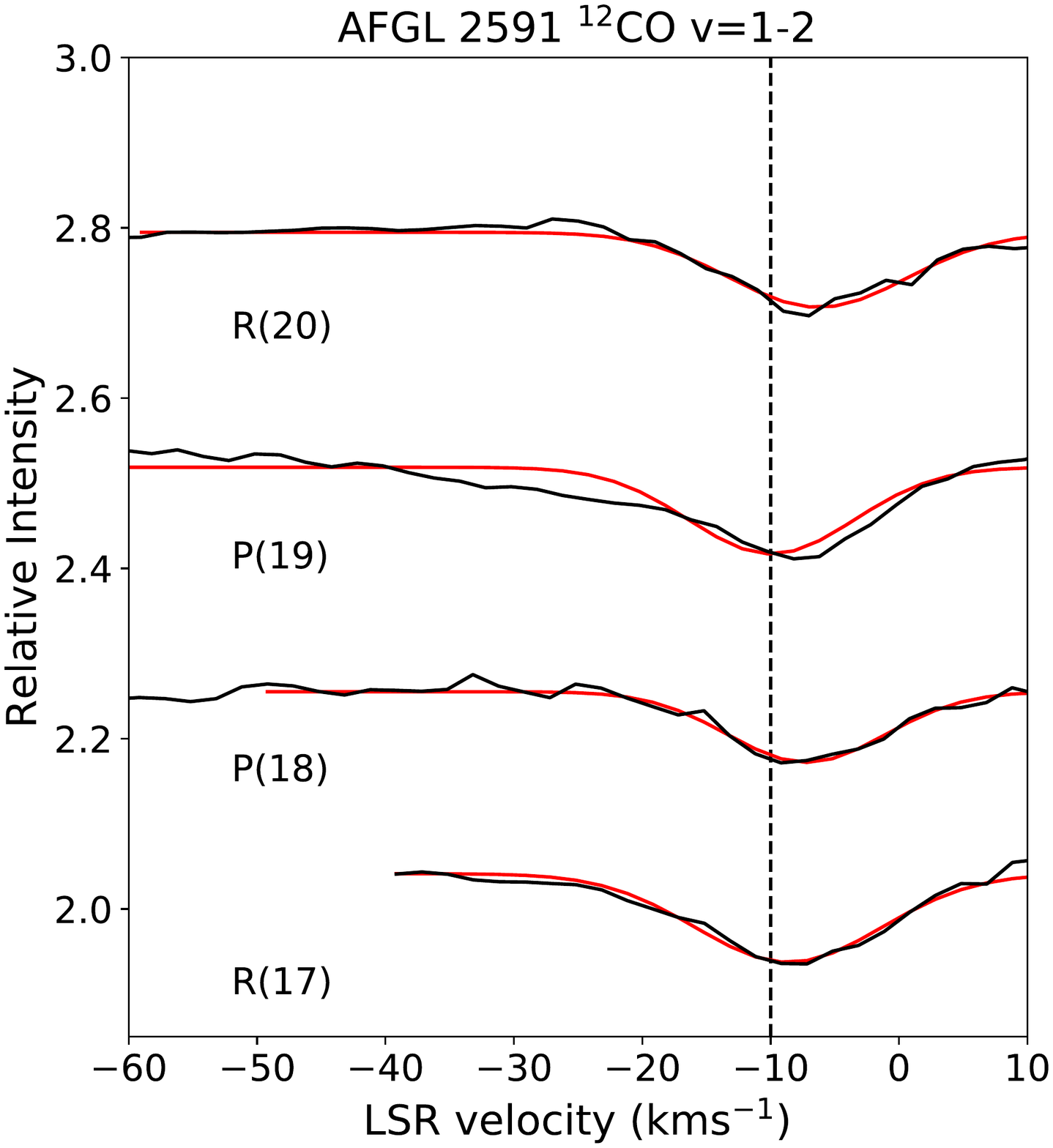}
\caption{Lines of $^{12}$CO v=1-2 and high J $^{13}$CO v=0-1 in AFGL 2591 detected with iSHELL at 4.5 ${\mu}m$. The blue dotted lines represent the individual velocity components and the red solid lines show the overall fit. The black dashed line denotes -10 kms$^{-1}$.}
\label{2591CO_prof}
\end{figure}

AFGL 2591 shows a complex structure, with five velocity components of $^{13}$CO detected. The scope of this paper will cover only the -10 kms$^{-1}$ component, which has been previously discussed by \citet{Barr2018}. These results are in agreement with a previous IR study of CO carried out by \citep{Mitchell1989}. We do not discuss the $^{12}$CO v=0-1 transitions as they are saturated up until J=9. The rotation diagram for $^{13}$CO at -10 kms$^{-1}$ is shown in Figure \ref{CO2591}. Furthermore, vibrationally excited $^{12}$CO is detected with a single temperature component at -10 kms$^{-1}$, and we derive a vibrational temperature of 623$\pm$292 K. A comparison of line profiles from low and high J level is displayed in Figure \ref{CO2591} and detected unblended transitions of $^{13}$CO v=0-1 and $^{12}$CO v=1-2 are presented in Figure \ref{2591CO_prof}. 

The rotation diagram of $^{13}$CO in \citet{Barr2018} and \citet{Mitchell1989} shows two temperature components at -10 kms$^{-1}$. These two temperature components are also distinct physical components. This can be seen by comparing low and high J transitions around the dashed line in Figure \ref{CO2591}. The line profile of the R(3) line is dominated by a narrow velocity component with a width of 1.5 kms$^{-1}$ and peak velocity of -9.2 kms$^{-1}$. This component disappears past J=6 which is reflected in the R(9) transition (Fig \ref{CO2591}) where we see a broad component uncovered with a width of 10.5 kms$^{-1}$ centred on a velocity of -12 kms$^{-1}$. The temperatures of the narrow and broad components are 49 K and 671 K, respectively.  Therefore we only show the hot component in the rotation diagram in Figure \ref{CO2591} and discuss only the high temperature component in the rest of the paper, the temperature of which is consistent with that of other species at the same velocity.

\subsubsection{CS}

\begin{figure}[h!]
\centering
\begin{tabular}{@{}cccc@{}}
\includegraphics[width=.45\textwidth]{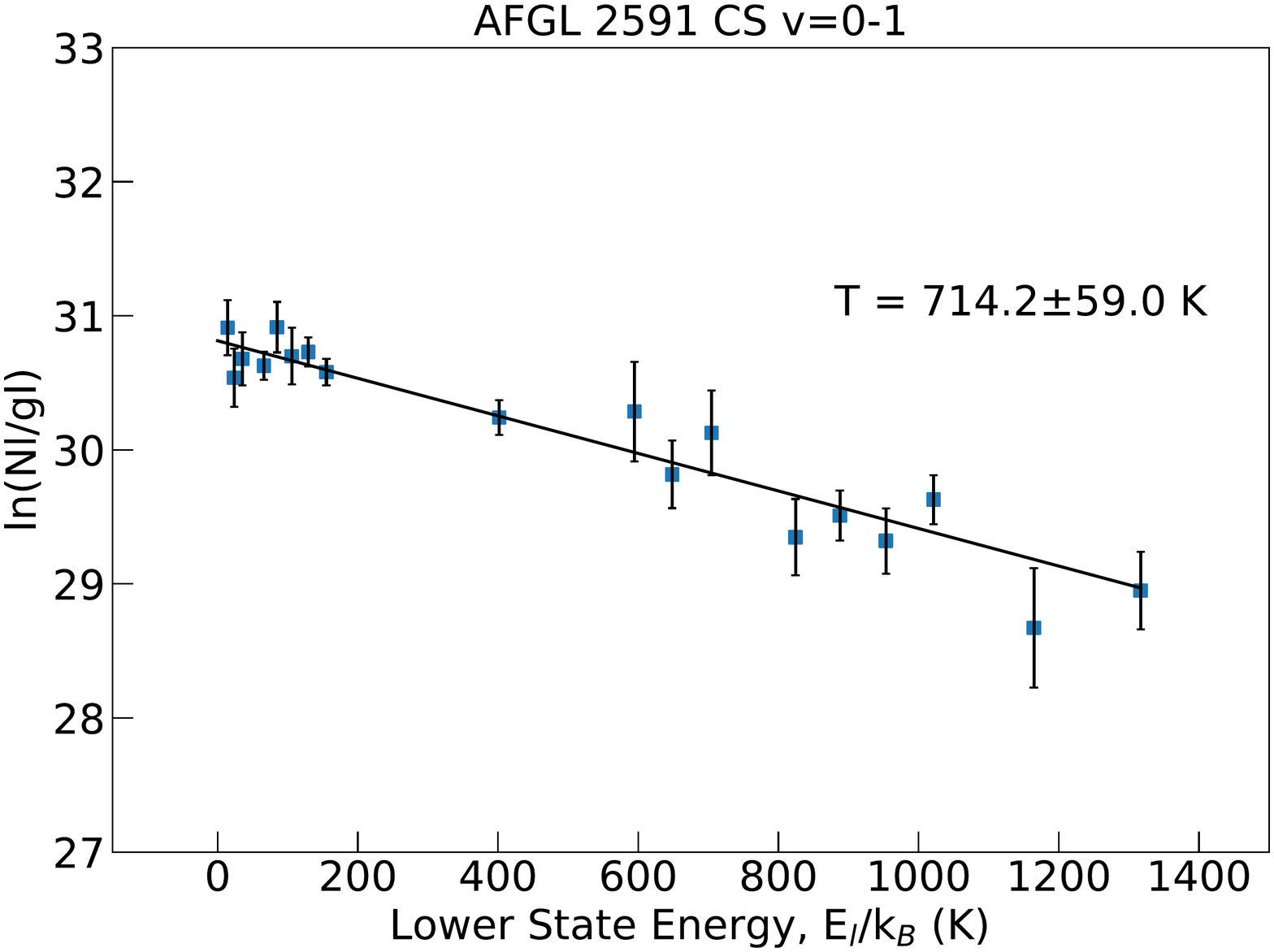} 
\includegraphics[width=.45\textwidth]{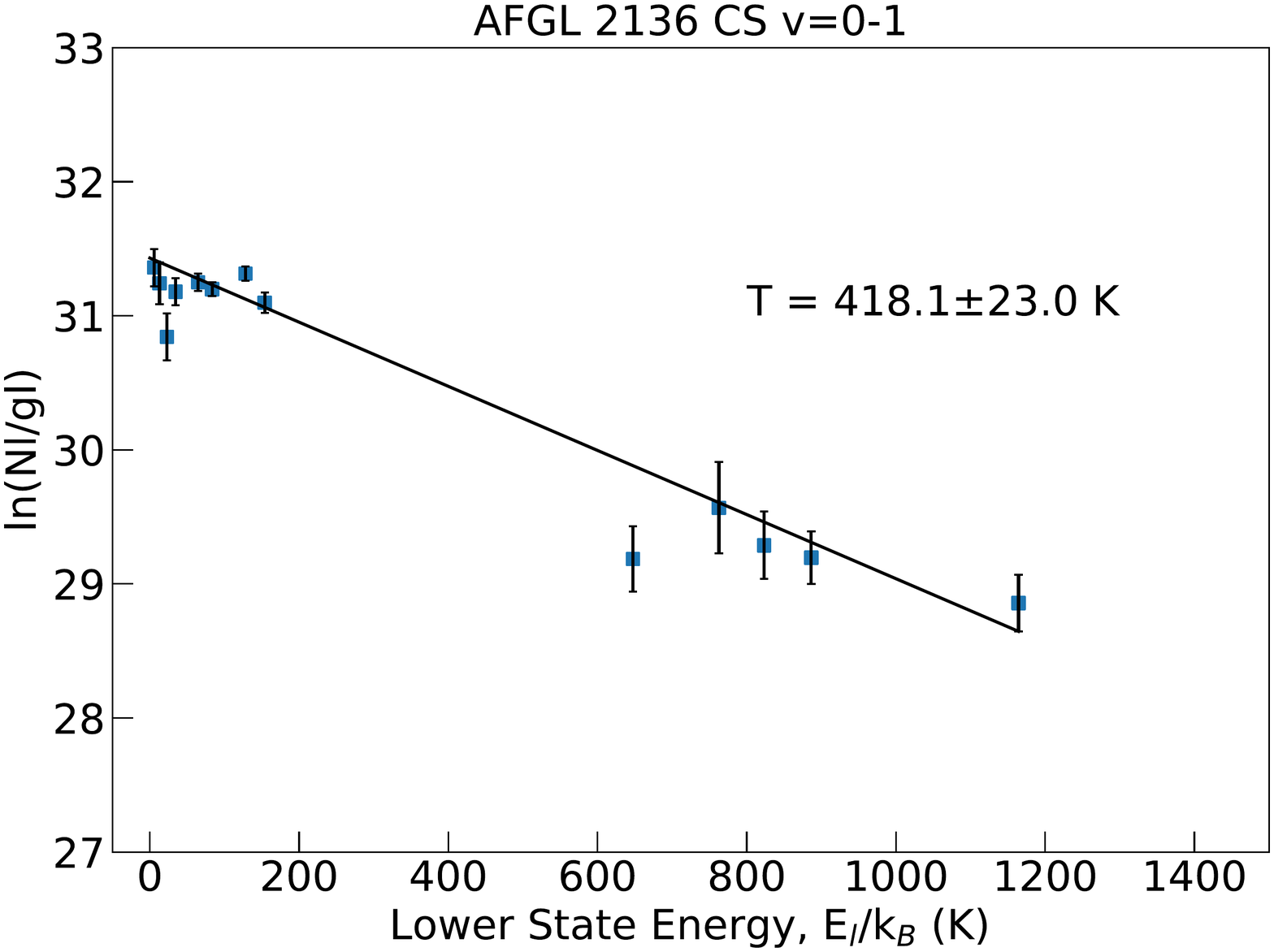} 
\end{tabular}
\caption{CS rotation diagrams for transitions in AFGL 2591 and AFGL 2136 detected with EXES. Rotational temperatures are shown.}
\label{CS}
\end{figure}

CS is detected towards AFGL 2591 and has been discussed extensively by \citet{Barr2018}. One velocity component is detected around -10 kms$^{-1}$ and corresponds with one of the velocity components of $^{13}$CO and the velocity of vibrationally excited $^{12}$CO. Figure \ref{CS} shows that the rotation diagram reveals a CS temperature of 714$\pm$59 K which is consistent with the hot component of $^{13}$CO at -10 kms$^{-1}$ and $^{12}$CO v=1-2. Line widths are in agreement for CS and hot $^{13}$CO for equivalent energy level. Therefore CS and CO trace the same region of the hot core.

\subsubsection{HCN}
 
20 lines of the $\nu_2$ band of HCN are detected in the v$_2$=0-1 transition with TEXES, spanning an energy range of 2500 K. From the rotation diagram, we derive a high temperature of 675$\pm$32 K. All line profile parameters are summarised in Tables \ref{sum2591} and \ref{sum2136}.

\begin{figure}[h!]
\centering
\begin{tabular}{@{}cccc@{}}
\includegraphics[width=.45\textwidth]{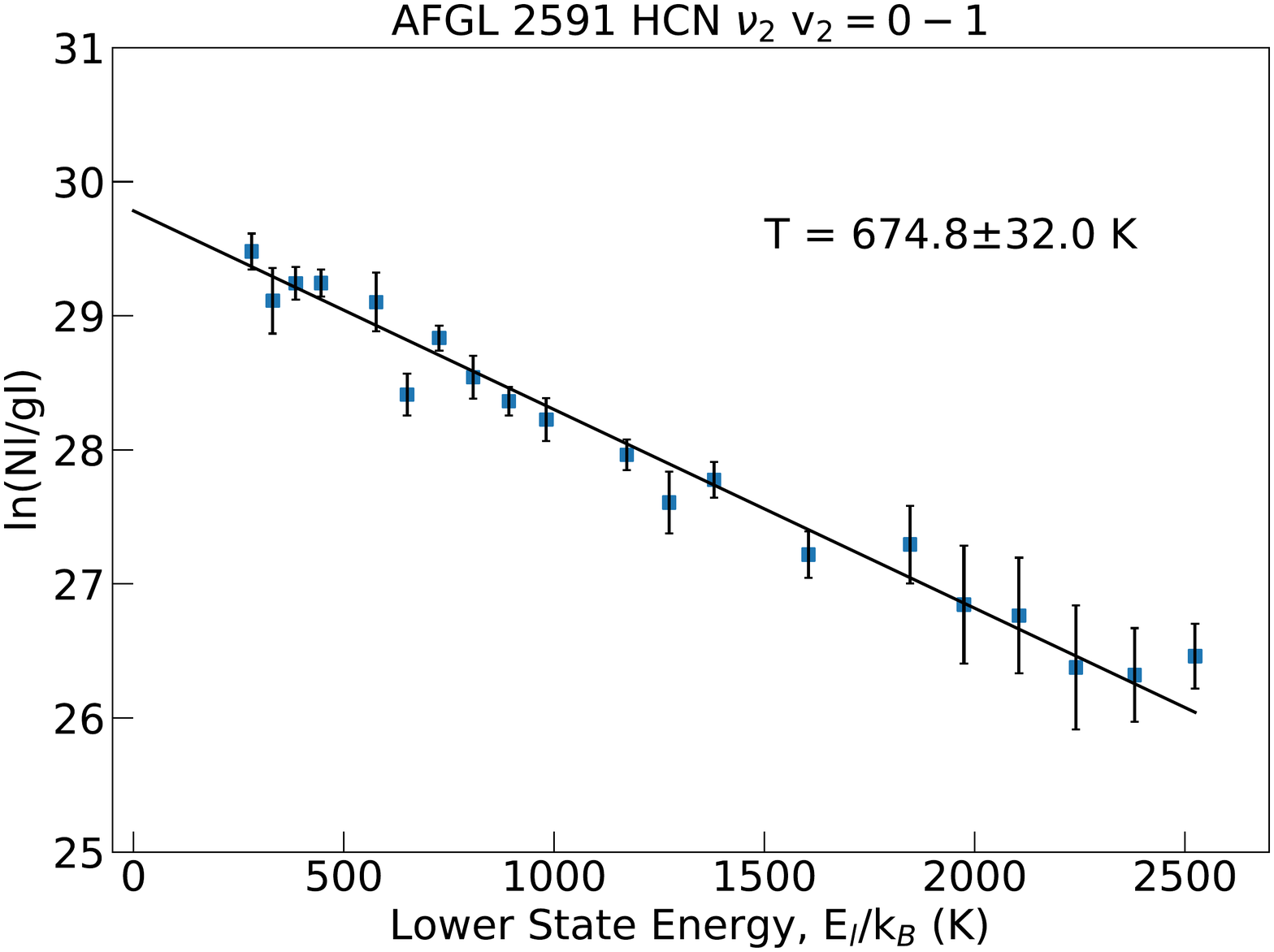} 
\includegraphics[width=.45\textwidth]{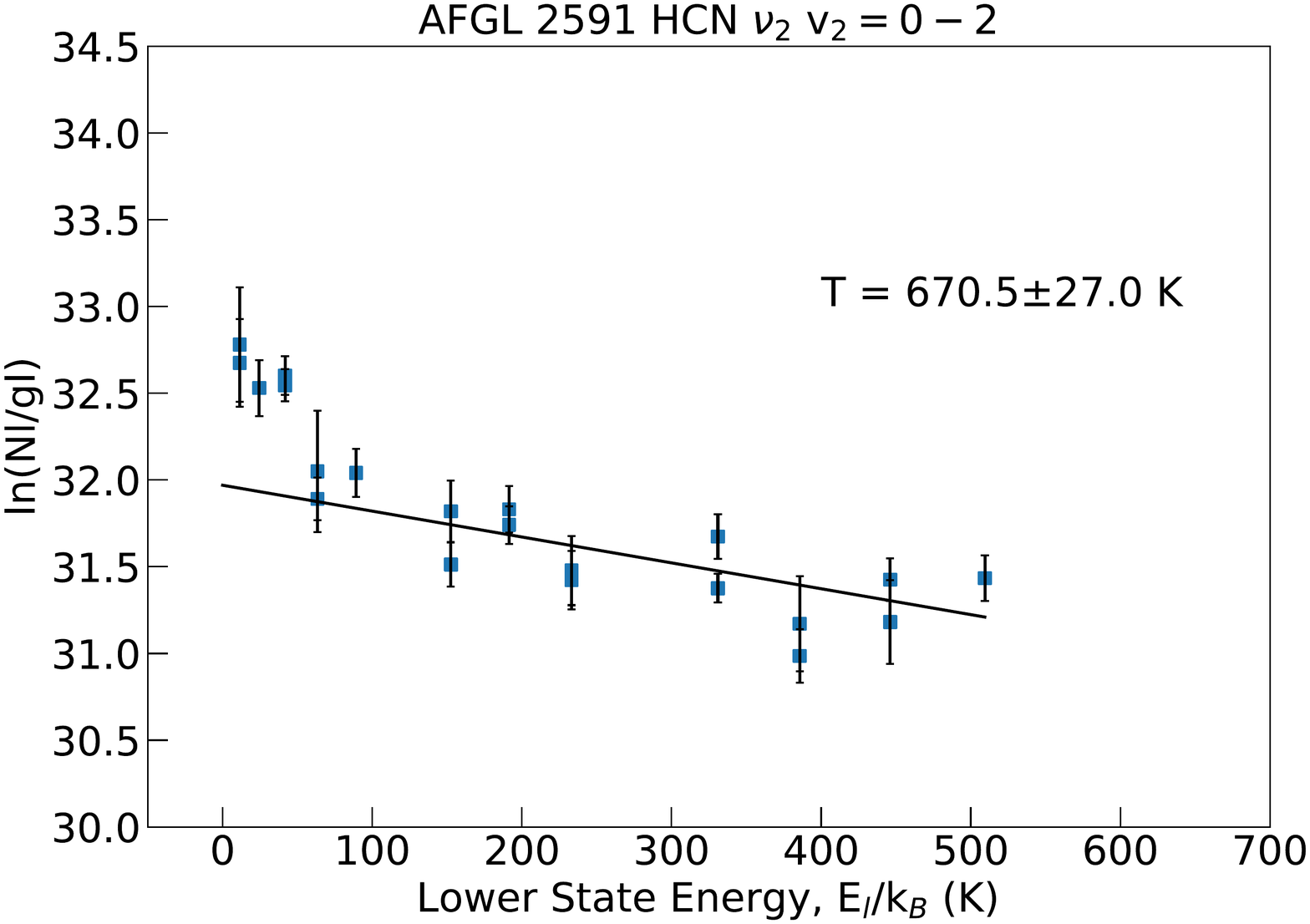} \\
\includegraphics[width=.45\textwidth]{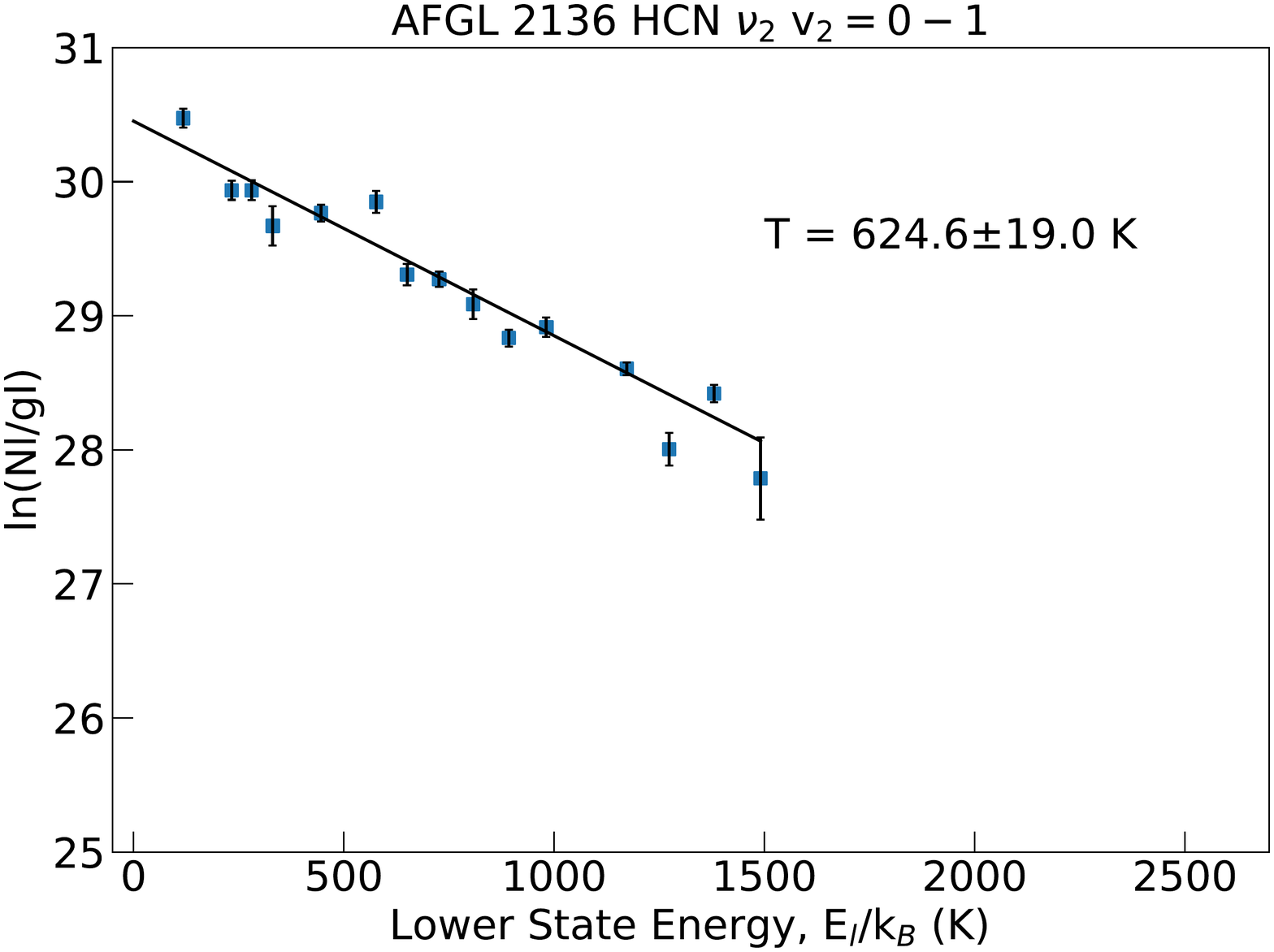} 
\includegraphics[width=.45\textwidth]{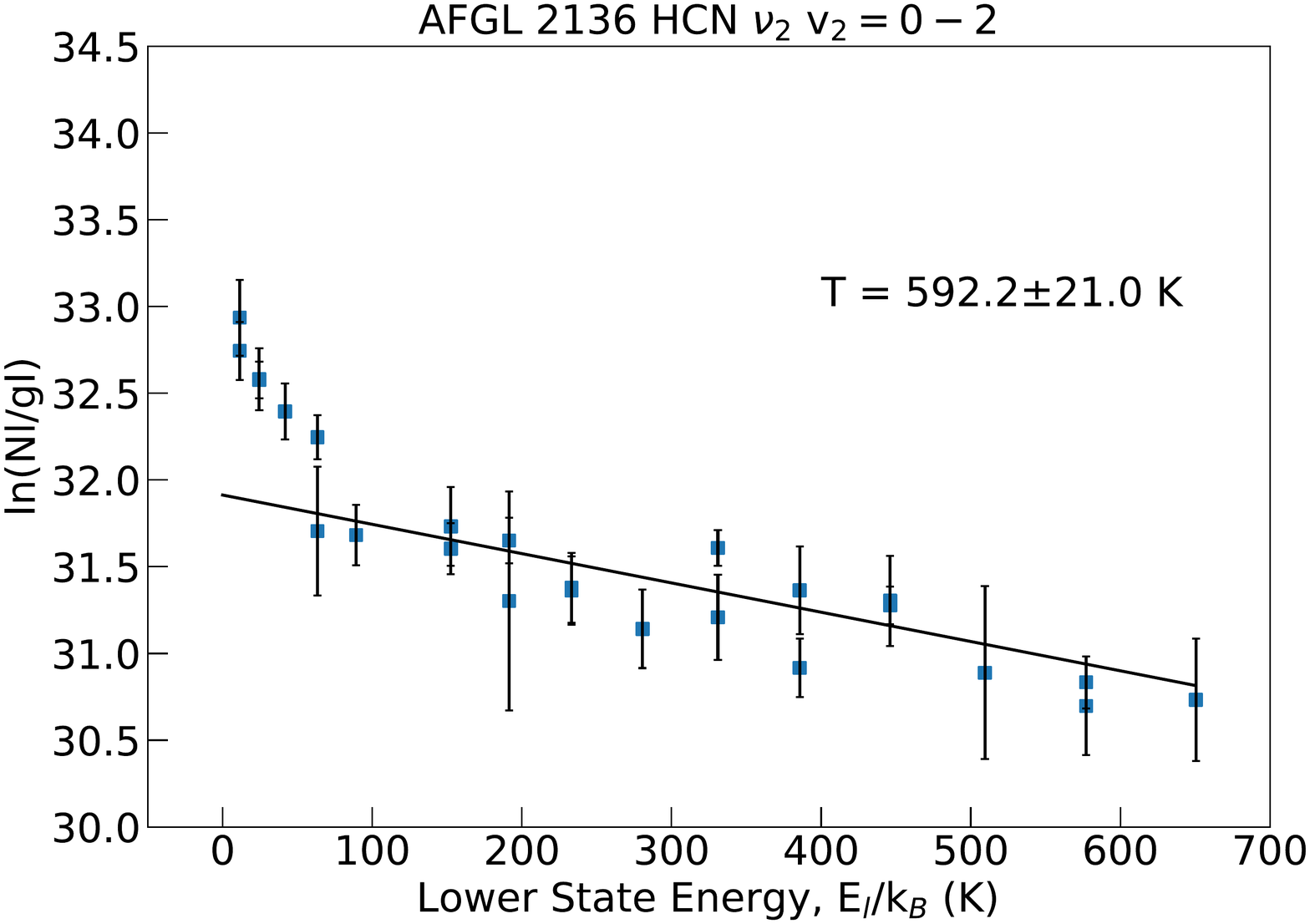} \\
\includegraphics[width=.45\textwidth]{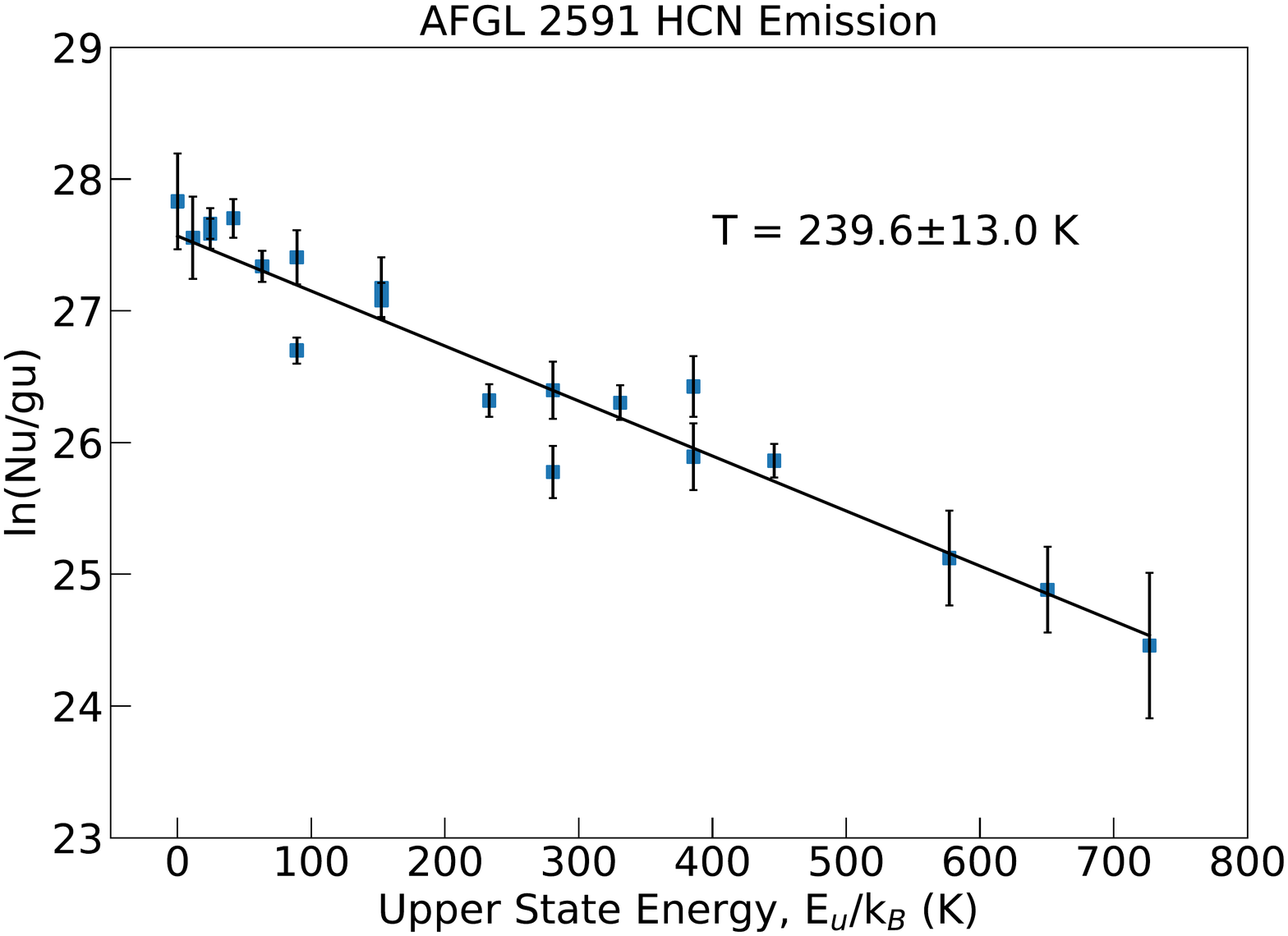}
\end{tabular}
\caption{HCN rotation diagrams for AFGL 2591 and AFGL 2136. The v$_2$=0-1 transition is detected at longer wavelengths with TEXES. The v$_2$=2-0 transition comes from the EXES data. The rotation diagram for the HCN $\nu_1$ band seen in emission is also shown which is detected with iSHELL.}
\label{HCN}
\end{figure}

A second vibrational transition of HCN is detected at shorter wavelengths with EXES, the v$_2$=0-2 transition in the $\nu_2$ band. From the 22 lines, the rotation diagram reveals the presence of a temperature gradient (Figure \ref{HCN}) by a curving of the rotation diagram at low J level, as line profiles are consistent from low to high J (Fig \ref{Lineprof2591}). Thus the physical conditions derived from the rotation diagram may not be reflective of the true values for this species. This is observed in the v$_2$=0-2 transition of HCN while it is not seen in the v$_2$=0-1 transition. This may be because, in the v$_2$=0-1 transition, only lines with J$\geq$7 are observed as our observations did not extend to long enough wavelengths. As no low J transitions are observed in the HCN v$_2$=0-1 band, we cannot ascertain the presence (or absence) of an upward curvature in the rotation diagram. Due to the non-linearity of the rotation diagrams, we derive physical conditions from a fit to the flat, or high energy, part of the rotation diagram. A temperature of 670$\pm$118 K is found, in good agreement with the rotational temperature derived for the v$_2$=0-1 transition of HCN, as well as CS and CO. The line widths of CO, CS and HCN are also in agreement.

The column density measured in the v$_2$=0-2 transition at 7 ${\mu}m$ is larger than the column density of the v$_2$=0-1 transition at 13 ${\mu}m$ by one order of magnitude, despite the fact that these two transitions both trace the ground state. Compared to the v$_2$=0-1 transition, the line widths agree but the velocity is blue-shifted by a few kms$^{-1}$, which is around the instrumental spectral resolution.

The R(0) line in the HCN v$_2$=0-2 band is offset in velocity from the other HCN lines and appears to be tracing cold foreground absorption. Hot HCN has been detected towards AFGL 2591 before via MIR absorption with Infrared Space Observatory Short Wavelength Spectrometer (ISO-SWS) \citep{Lahuis2000}. The derived temperature of 650$^{+75}_{-50}$ K is in agreement with what we find at high spectral resolution.

HCN has also been detected in emission in the L-band spectrum of AFGL 2591. 21 lines tracing the $\nu_1$ band are detected and the rotation diagram is also plotted in Figure \ref{HCN}. The velocity and line width are not in agreement with the HCN in absorption. The rotation diagram reveals a temperature of only 240$\pm$13 K, lower than the other two transitions seen in absorption.

\subsubsection{C$_2$H$_2$}

Several bands of C$_2$H$_2$ have been detected in AFGL 2591 with TEXES. Rotation diagrams have been constructed using statistical weights from the HITRAN database, which incorporates a factor of 3 difference between the statistical weights for the ortho- and para- states, due to nuclear spin degeneracy. Therefore the statistical weights of para- lines are 2J+1 however the statistical weights of the ortho- lines are 3(2J+1). We also use the partition function, Q, given in HITRAN, and incorporate the difference in statistical weights by taking Q(ortho) = 3/4 Q and Q(para) = 1/4 Q. Ortho- and para- states have been split up and treated as separate molecular species. All line parameters are summarised in Tables \ref{sum2591} and \ref{sum2136}.

\begin{figure}[htb]
\centering
\begin{tabular}{@{}cccc@{}}
\includegraphics[width=.45\textwidth]{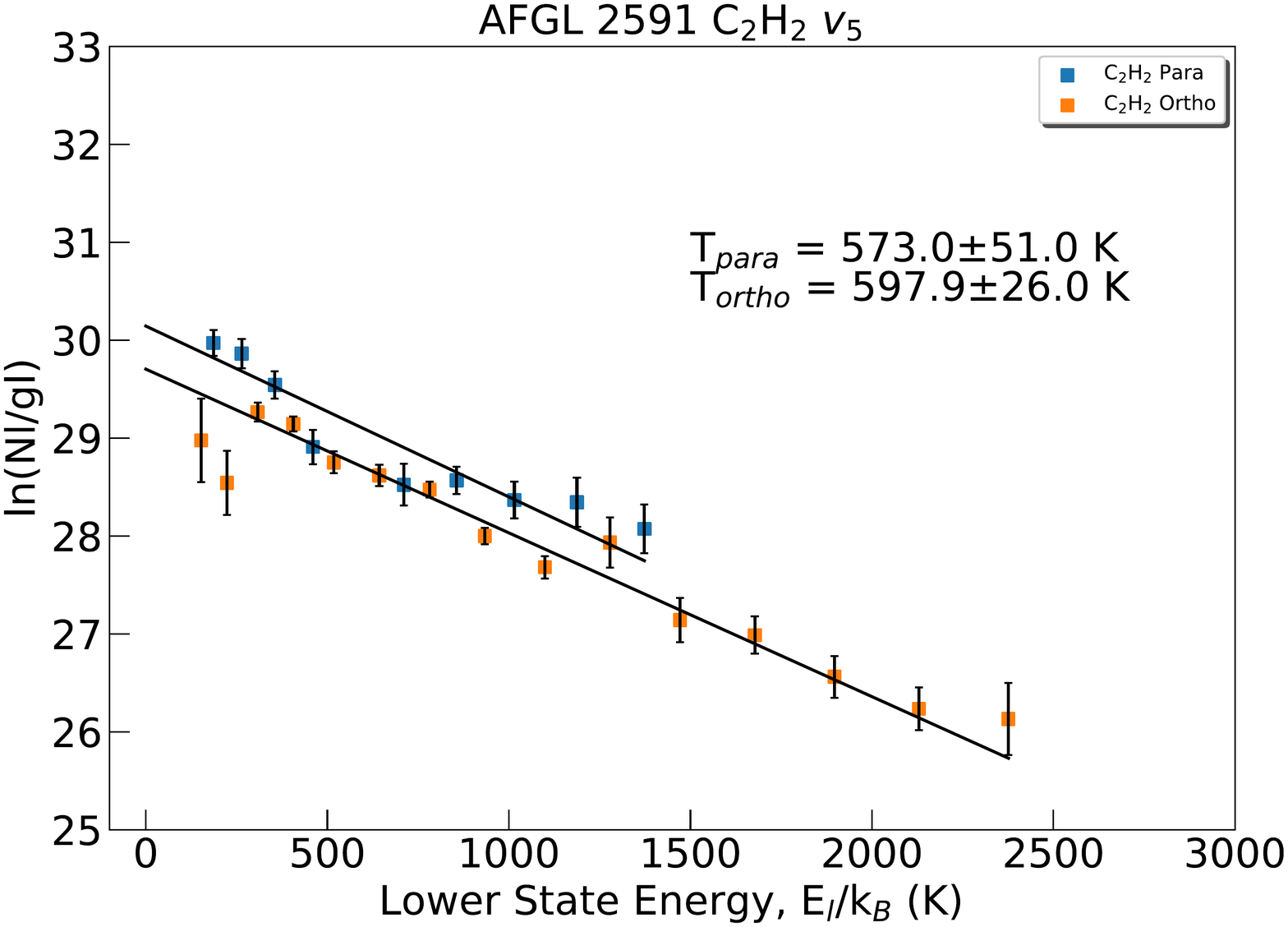}
\includegraphics[width=.45\textwidth]{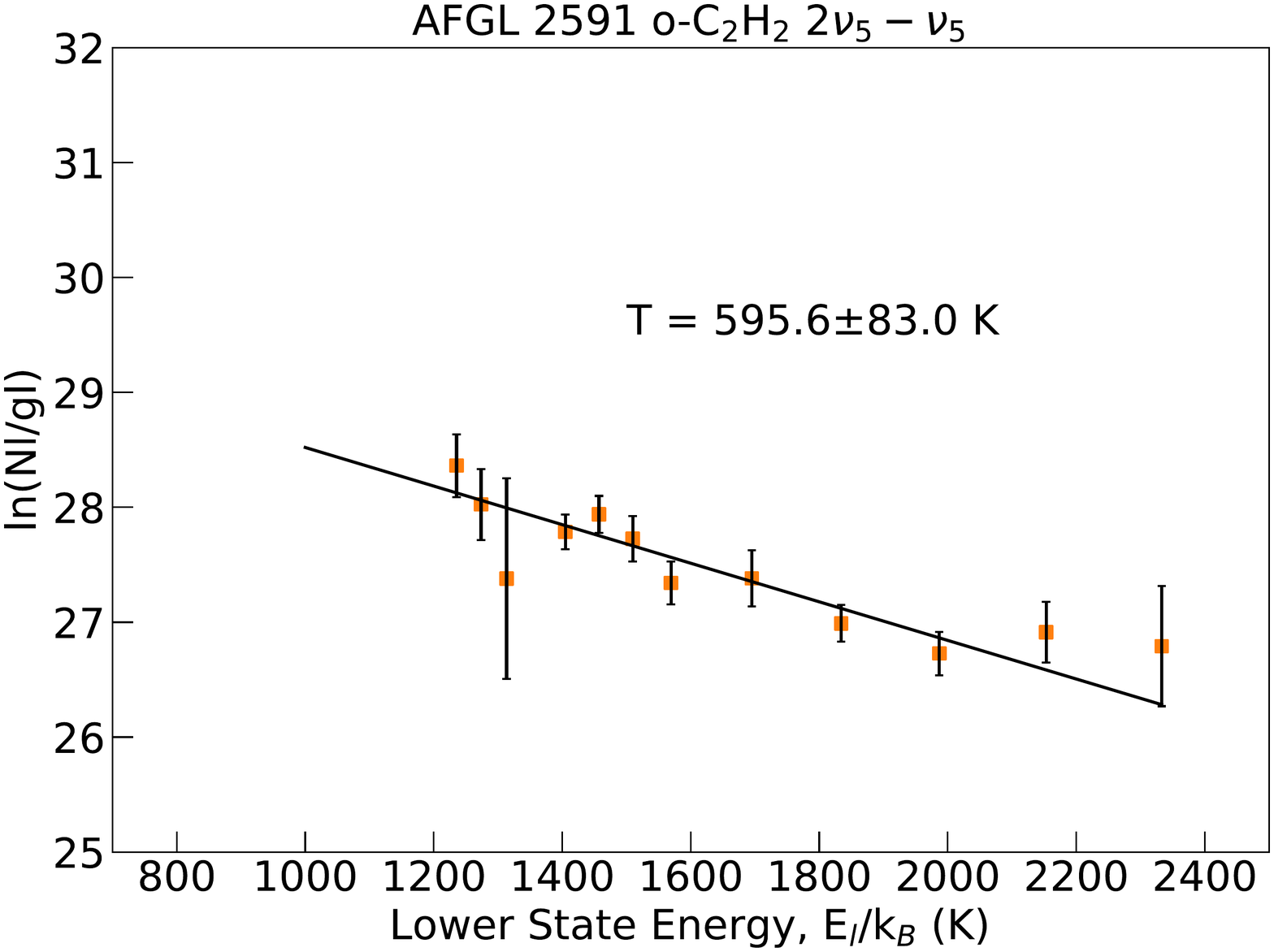} \\
\includegraphics[width=.45\textwidth]{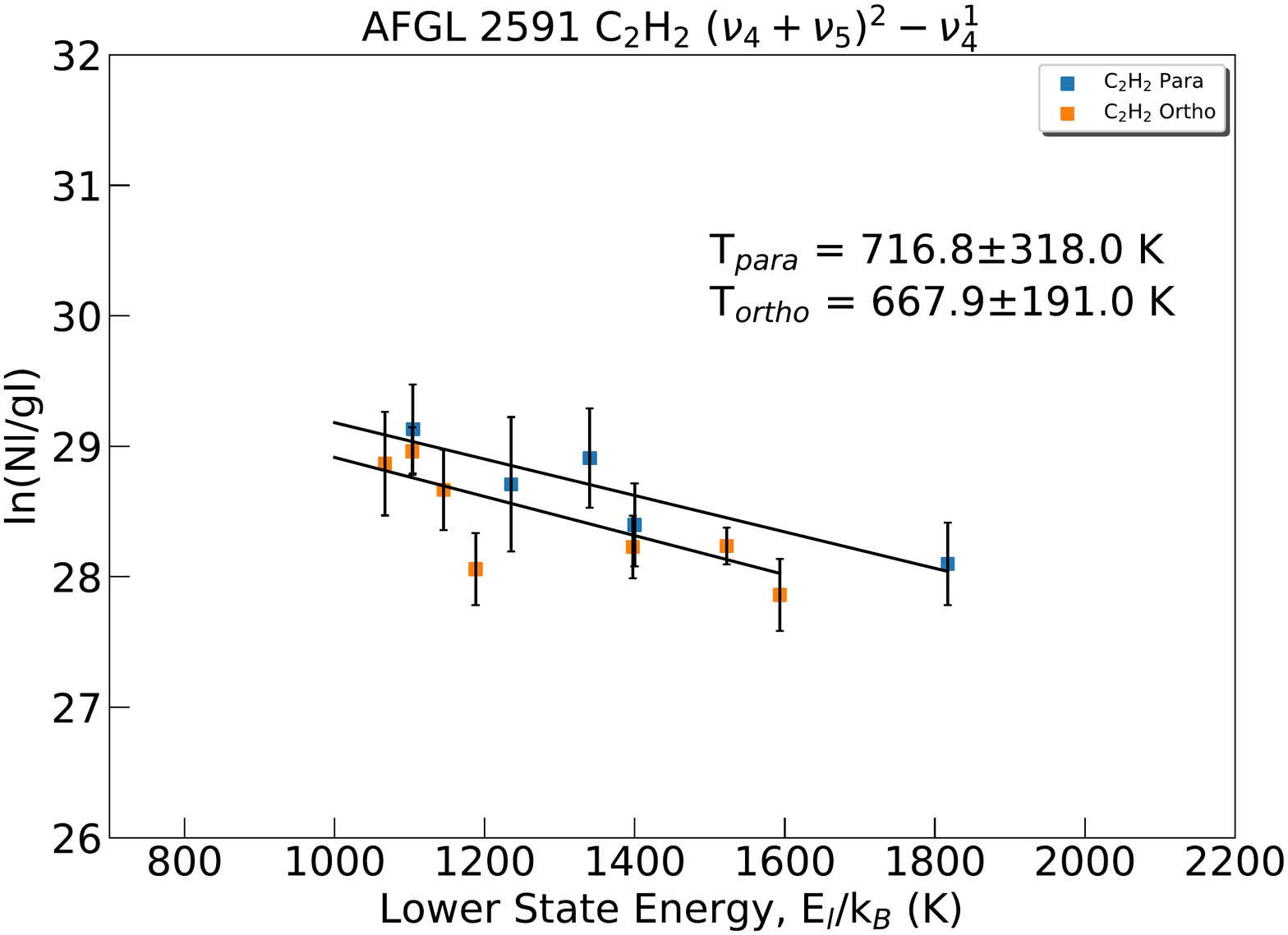}
\includegraphics[width=.45\textwidth]{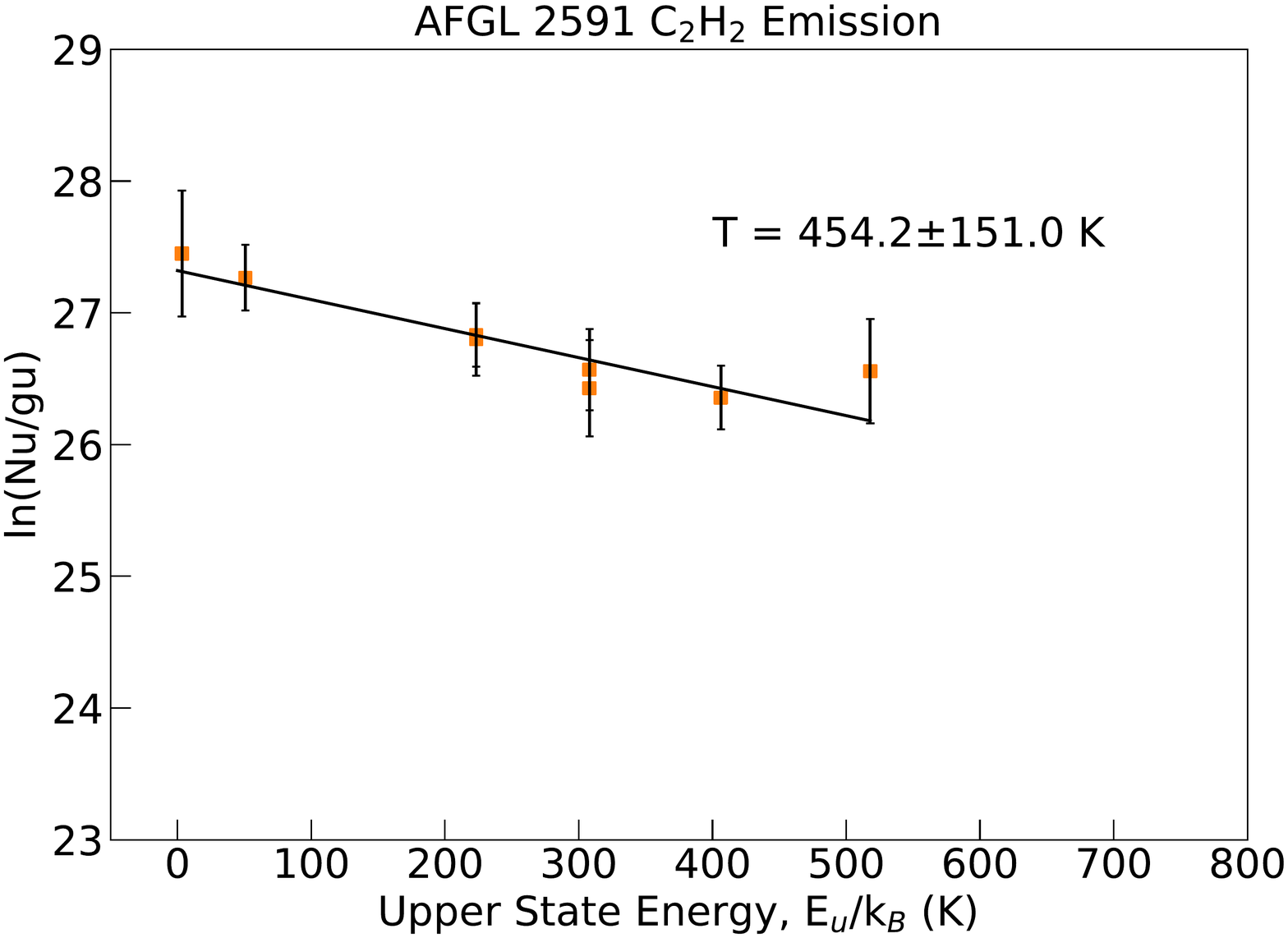}  \\ 
\end{tabular}
\caption{Acetylene rotation diagrams for the C$_2$H$_2$ bands detected in AFGL 2591 with TEXES. Ortho- and Para-C$_2$H$_2$ states are split into their corresponding ladders, and are shown as orange and blue respectively. The rotational temperatures for each are also given.}
\label{C2H2GL2591}
\end{figure}

In AFGL 2591, 26 lines of the $\nu_5$ fundamental band of C$_2$H$_2$ are detected. The temperature ($\simeq 600$ K) derived from the rotation diagrams is consistent with the other species. This suggests that o-C$_2$H$_2$ and p-C$_2$H$_2$ in this band come from the same part of the hot core. We derive an apparent ortho-to-para ratio (OPR) of 2.1$\pm$0.3, and this will be discussed further in section 5.2. The ortho- lines are seen up to higher J level, which is reflected in the rotation diagram.

Two vibrationally excited bands of C$_2$H$_2$ are marginally detected in AFGL 2591, the ($\nu_4+ \nu_5)^2 - \nu_4^1$ and the 2$\nu_5^2 - \nu_5^1$ band, probing the $\nu_4$ and $\nu_5$ levels respectively. The rotation diagrams are shown in Figure \ref{C2H2GL2591}. For the ($\nu_4+ \nu_5)^2 - \nu_4^1$ band, the line widths of p-C$_2$H$_2$ and o-C$_2$H$_2$ agree well however the velocity of p-C$_2$H$_2$ is slightly lower than o-C$_2$H$_2$. The parameters derived for this band have a large uncertainty due to the weakness of the lines.

For the vibrationally excited 2$\nu_5 - \nu_5$ band, we only observe o-C$_2$H$_2$ lines. We derive a rotational temperature of 596 K, in good agreement with the rotational temperature for the $\nu_5$ band. The line profiles of this vibrationally excited band are consistent with those of the $\nu_5$ fundamental band. 

The $\nu_2 + (\nu_4 + \nu_5)$ band of C$_2$H$_2$ was detected in emission in the L-band towards AFGL 2591. The lines are narrower than the other C$_2$H$_2$ lines we observe in absorption by 4 kms$^{-1}$. Also, the velocity is clearly offset from the v$_{lsr}$ of other species. The rotation diagram is shown in Figure \ref{C2H2GL2591}. The rotation diagram gives a temperature of 454$\pm$151 K. Only o-C$_2$H$_2$ is detected in emission as the p-C$_2$H$_2$ lines are too weak. 

The blended Q-branch of hot C$_2$H$_2$ has been detected in AFGL 2591 with ISO-SWS \citet{Lahuis2000}. The derived excitation temperature (900 K) is higher than what we observe with TEXES (Table \ref{sum2591}). This may reflect the limited spectral resolution of ISO-SWS - hampering extraction of physical conditions from this blended line - and the severe fringing inherent to this instrument.

\subsubsection{NH$_3$}

For NH$_3$, we also treat ortho- and para- states as separate molecular species. We also take different partition functions as in the case for C$_2$H$_2$. For NH$_3$ the ortho states have K as a multiple of 3, and the para states have K=3n$\pm$1. The statistical weights of the o-NH$_3$ lines are a factor of 2 higher than the p-NH$_3$ lines \citep{Simeckova2006}. Since there are twice as many options for J for para states, this results in an equal number of ortho and para molecules. Thus we implement the partition functions as Q(ortho) = Q(para) = Q/2, where Q is the partition function for NH$_3$ given in HITRAN. For NH$_3$ the value of the OPR for high temperatures is equal to 1 \citep{Faure2013}. 

For AFGL 2591, NH$_3$ is harder to detect than in AFGL 2136. This reflects the poorer data quality due to fringing and lower signal to noise. AFGL 2136 was observed with TEXES on Gemini and used Orion BN as a standard star to improve atmospheric and standing wave correction, while AFGL 2591 was observed with the IRTF without a standard star. Since the temperature that we measure for NH$_3$ gas is hotter than in AFGL 2136, the energy spreads out over more, weaker transitions. Then, since the column densities of NH$_3$ derived for the two sources are the same, the lines in AFGL 2591 will be weaker than those in AFGL 2136. 

\begin{figure}[h!]
\centering
\begin{tabular}{@{}cccc@{}}
\includegraphics[width=.45\textwidth]{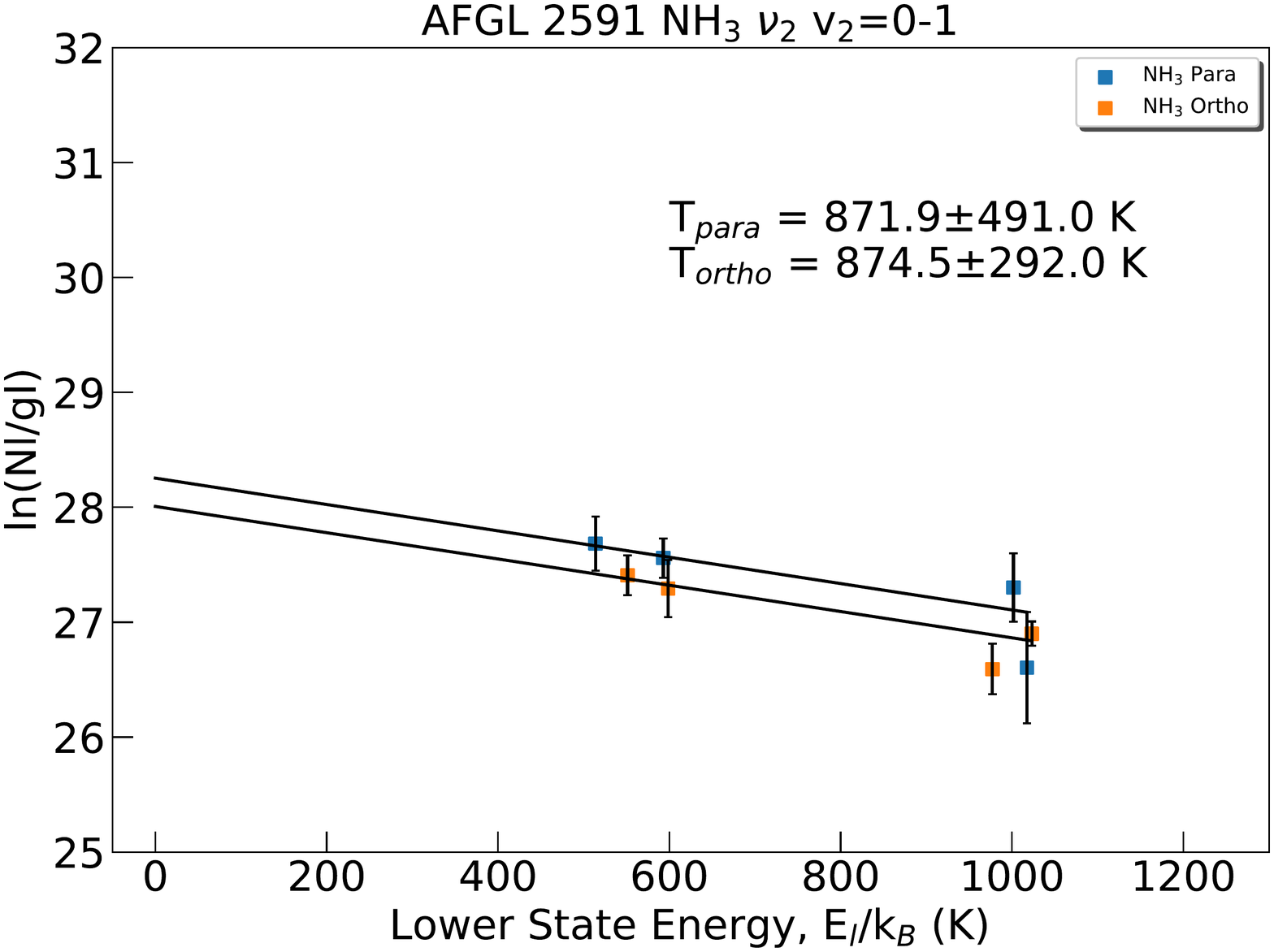} 
\includegraphics[width=.45\textwidth]{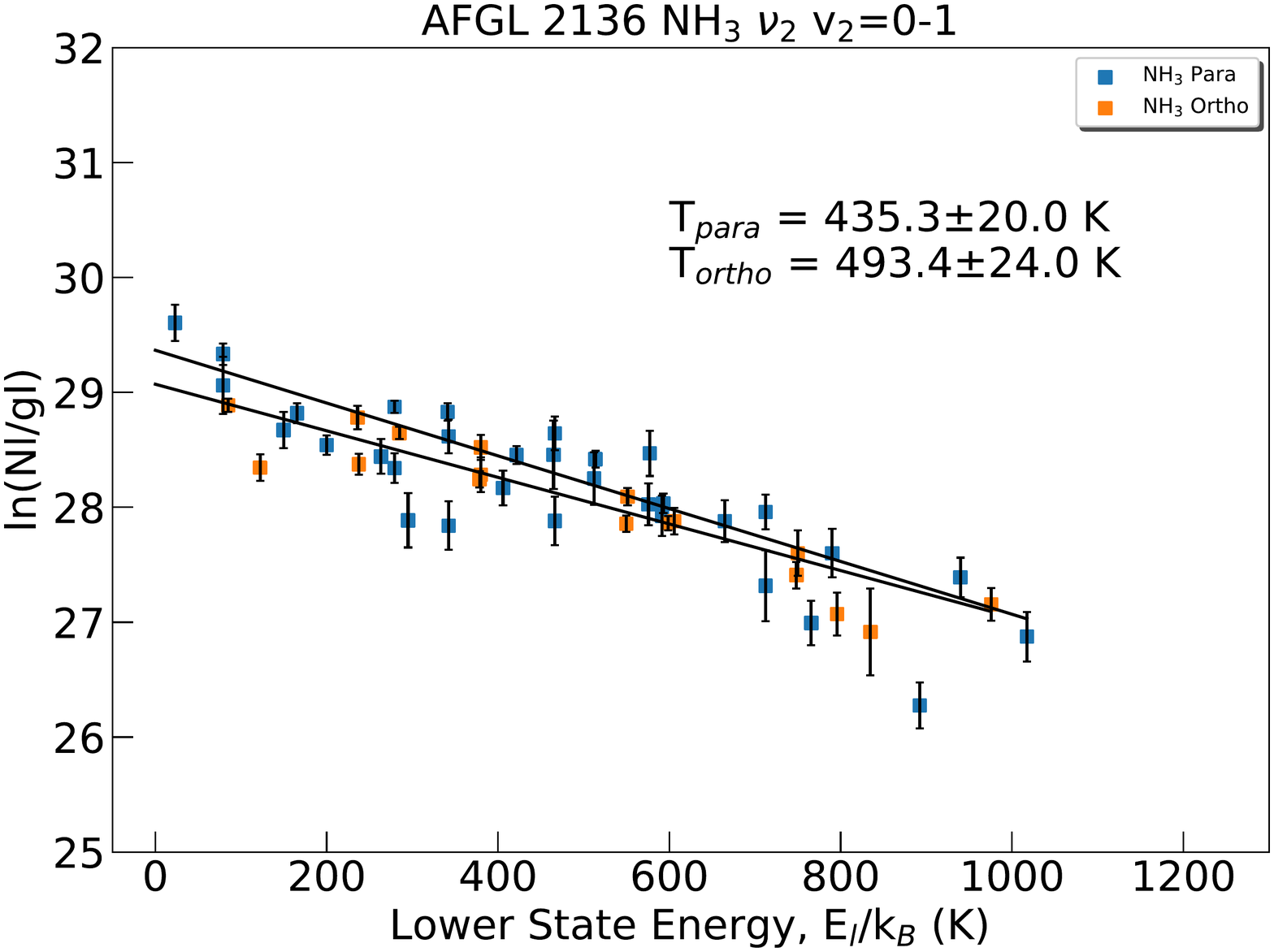}
\end{tabular}
\caption{Rotation diagrams of the v$_2$=0-1 transition in the $\nu_2$ band of NH$_3$ in AFGL 2591 and AFGL 2136 detected with TEXES. Both ortho- and para- states are detected and they are plotted in orange and blue respectively.}
\label{NH3}
\end{figure}

Both o-NH$_3$ and p-NH$_3$ lines are detected in the v$_2$=0-1 transition of the $\nu_2$ band. The peak velocities are in agreement, however the line width of o-NH$_3$ is almost twice as large as p-NH$_3$. Temperatures are in agreement and a high temperature of around 875 K is derived. An OPR=1$\pm$0.4 is found for AFGL 2591, consistent with the high rotational temperature. The rotation diagram is shown in Figure \ref{NH3}.

\subsection{AFGL 2136}

\subsubsection{CO}

AFGL 2136 shows a much simpler morphology compared to AFGL 2591, with only two velocity components of CO observed, which correspond to two separate physical components. For $^{13}$CO, one velocity component is detected at 22.3$\pm$0.2 kms$^{-1}$ with a line width of 3.4$\pm$0.7 kms$^{-1}$. This velocity component is superimposed on a broader underlying velocity component centred at 26.5$\pm$0.3 kms$^{-1}$, which has a width of 13.2$\pm$0.9 kms$^{-1}$. In C$^{18}$O, the narrow and broad velocity components are also observed, with line widths and velocities consistent with $^{13}$CO. $^{12}$CO v=0-1 lines are saturated for J$\lesssim6$. 

\begin{figure}[h!]
\centering
\begin{tabular}{@{}cccc@{}}
\includegraphics[width=0.45\textwidth]{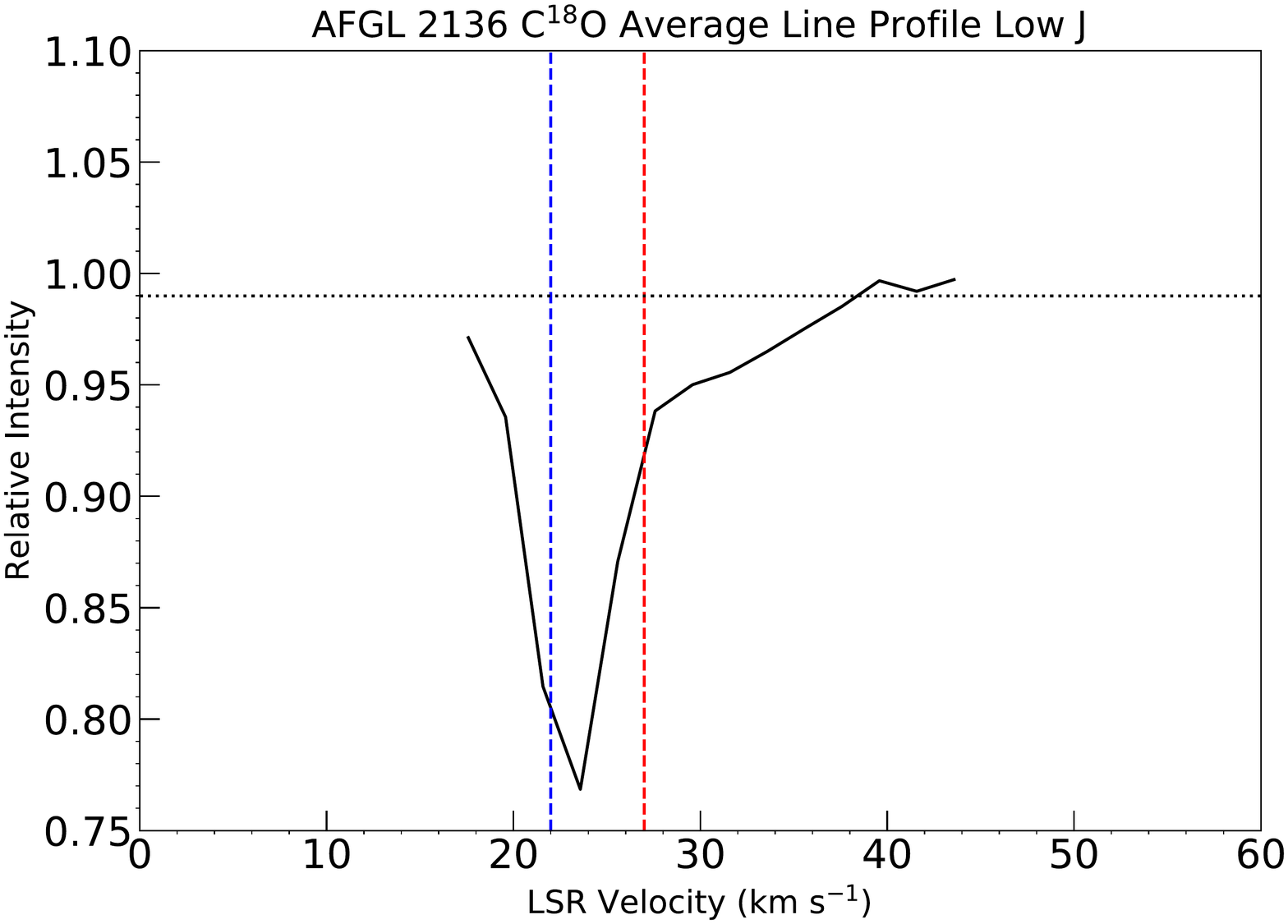}
\includegraphics[width=0.45\textwidth]{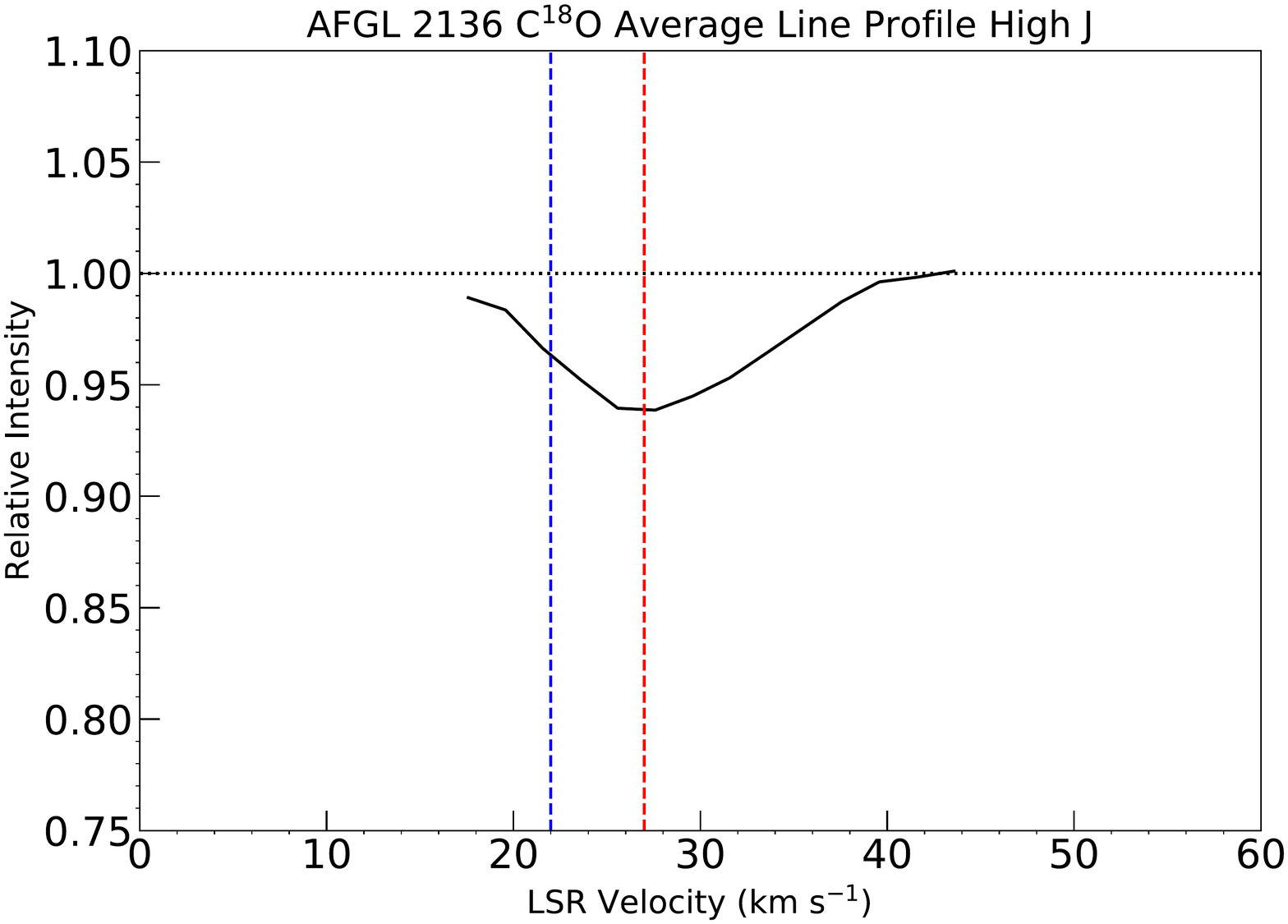} \\ 
\includegraphics[width=.45\textwidth]{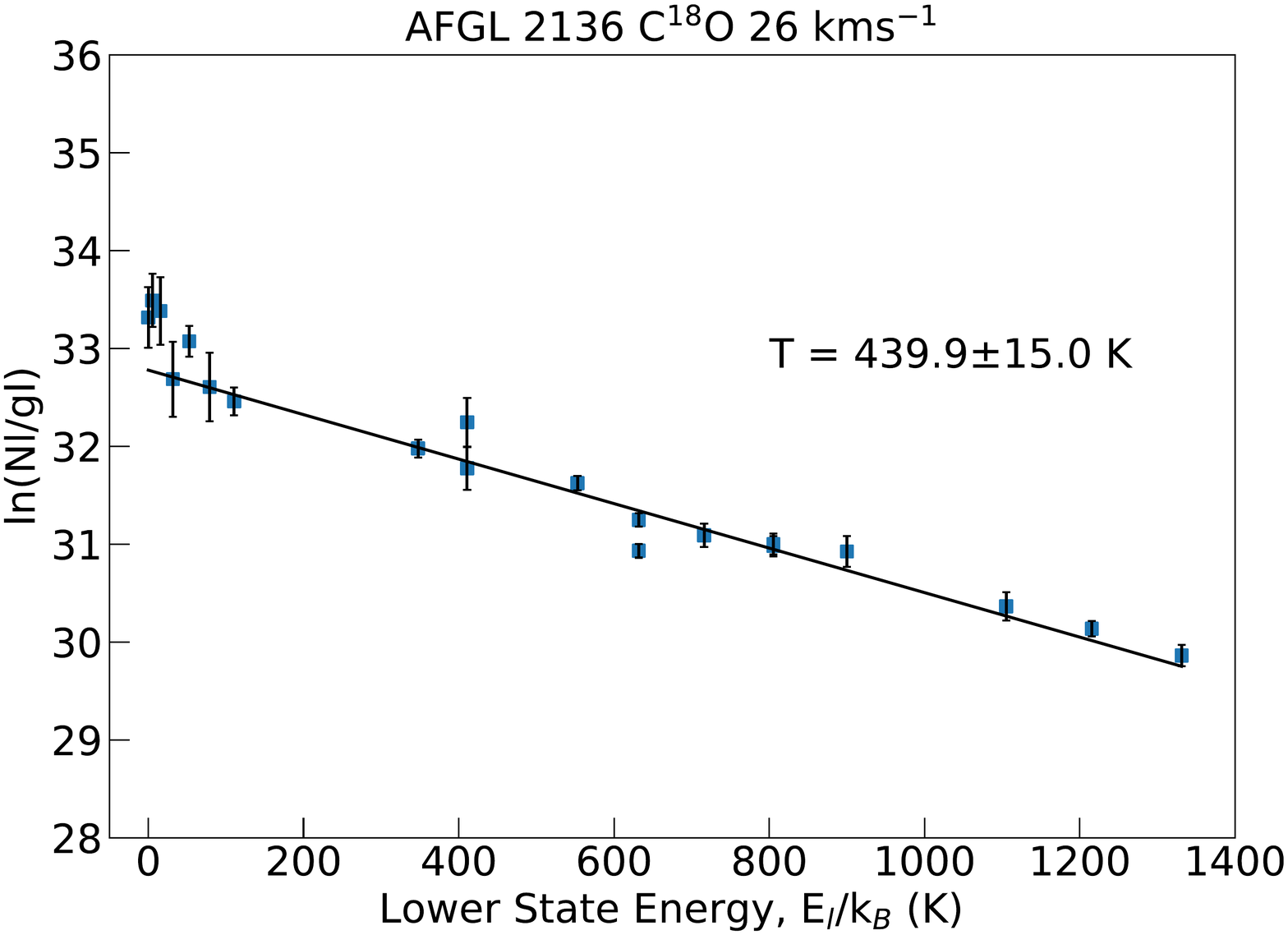}  
\includegraphics[width=.45\textwidth]{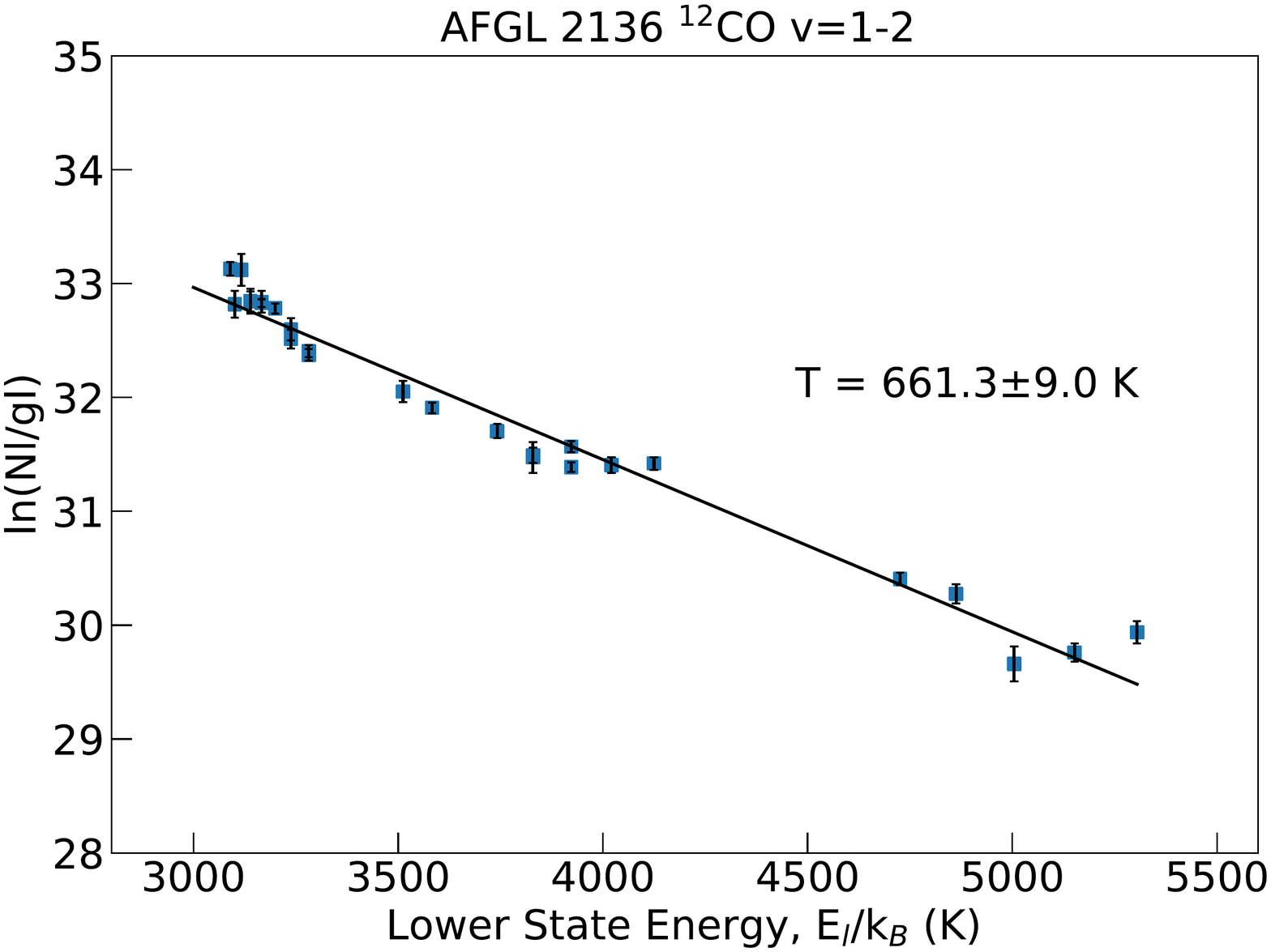} \\ 
\includegraphics[width=.45\textwidth]{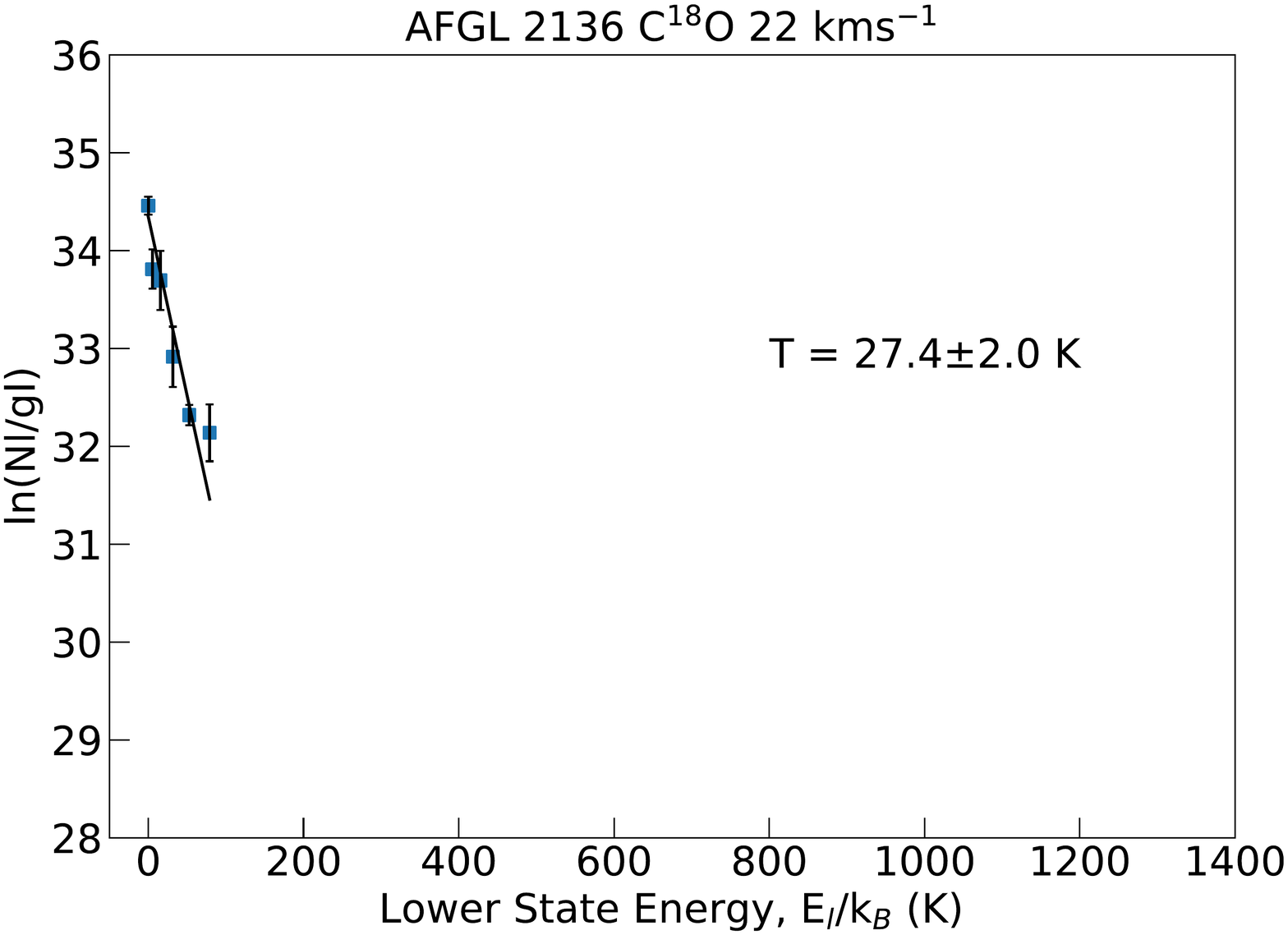}  
\includegraphics[width=.45\textwidth]{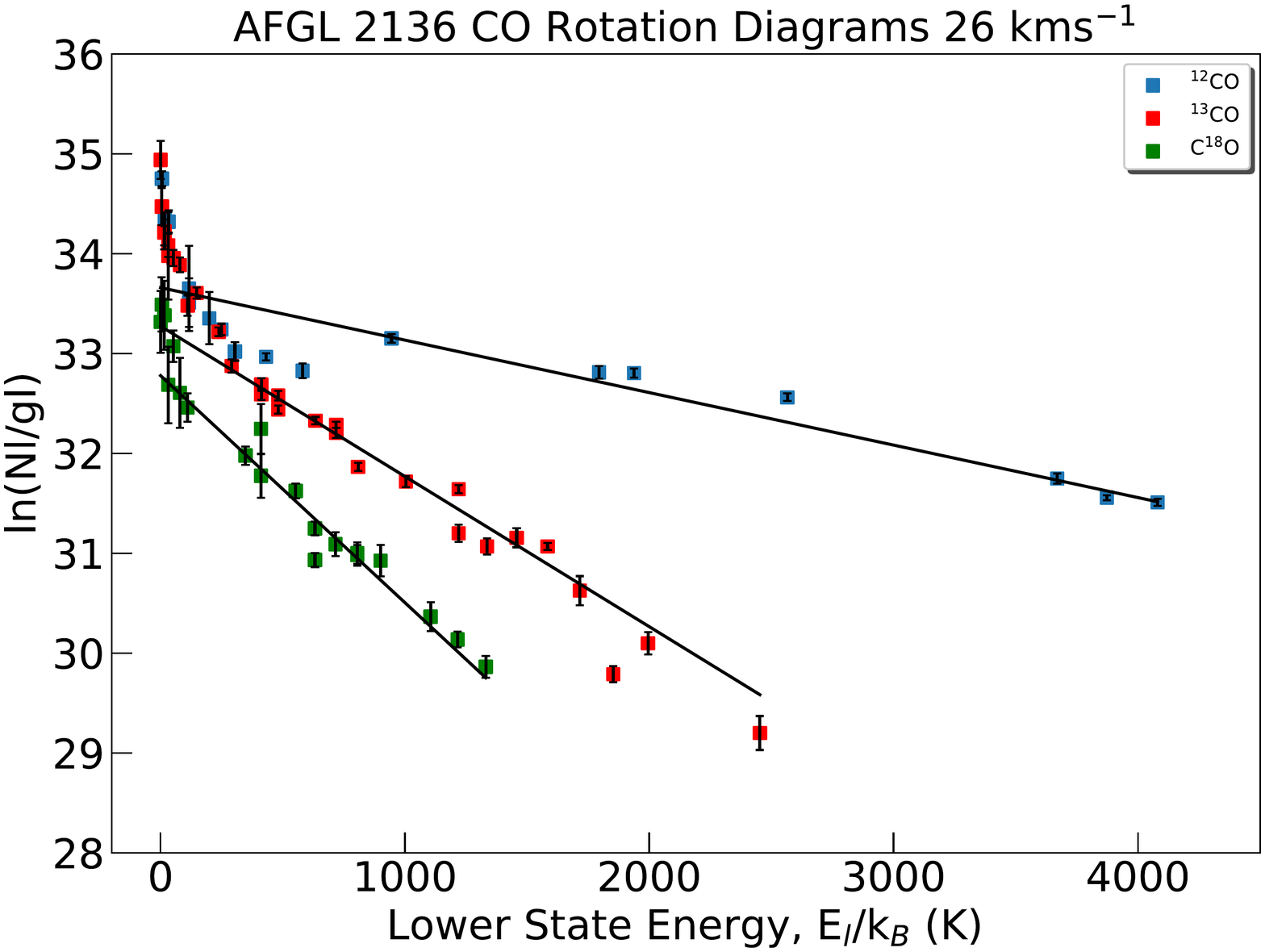}  
\end{tabular}
\caption{C$^{18}$O v=0-1 rotation diagrams and average line profiles for high and low J, along with the rotation diagram for $^{12}$CO v=1-2 in AFGL 2136. An additional plot for the hot component in AFGL 2136 is shown with $^{12}$CO v=0-1, $^{13}$CO v=0-1 and C$^{18}$O v=0-1 rotation diagrams plotted on the same panel. Blue and red dashed lines correspond to the systemic velocity of the source and velocity of IR observations, respectively, which are 22 and 27 kms$^{-1}$, respectively. The black dotted line denotes the continuum level}
\label{CO2136}
\end{figure}

Separate rotation diagrams are made for each velocity component. Figure \ref{CO2136} shows the rotation diagrams for the broad component seen in $^{12}$CO, $^{13}$CO and C$^{18}$O plotted in the same panel. The four lines in the $^{12}$CO v=0-1 rotation diagram between 1000 and 3000 K are overproduced since they are difficult to disentangle from the component at 22 kms$^{-1}$. For $^{12}$CO v=0-1, the transitions are optically thick since the optical depth at line centre is $>1$.

This is also the case for $^{13}$CO, where low J lines of  $^{13}$CO and $^{12}$CO overlap in the rotation diagram. For $E_l > 500$ K the $^{12}$CO and $^{13}$CO rotation diagrams start to separate, however the higher J lines for $^{13}$CO remain to have an optical depth above 2/3.

\begin{figure}[h!]
\centering
\begin{tabular}{@{}cccc@{}}
\includegraphics[width=0.25\textwidth]{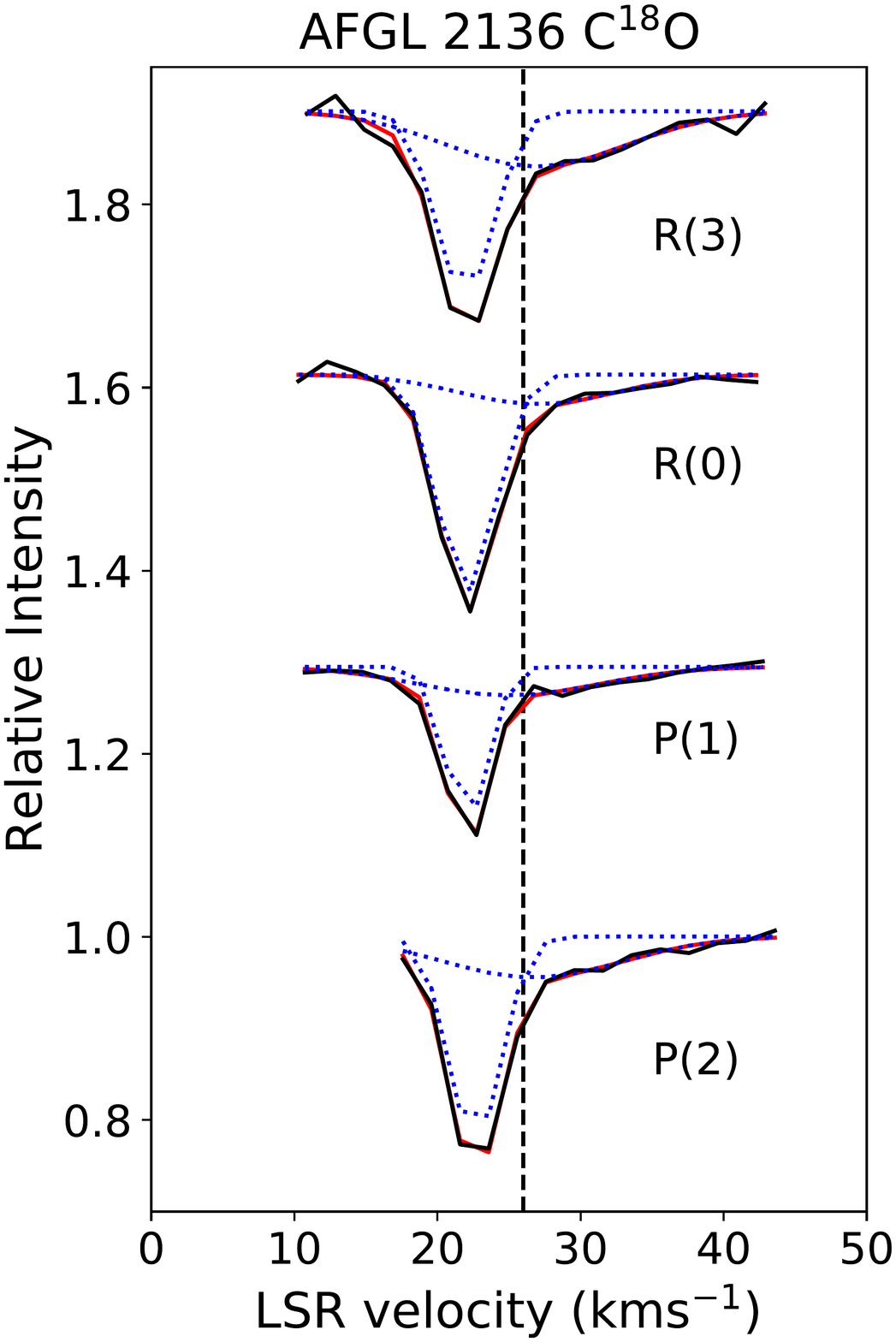}
\includegraphics[width=0.25\textwidth]{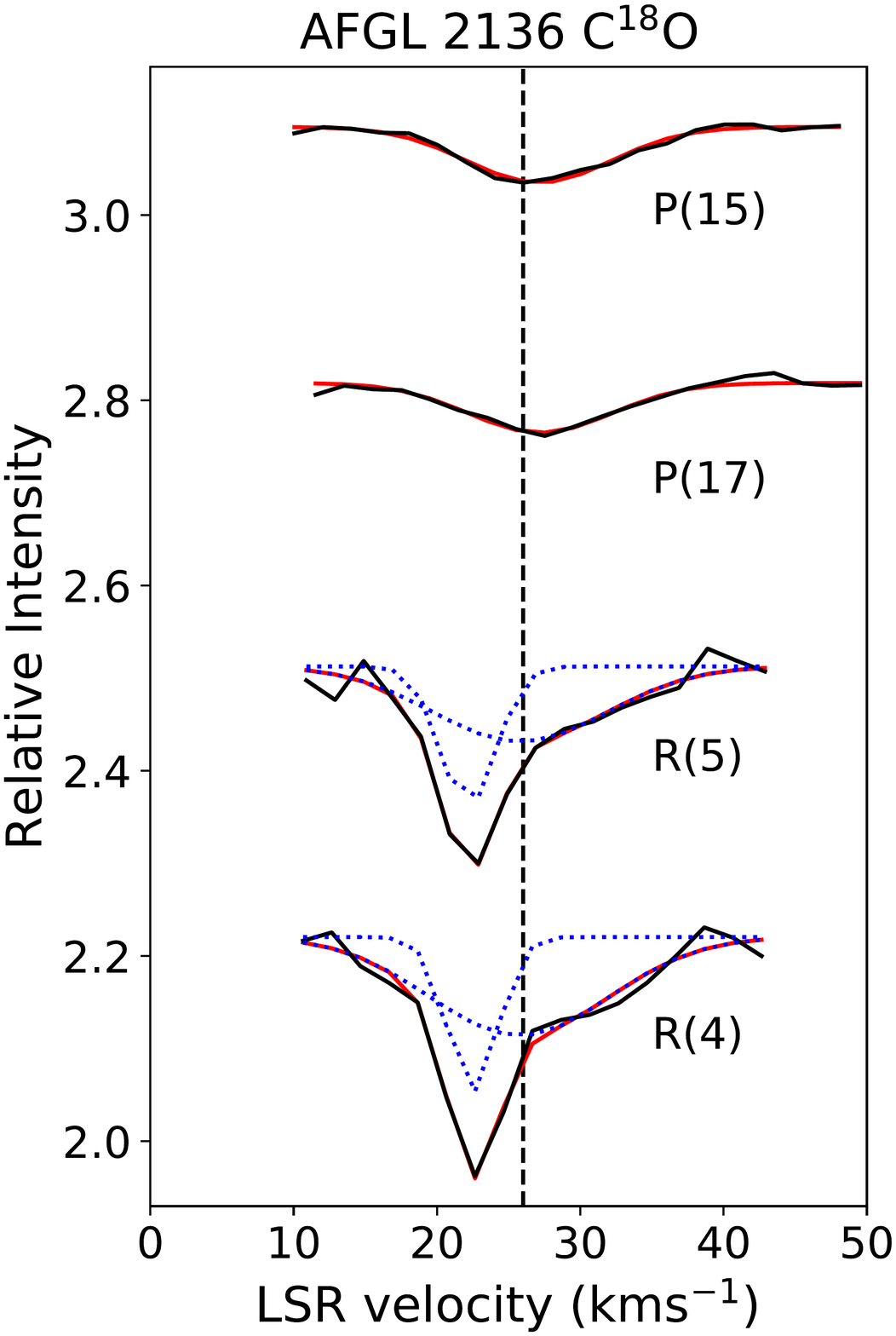}
\includegraphics[width=0.25\textwidth]{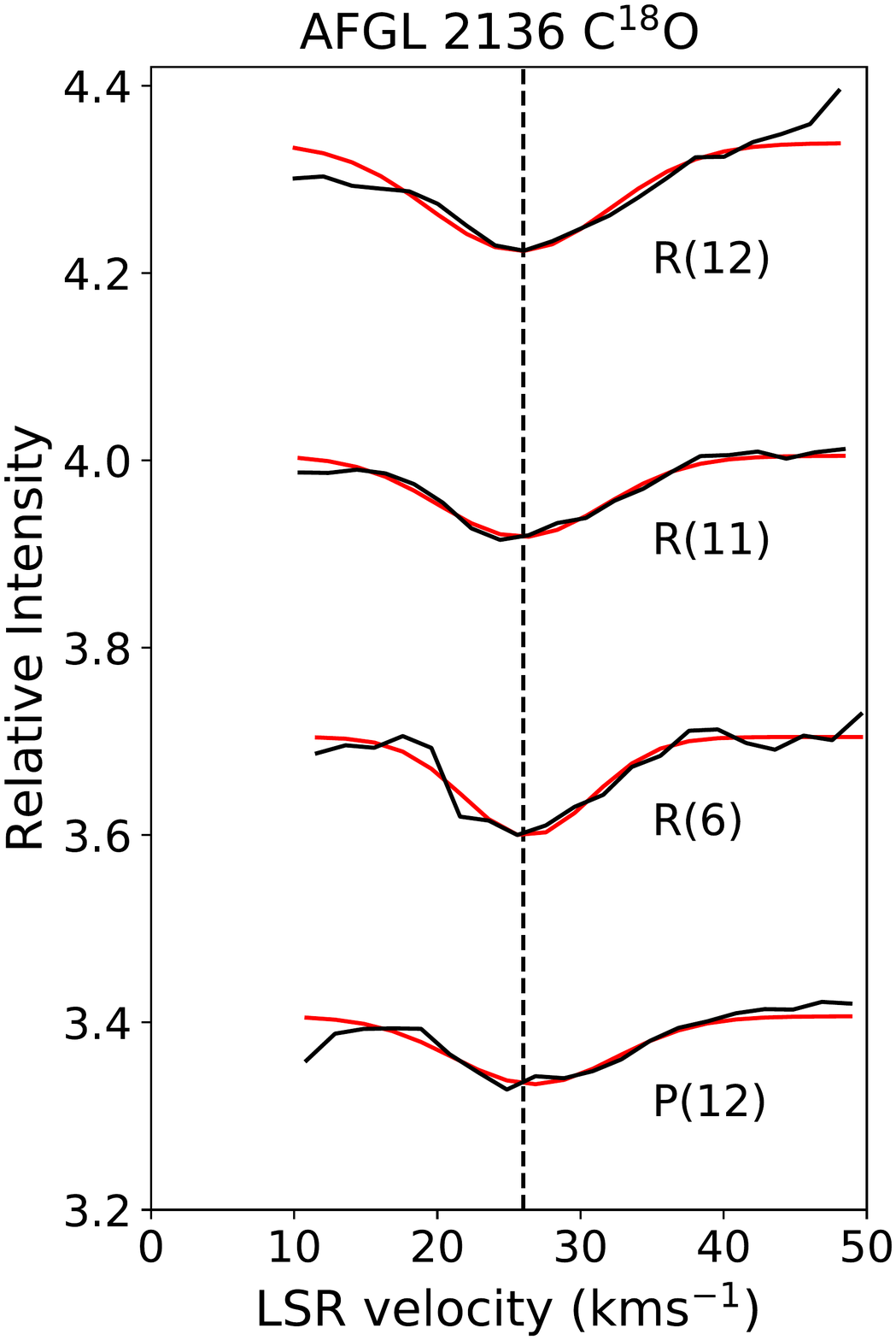} \\
\includegraphics[width=0.25\textwidth]{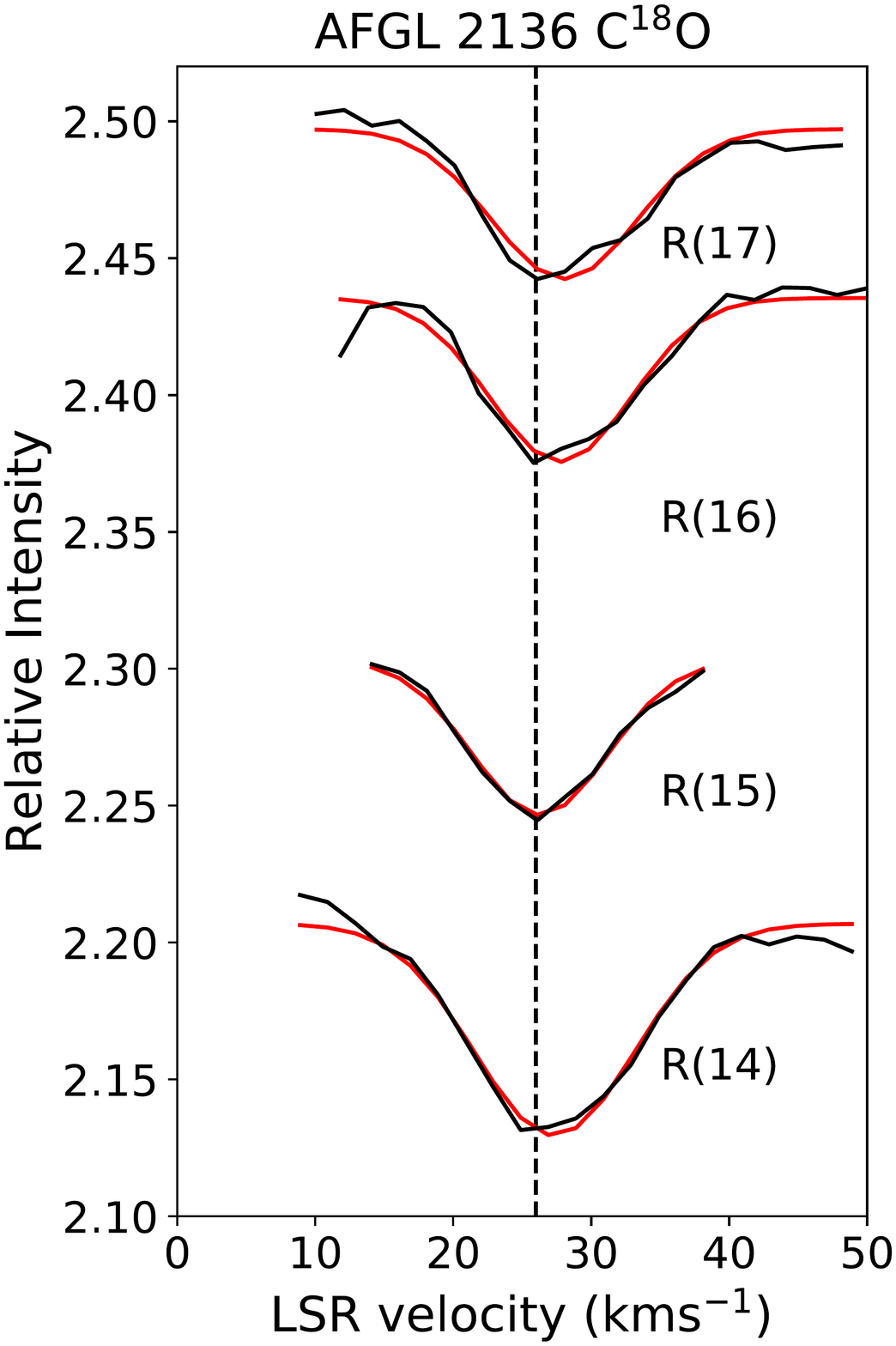}
\includegraphics[width=0.25\textwidth]{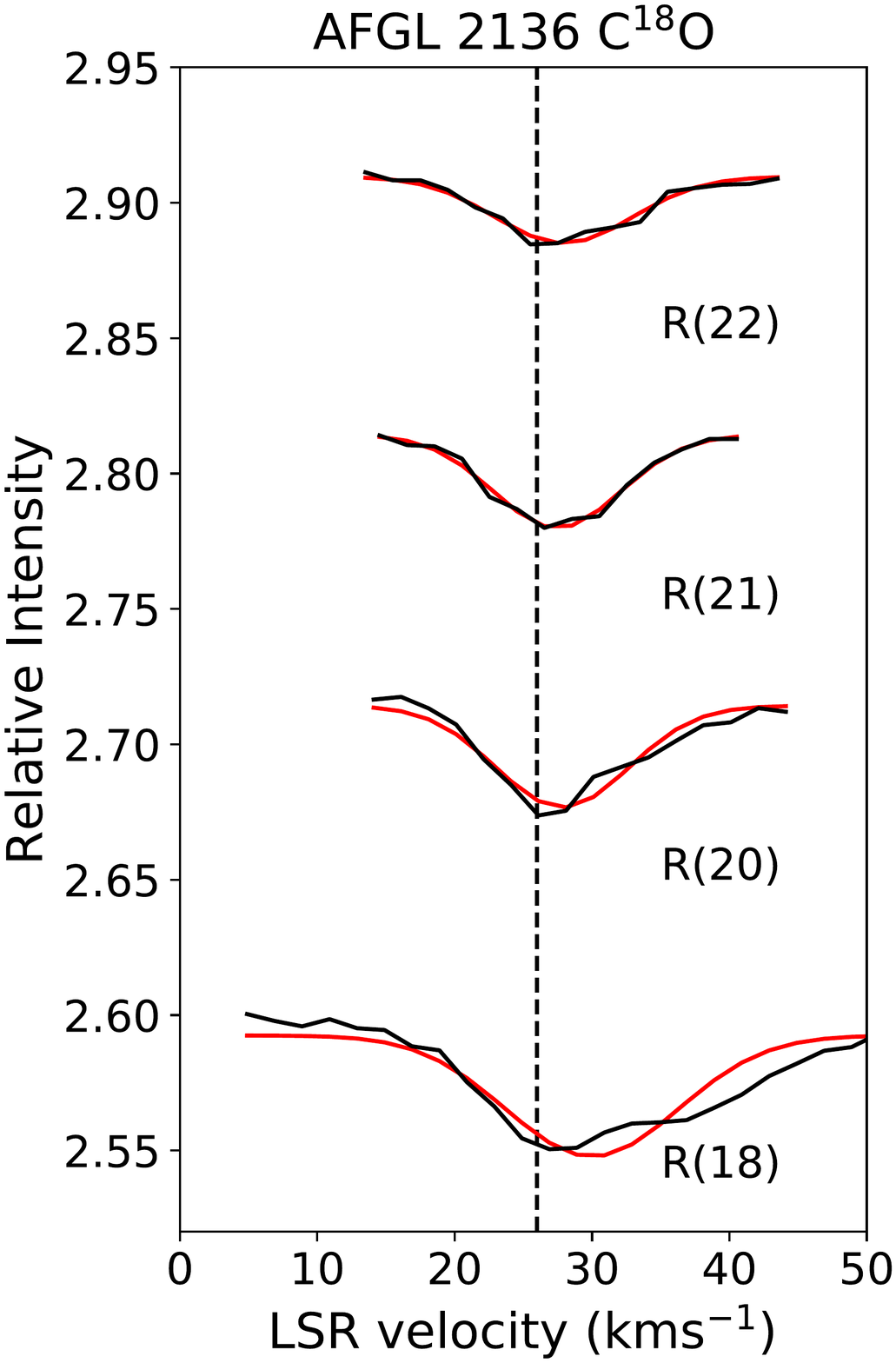}
\includegraphics[width=0.25\textwidth]{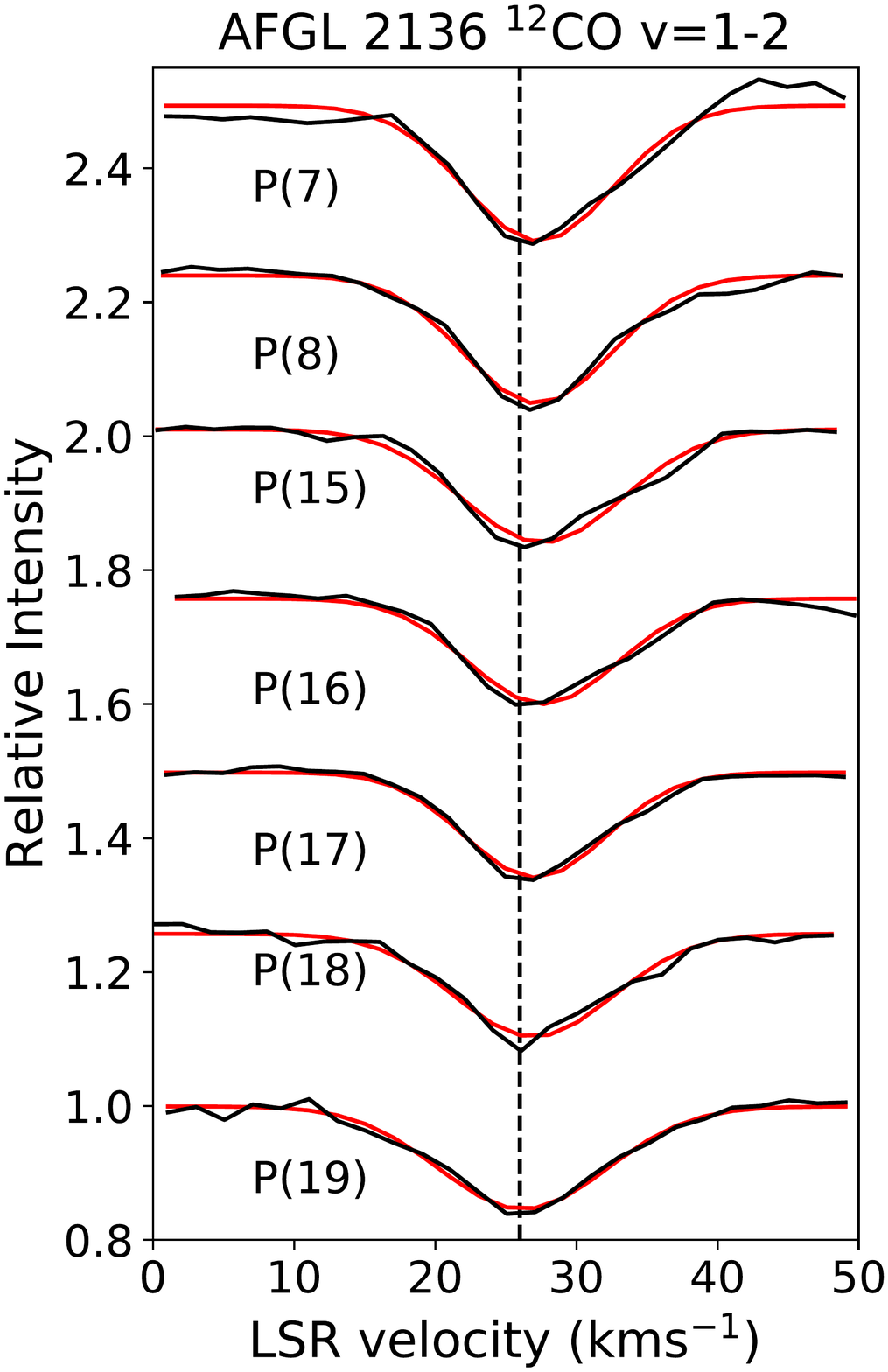}\\
\includegraphics[width=0.25\textwidth]{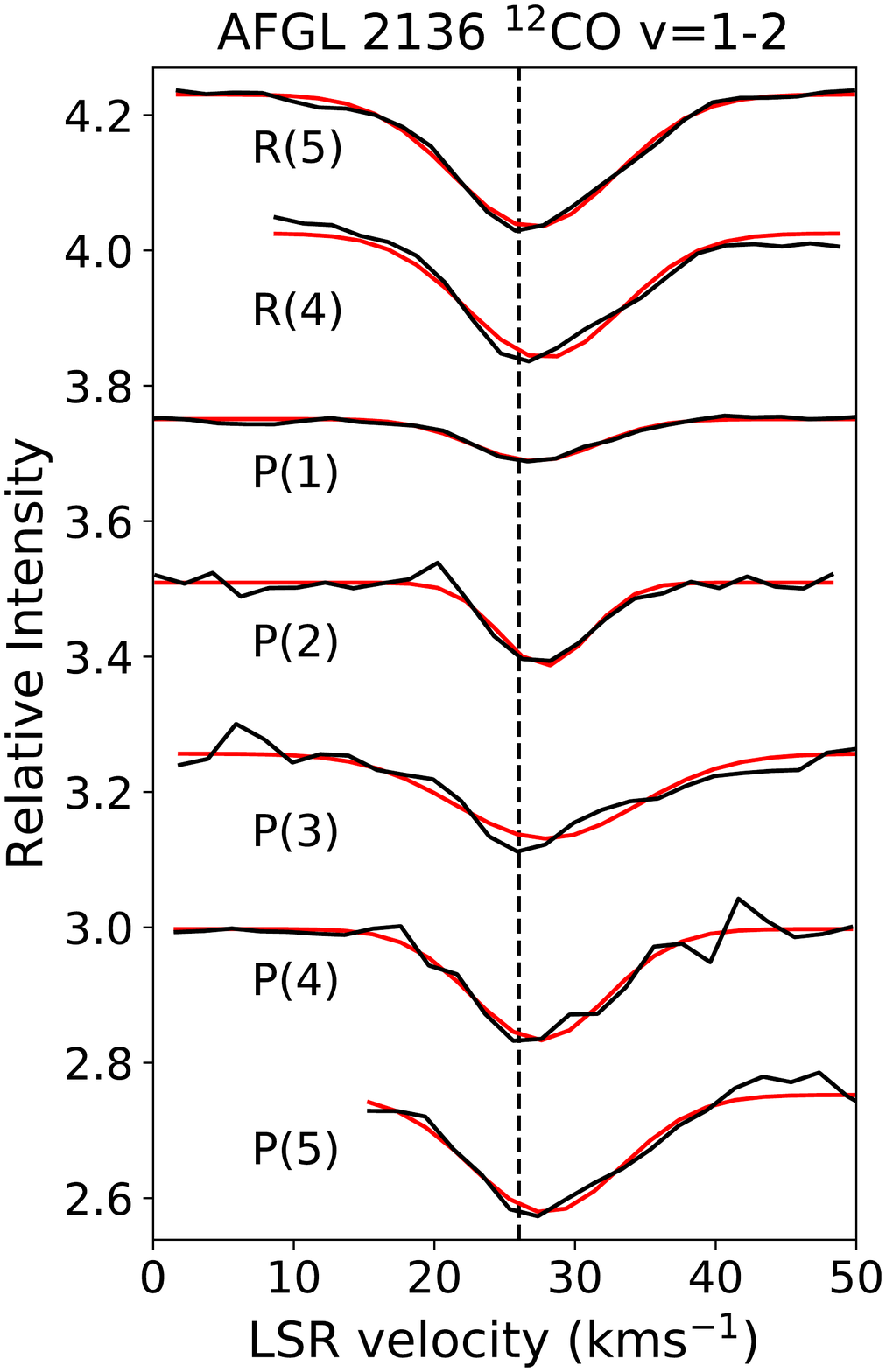}
\includegraphics[width=0.25\textwidth]{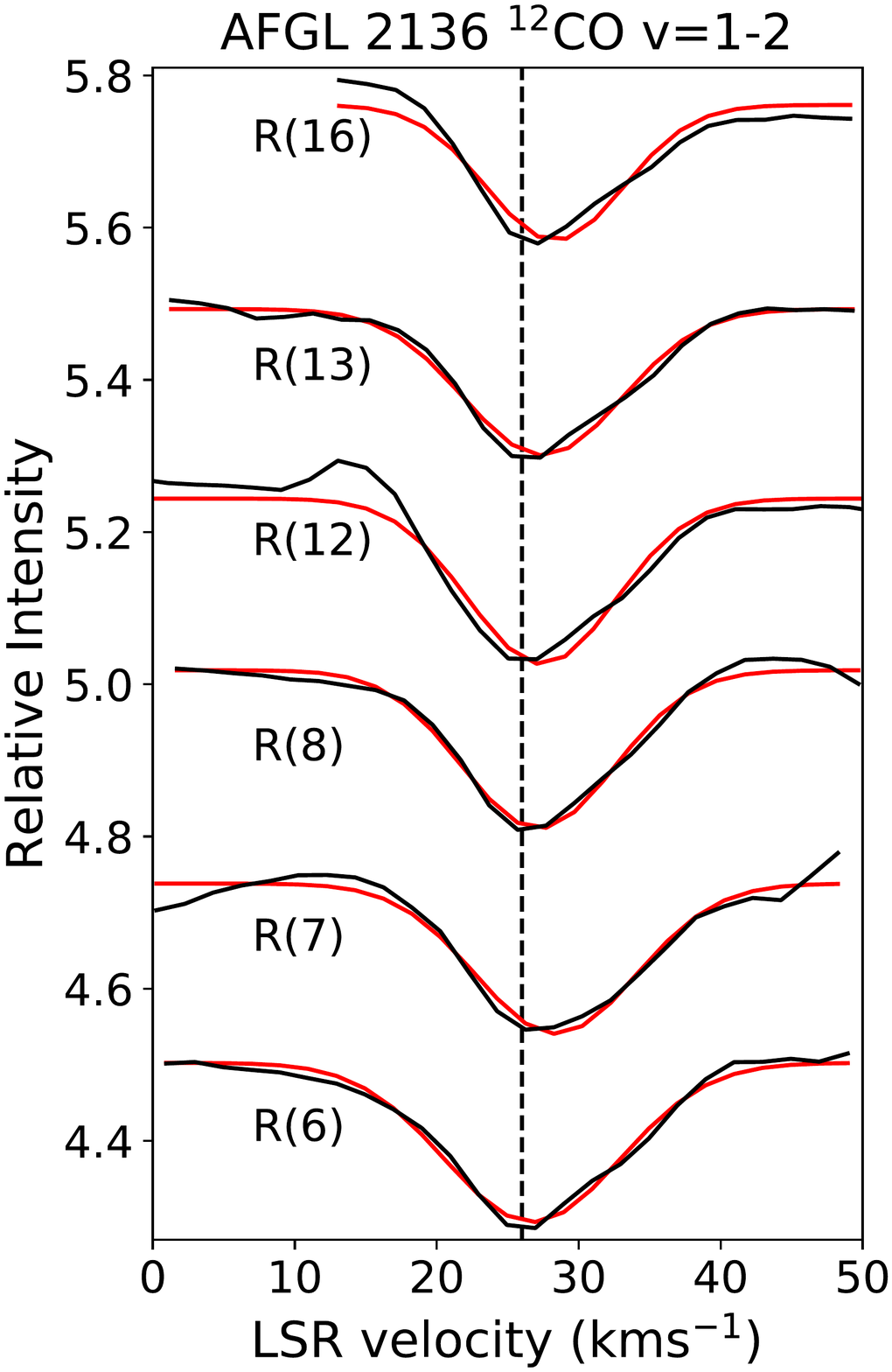}
\includegraphics[width=0.25\textwidth]{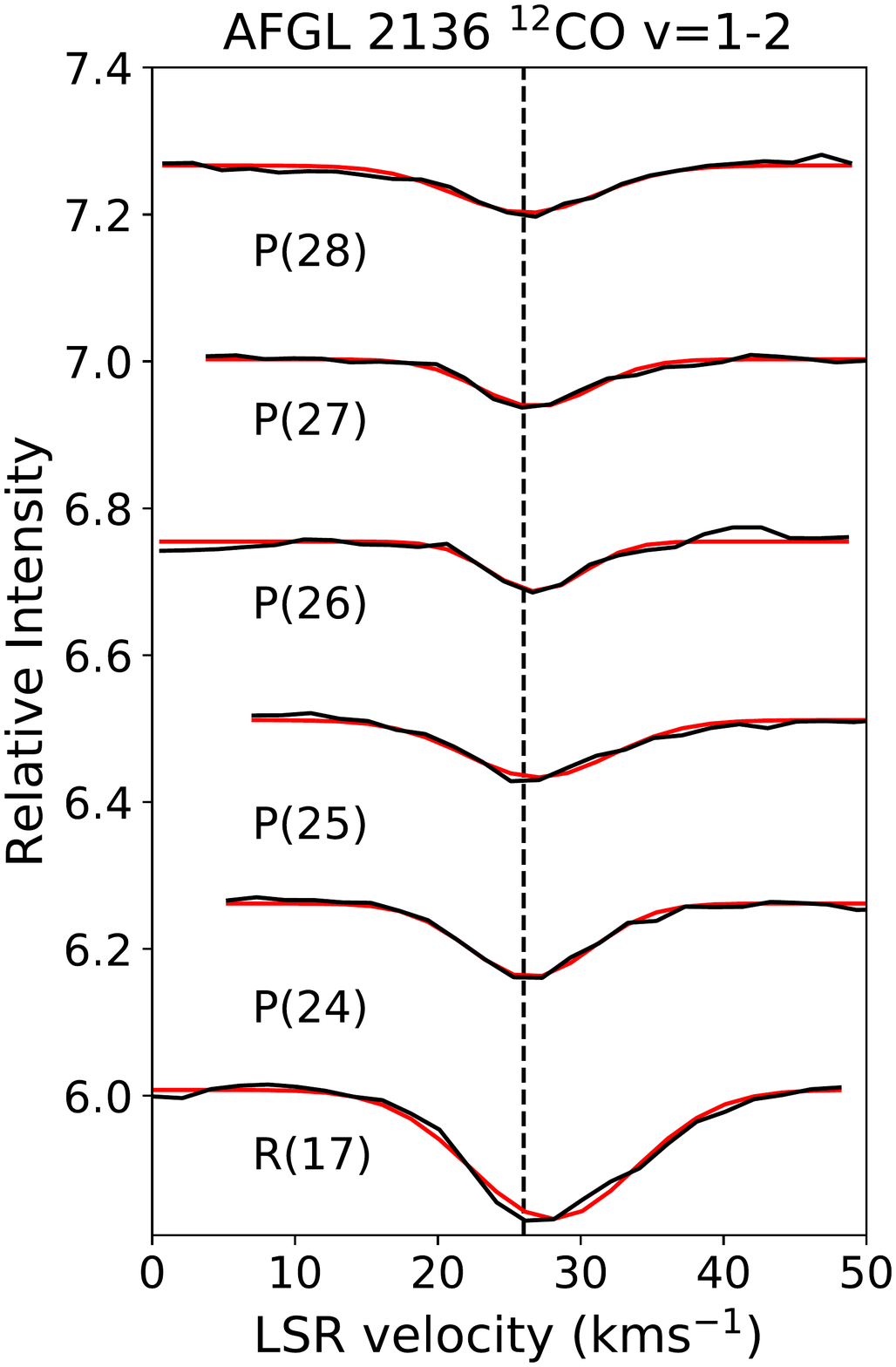}
\end{tabular}
\caption{Multi-gaussian fitting of C$^{18}$O v=0-1 and $^{12}$CO v=1-2 line profiles in AFGL 2136 detected with iSHELL at 4.5 ${\mu}m$. The dashed line represents 27 kms$^{-1}$. The overall fit is shown in red and the individual gaussian components are in blue dotted lines. }
\label{2136CO_prof}
\end{figure}

As a result of this, we instead focus on the C$^{18}$O transitions which are optically thin, and the fits to the C$^{18}$O lines are shown in Figure \ref{2136CO_prof}. Comparing the line profiles between high and low J in C$^{18}$O (Fig \ref{CO2136}) the broad component centred on the red dashed line is present from low to high J, whilst the narrow component is present only at low J. This is a result of the fact that the two velocity components are also distinct in temperature, with the broad and narrow velocity components being hot and cold respectively. This distinction is seen also in $^{12}$CO and $^{13}$CO. Therefore the two velocity components trace different physical components. The focus of the rest of this article will be the broad component at 26 kms$^{-1}$, the line parameters of which are presented in Table \ref{gl2136}.

C$^{18}$O exhibits a straight line which suggests that the transitions are optically thin and that local thermodynamic equilibrium (LTE) is a good approximation. A temperature of 27$\pm$2 K and 440$\pm15$ K is derived for the cold and hot component, respectively. In the rotation diagram, several of the low J level lines in C$^{18}$O appear overproduced. This is a result of blending between the 22 kms$^{-1}$ and 26 kms$^{-1}$ velocity components which are difficult to disentangle since they are only separated by a small amount in velocity space, whereas for the higher energy lines, the narrow cold component at 22 kms$^{-1}$ is not observed. Therefore these lines give a better handle on the real temperature of the gas. 

26 lines of vibrationally excited $^{12}$CO are also detected in AFGL 2136, and are presented in Figure \ref{2136CO_prof}. The transitions show a single velocity component at 27.1$\pm$0.3 kms$^{-1}$ with a width of 12.1$\pm$0.7 kms$^{-1}$. This is in good agreement with hot C$^{18}$O. From Figure \ref{CO2136} we see that the rotation diagram is a straight line implying that the lines are optically thin and in LTE, with a rotational temperature of 661$\pm$9 K. We derive a vibrational temperature of 490$\pm$39 K.

The velocity derived from our observations of the hot CO component at 26 kms$^{-1}$ is in agreement with what is found by \citet{Mitchell1990} for the v=0-1 band. They do not resolve the two velocity components in the CO line profiles. Therefore they fit two temperature components at the same velocity in the rotation diagram. The temperature of their hot component is consistent with what we find for fitting the $^{13}$CO lines with E$_l$ above 100 K. The temperature that we derive for the narrow component at 22 kms$^{-1}$ is higher than the temperature that they derive for their cold component. \citet{Mitchell1990} neglect the $^{12}$CO v=0-1 transitions as they are strongly saturated and they do not discuss C$^{18}$O.

The two velocity components are also observed in the CO v=0-2 band at 2.3-2.4 ${\mu}m$\citep{Goto2019}. The temperatures for the hot component are in agreement within the errors, however for the cold component, \citet{Goto2019} measure a temperature of 16.3$\pm$2.4 K while we derive a temperature of 27$\pm$2 K for the v=0-1 band. This may be a result of uncertainty in disentangling the two velocity components since they are heavily blended. It is difficult to gauge whether or not the single gaussian fit to the 26 kms$^{-1}$ component that we use is adequate for deriving state-specific column densities (Figs \ref{CO2136} \& \ref{2136CO_prof}). This can be further seen from the line profiles of the CO v=0-2 band, where \citet{Goto2019} fit two velocity components to the hot CO component. In this case the high J CO lines may be doubly peaked as well, which is also true for the H$_2$O lines at 2.5 ${\mu}m$ in this source \citep{Indriolo2020}.

\subsubsection{CS}

13 lines of CS are detected towards AFGL 2136 spanning an energy range of 7 to 1165 K. It is one of the species most readily detected in our spectra. One velocity component is detected at 26.1$\pm$0.4 kms$^{-1}$ with a line width of 8.0$\pm$1.1 kms$^{-1}$. All lines are accounted for from J=2 to J=31 with non-detected lines being lost in strong telluric absorption or blending with other hot core lines. The CS velocity is consistent with that derived in CO, however the line width is narrower for equivalent energy levels.

Temperatures and column densities for CS are derived from the rotation diagram shown in Figure \ref{CS}. A temperature of 418$\pm$23 K is measured, in good agreement with the temperature found for C$^{18}$O.

\subsubsection{HCN}

For AFGL 2136, 15 lines of the $\nu_2$ band in the v$_2$=0-1 transition are detected spanning an energy range of only 1500 K. The rotation diagram gives a temperature of 625$\pm$19 K. 26 lines of HCN of the $\nu_2$ band in the v$_2$=0-2 transition are detected in AFGL 2136. The linear part of the rotation diagram gives a temperature of 605$\pm$96 K. This is in good agreement with the temperature of the v$_2$=0-1 transition of HCN. The column density derived for the v$_2$=0-2 transition is a factor of 4 larger than the v$_2$=0-1 transition. The line profiles for the v$_2$=0-1 and v$_2$=0-2 transition are similar in AFGL 2136.

 As for AFGL 2591, evidence for a temperature gradient is observed in the v$_2$=0-2 transition of HCN while it is not seen in the v$_2$=0-1 transition. The line profiles in the v$_2$=0-2 transition are consistent from low to high J (Fig \ref{Lineprof2136}). As in AFGL 2591, the R(0) line of the HCN v$_2$=0-2 band is offset in velocity from the other HCN lines and appears to be tracing cold foreground absorption. \citep{Lahuis2000} detected HCN with ISO towards AFGL 2136 and the derived temperature of 650$^{+75}_{-50}$ K agrees between low and high spectral resolution.

\begin{figure}[htb]
\centering
\begin{tabular}{@{}cccc@{}}
\includegraphics[width=.45\textwidth]{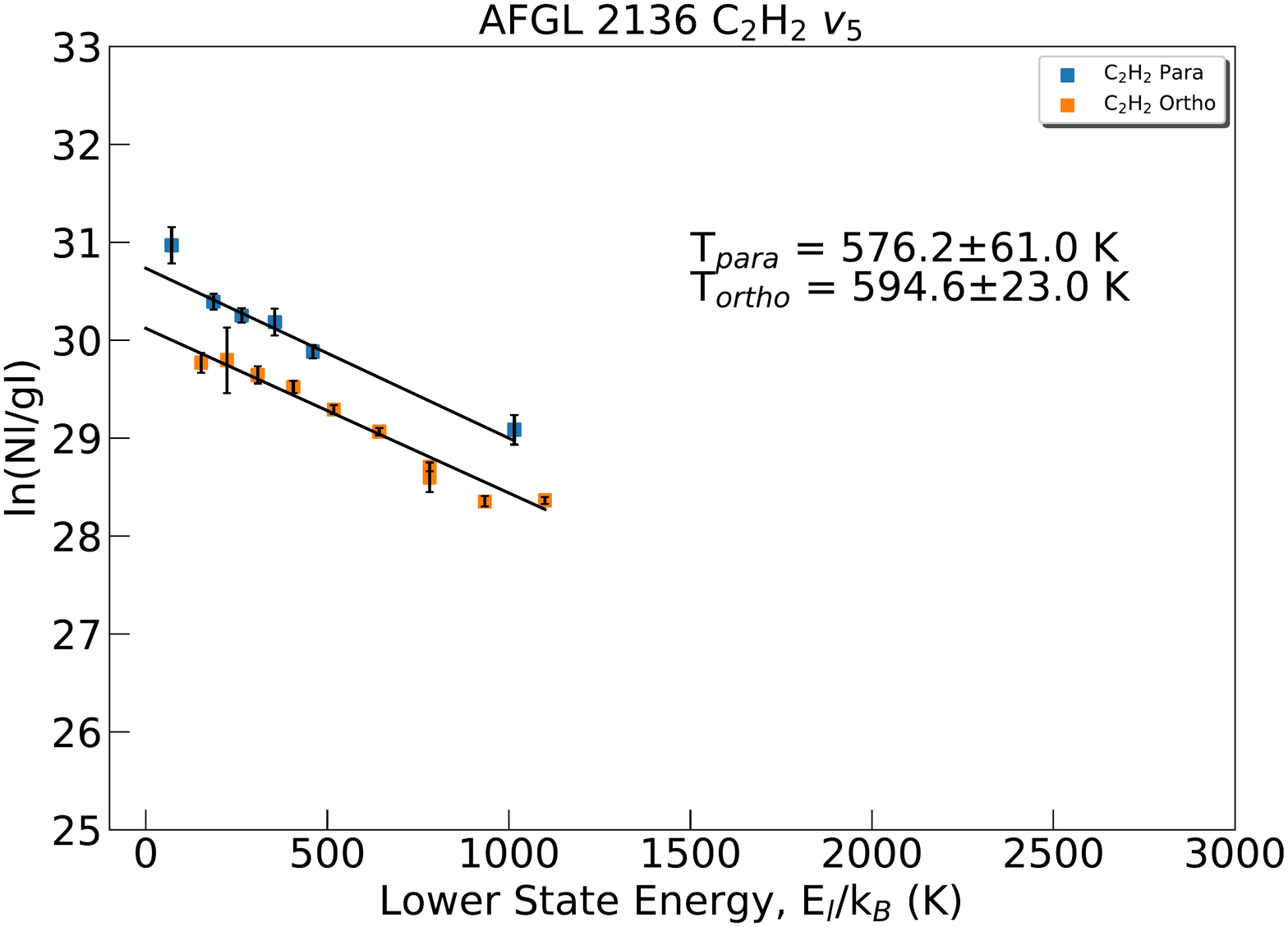}
\includegraphics[width=.45\textwidth]{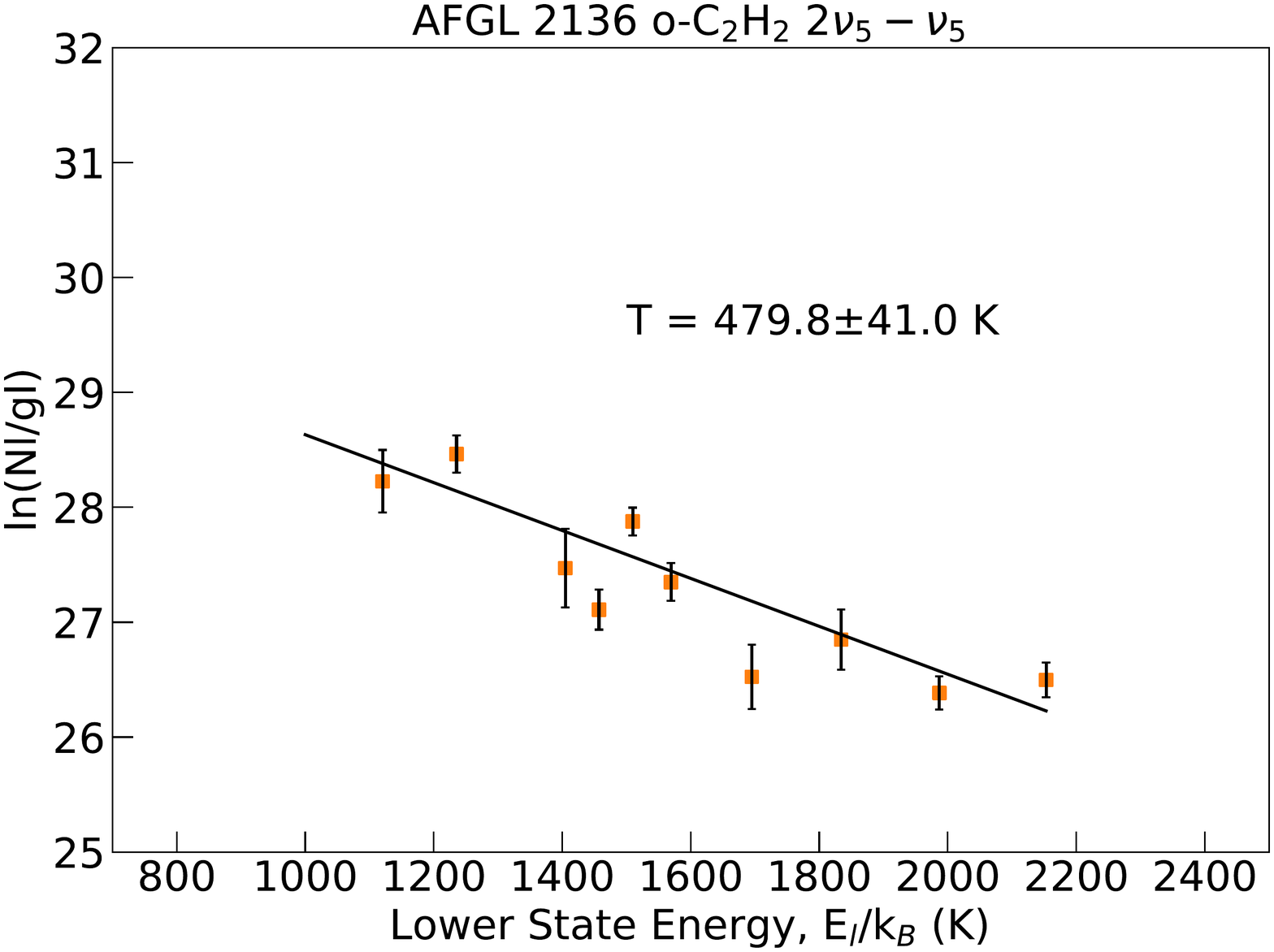}\\ 
\includegraphics[width=.45\textwidth]{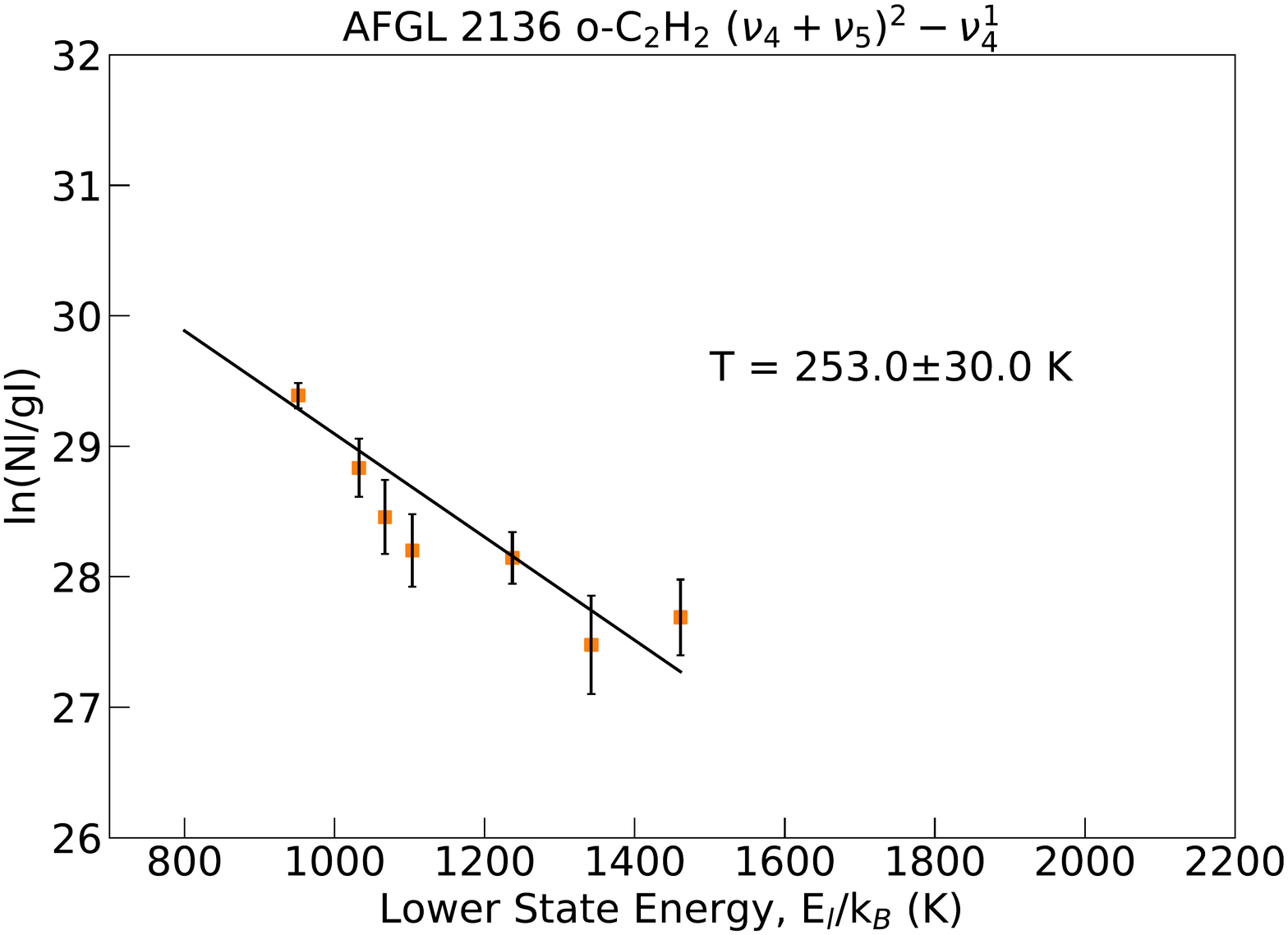} 
\includegraphics[width=.45\textwidth]{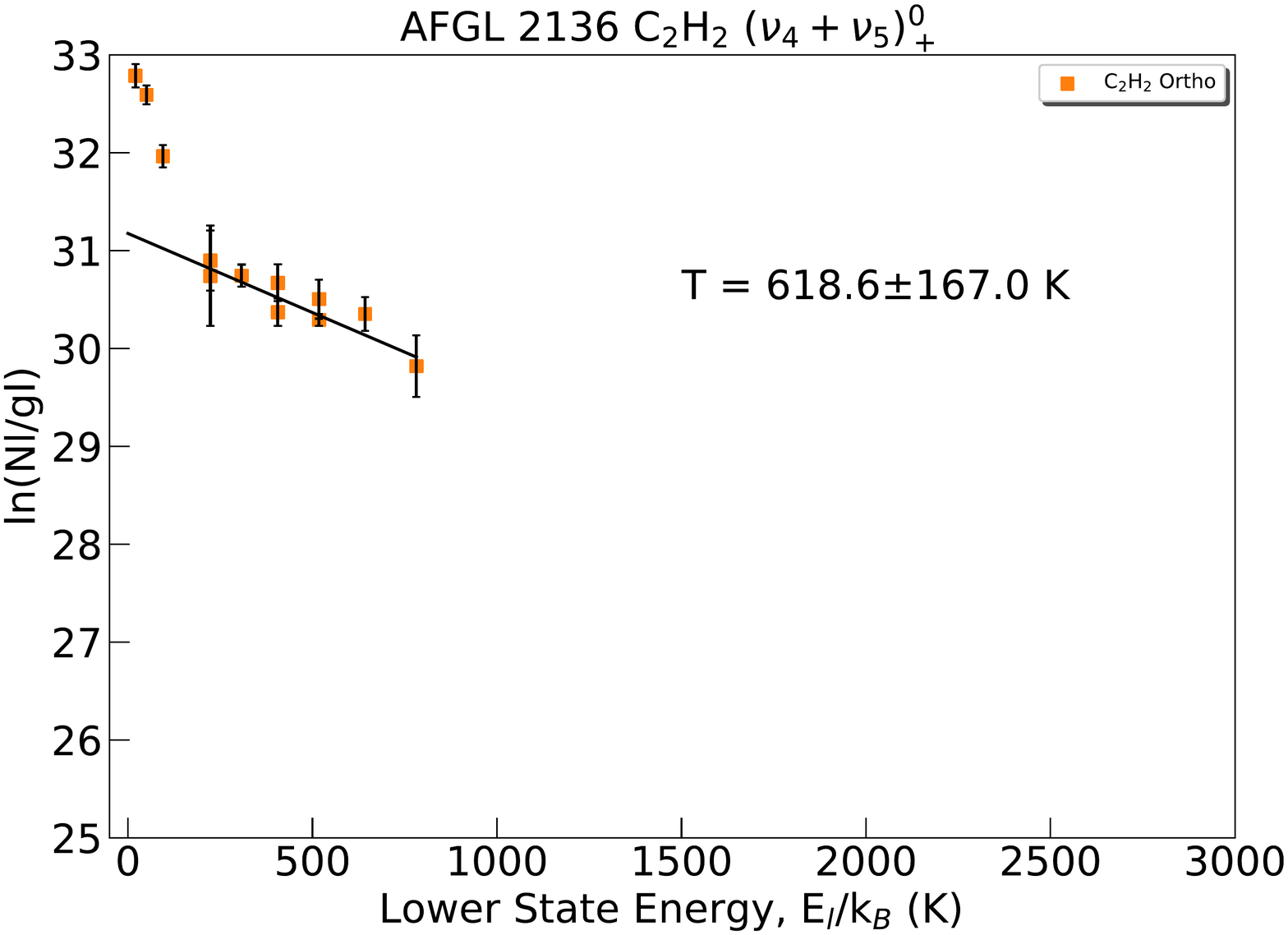} \\
\includegraphics[width=.45\textwidth]{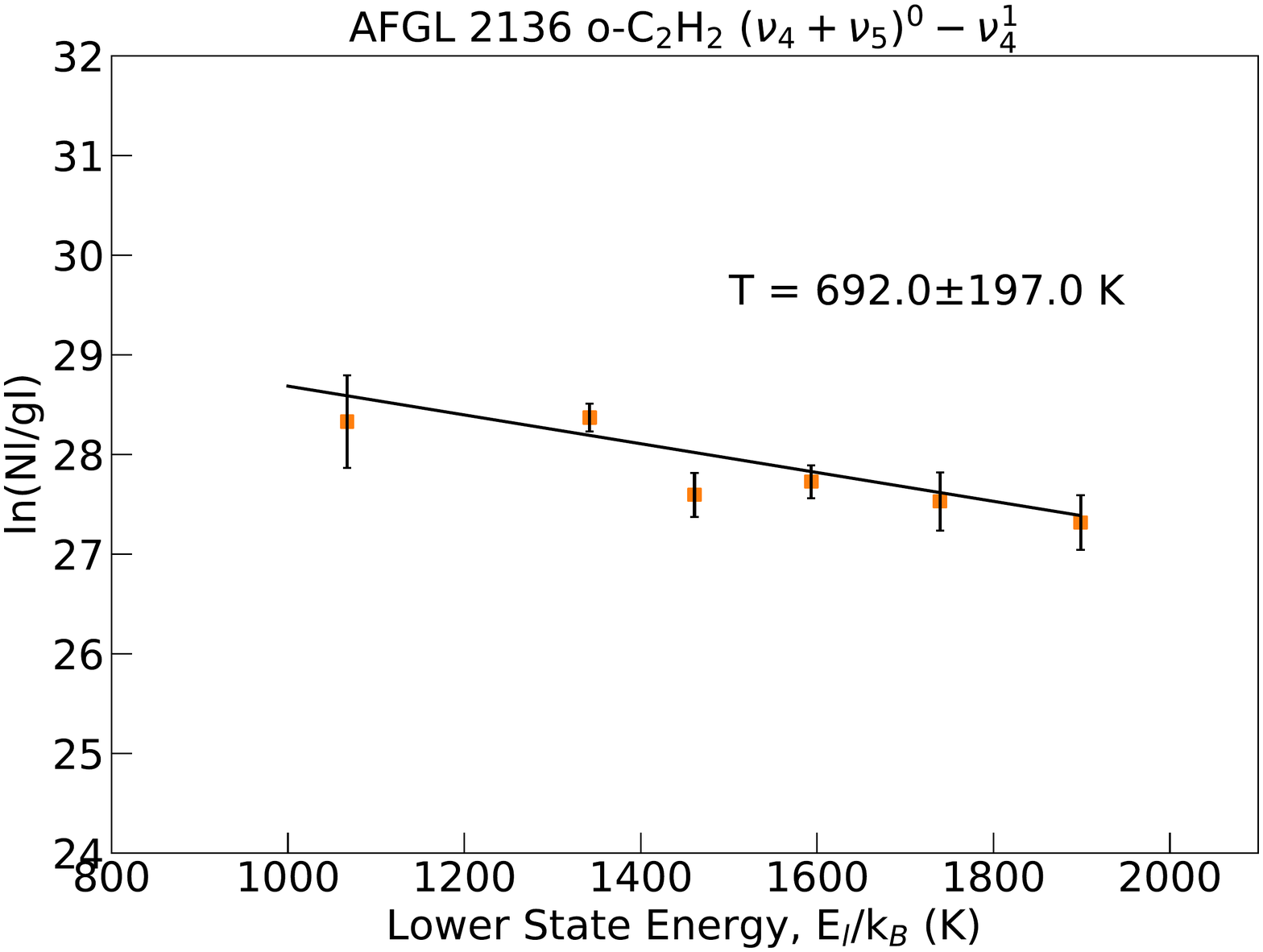} 
\end{tabular}
\caption{Acetylene rotation diagrams for the C$_2$H$_2$ bands detected in AFGL 2136 with TEXES. Ortho- and Para-C$_2$H$_2$ states are split into their corresponding ladders, and are shown as orange and blue respectively. The rotational temperatures for each are also given.}
\label{C2H2GL2136}
\end{figure}

\subsubsection{C$_2$H$_2$}

For the $\nu_5$ fundamental band of C$_2$H$_2$ in AFGL 2136, 16 lines are detected. The temperature and line profiles are consistent with the other species in this source suggesting that they are co-spatial in this hot core. We derive an apparent OPR of 1.8$\pm$0.2. 

We also detect vibrationally excited C$_2$H$_2$ in AFGL 2136 and rotation diagrams are shown in Figure \ref{C2H2GL2136}. Similar to AFGL 2591, we only observe o-C$_2$H$_2$ in the 2$\nu_5^2 - \nu_5^1$ band. A temperature of 480$\pm$41 K is derived, slightly lower than the $\nu_5$ band but within the errors. 

In AFGL 2136 we see an additional band, the ($\nu_4+ \nu_5)^0 - \nu_4^1$ band. The line parameters are consistent with the ($\nu_4+ \nu_5)^2 - \nu_4^1$ band of C$_2$H$_2$. The temperature of this band is comparable to the $\nu_5$ band, with 692$\pm$197 K derived from the rotation diagram. The physical conditions derived from this band however are very uncertain.
 
 In AFGL 2136, the $(\nu_4 + \nu_5)$ band of o-C$_2$H$_2$ is detected in the EXES data around 7 ${\mu}m$. A column density of 5.0$\pm$0.1$\times$10$^{16}$ cm$^{-2}$ is derived. This is a factor of 4 higher than expected based on the column density of the $\nu_5$ band, despite the fact that these bands trace the same lower vibrational energy level. The temperature derived for the $(\nu_4 + \nu_5)$ band is measured only from the higher energy lines, since there is non-linearlity for the low energy lines, analogous to HCN at 7 ${\mu}m$. The derived temperature is equivalent to that found for the $\nu_5$ band. Similar behaviour is observed in HCN and the difference of the column density between the 7 and 13 ${\mu}m$ C$_2$H$_2$ is  of the same order as that found for HCN in AFGL 2136 (a factor of 4). We expect that non-linearity is also present in the $\nu_5$ band, however we do not observe to long enough wavelengths to observe this trend. However it does appear that the R(6e) line is slightly overproduced in the $\nu_5$ band. The $(\nu_4 + \nu_5)$ band of C$_2$H$_2$ is observed in AFGL 2136, but not in AFGL 2591. 

C$_2$H$_2$ has been detected in AFGL 2136 with ISO-SWS with a temperature of 800 K \citep{Lahuis2000}, higher than what we observe with EXES/TEXES. This may be a reflection of the much lower spectral resolution of ISO-SWS and the fringing inherent to this instrument.

\subsubsection{NH$_3$}

Figure \ref{NH3} shows the rotation diagram for the v$_2$=0-1 transition in the $\nu_2$ band of NH$_3$ in AFGL 2136. Approximately 50 lines are detected including both ortho and para states and the transitions cover an energy level range of over 1000 K. p-NH$_3$ and o-NH$_3$ have peak velocities and line widths that are in agreement, however the temperature derived from the rotation diagrams are slightly different with the o-NH$_3$ being higher. An OPR of 1$\pm$0.5 is derived, consistent with the high rotational temperature. NH$_3$ in AFGL 2136 is a very good example of the power of this method to detect many lines by only observing a handful of spectral settings. Another band of NH$_3$, the $\nu_4$ band at 6 ${\mu}m$ is included in our spectral range, but no transitions were apparent, consistent with \citet{Indriolo2020}.

\begin{deluxetable*}{ccccccccc}[htb]
\tablecaption{Summary of Species in AFGL 2591 \label{tab:pars}}
\tablecolumns{8}
\tabletypesize{\scriptsize}
\tablehead{
\colhead{Species} & \colhead{Band} & \colhead{$\lambda_0$ (${\mu}m$)} & \colhead{Number of Lines}  & \colhead{Temperature (K)} & \colhead{N (cm$^{-2}$)} & \colhead{Abundance (w.r.t H)} & \colhead{v$_{lsr}$ (kms$^{-1}$)} & \colhead{$\Delta$V (kms$^{-1}$)}
}
\startdata
$^{13}$CO &  v=0-1 & -- & 8 &  49 $\pm$3 & 3.8$\pm0.2 \times10^{16}$ & 2.4$\pm0.2\times10^{-6}$ & -9.2$\pm$0.3 & 1.5\\
 &  v=0-1 &  -- & 8 & 671 $\pm$124 & 3.4$\pm0.2 \times10^{16}$ &  2.8$\pm1.0\times10^{-6}$ & -12.0$\pm$0.6 & 10.5 \\
$^{12}$CO &  v=1-2 & 4.7  & 8  & 664 $\pm$43 & 1.5$\pm0.6 \times10^{16}$ &  1.2$\pm0.4\times10^{-6}$ & -8.4$\pm$0.5 & 16.4$\pm$1.6 \\
H$_2$O & $\nu_2$ v$_2$=0-1 & 6.0  & 8 &  $^a$ 640 $\pm$ 80 & $^a$ 3.7$\pm0.8 \times10^{18}$ & -- & -11.0 & 12.4 \\
HCN &  $\nu_2$ v$_2$=0-2 & 7.0  &16  & 671 $\pm$ 118 & 2.4$\pm0.2 \times10^{17}$ & 2.0$\pm1.0\times10^{-5}$ & -11.7$\pm$0.5 & 10.8$\pm$1.3\\
 &  $\nu_2$ v$_2$=0-1 & 14.0 & 20  & 675 $\pm$32 & 2.7$\pm0.2 \times10^{16}$ & 3.4$\pm0.6\times10^{-6}$ & -8.9$\pm$0.5 & 8.2$\pm$1.3 \\
 &  $\nu_1$ v=1-0 & 3.0  & 21  & 240$\pm$13 & 7.9$\pm0.5 \times10^{15}$ & -- & -9.6$\pm$0.5 & 6.3$\pm$1.0 \\
CS & v=0-1 &  7.8  & 18  & 713 $\pm$ 59 & 1.6$\pm0.1 \times10^{16}$ & 1.5$\pm0.3\times10^{-6}$ & -10.4$\pm$0.5 & 8.4$\pm$1.7\\
p-C$_2$H$_2$  & $\nu_5$ & 13.7 & 9  & 573$\pm$51 & 4.8$\pm0.5 \times10^{15}$ & 5.8$\pm2.3\times10^{-7}$ &  -8.8$\pm$0.4 & 7.7$\pm$1.1 \\
  & ($\nu_4+ \nu_5)^2 - \nu_4^1$ & 13.7  & 5  & 717$\pm$318 & 4.2$\pm3.8 \times10^{15}$ & 4.9$\pm4.4\times10^{-7}$ & -9.6$\pm$0.9 & 8.8$\pm$2.5 \\
o-C$_2$H$_2$ & $\nu_5$ & 13.7 & 15  & 598$\pm$26 & 1.0$\pm0.1 \times10^{16}$ & 1.2$\pm2.3\times10^{-6}$ & -9.0$\pm$0.4 & 8.7$\pm$1.1\\
  &  2$\nu_5^2$ - $\nu_5^1$ & 13.7 & 12  & 596$\pm$83 & 2.7$\pm0.4 \times10^{15}$ & 3.5$\pm1.5\times10^{-7}$ & -8.7$\pm$0.6 & 8.3$\pm$1.8 \\
  & ($\nu_4+ \nu_5)^2 - \nu_4^1$ & 13.7 & 7  & 668$\pm$191 & 7.7$\pm6.9 \times10^{15}$ & 1.0$\pm0.9\times$10$^{-6}$  & -7.7$\pm$0.5  & 7.4$\pm$1.4\\
 &  $\nu_2 + (\nu_4 + \nu_5)$ & 3.0 & 8  & 454$\pm$151 & 6.0$\pm0.9 \times10^{15}$ & -- & -7.4$\pm$0.5 & 3.7$\pm$1.2 \\
p-NH$_3$ & $\nu_2$ v$_2$=0-1 & 9.5 & 4  & 872$\pm$491 & 1.3$\pm0.7 \times10^{16}$ & 1.6$\pm0.9\times10^{-6}$  & -6.9$\pm$0.6 & 6.2$\pm$1.5\\
o-NH$_3$   & $\nu_2$ v$_2$=0-1 & 9.5 & 4  & 875$\pm$292 & 1.0$\pm0.4 \times10^{16}$ & 1.4$\pm0.5\times10^{-6}$  & -7.3$\pm$0.4 & 7.8$\pm$1.1\\
\enddata
\label{sum2591}
\tablenotetext{}{Note: no abundances relative to H are given for H$_2$O because stellar atmosphere theory is not used in the analysis of \citet{Indriolo2015}.}
\tablenotetext{}{Note: no abundances for the HCN $\nu_1$ and C$_2$H$_2$ $\nu_2 + (\nu_4 + \nu_5)$ emission bands are given since we measure column densities for these bands which are given in section 5.3.}
\tablenotetext{a}{Indriolo et al. (2015)}
\end{deluxetable*}

\begin{deluxetable*}{ccccccccc}[htb]
\tablecaption{Summary of Species in AFGL 2136 \label{tab:pars}}
\tablecolumns{8}
\tabletypesize{\scriptsize}
\tablehead{
\colhead{Species} & \colhead{Band} & \colhead{$\lambda_0$ (${\mu}m$)} & \colhead{Number of Lines}  & \colhead{Temperature (K)} & \colhead{N (cm$^{-2}$)} & \colhead{Abundance (w.r.t H)} & \colhead{v$_{lsr}$ (kms$^{-1}$)} & \colhead{$\Delta$V (kms$^{-1}$)}
}
\startdata
C$^{18}$O &  v=0-1 &  -- & 6  & 27$\pm$ 2 & 8.7$\pm0.7 \times10^{15}$ & 2.8$\pm0.1\times10^{-6}$ & 22.2$\pm$0.3 & 2$\pm$0.8 \\
 &  v=0-1 & -- & 20  & 440$\pm$ 15 & 2.9$\pm0.2 \times10^{16}$ & 1.9$\pm0.1\times10^{-6}$ & 27.1$\pm$0.6 & 12.3$\pm$1.5 \\
$^{12}$CO & v=1-2   & 4.7 & 26 & 661$\pm9$ & 2.7$\pm0.2 \times10^{16}$ & 1.7$\pm0.4\times10^{-6}$ & 27.1$\pm$0.3 & 12.1$\pm$0.7 \\
H$_2$O & $\nu_1$ / $\nu_3$ & 2.5  & 34   & $^a$502 $\pm$ 12 & $^a$8.25$\pm$0.95$\times10^{18}$  & -- &  24.6$\pm$1.1 & 13.2$\pm$2.5 \\
HCN &  $\nu_2$ v$_2$=0-2 & 7.0 & 18  & 592 $\pm$ 21 & 1.8$\pm0.2 \times10^{17}$ & 1.6$\pm0.8\times10^{-5}$ & 26.2$\pm$0.5 & 8.5$\pm$1.6\\
 &  $\nu_2$ v$_2$=0-1 & 14.0 & 15  & 625$\pm$19 & 4.6$\pm0.2 \times10^{16}$ & 5.3$\pm0.4\times10^{-6}$ & 26.1$\pm$0.5 & 11.0$\pm$1.5\\ 
CS & v=0-1 & 7.8 & 13  & 418 $\pm$ 23 & 1.6$\pm0.1 \times10^{16}$ & 1.2$\pm0.1\times10^{-6}$ & 26.1$\pm$0.4 & 8.0$\pm$1.1\\
p-C$_2$H$_2$  &  $\nu_5$ & 13.7 & 6 & 576$\pm$61  & 8.8$\pm0.7 \times10^{15}$ & 1.0$\pm0.4\times10^{-6}$ & 27.0$\pm$0.3 & 10.9$\pm$1.1 \\
o-C$_2$H$_2$ &  $\nu_5$ &  13.7  & 10  & 595$\pm$23 & 1.6$\pm0.1 \times10^{16}$ & 1.7$\pm0.1\times10^{-6}$ & 27.2$\pm$0.3 & 11.3$\pm$0.9  \\
 &  2$\nu_5^2$ - $\nu_5^1$ & 13.7 & 10 & 480$\pm$41 & 1.6$\pm1.0 \times10^{15}$ & 2.3$\pm0.1\times10^{-7}$ & 26.6$\pm$0.4 & 7.1$\pm$1.2 \\
 &  ($\nu_4+ \nu_5)^2 - \nu_4^1$ & 13.7  & 7  & 253$\pm$30 & 1.7$\pm1.5 \times10^{15}$ & 1.3$\pm1.3\times10^{-7}$ & 26.9$\pm$0.6 & 8.1$\pm$1.8 \\
  &  ($\nu_4+ \nu_5)^0 - \nu_4^1$ & 13.7 & 6  & 692$\pm$197 & 2.1$\pm0.2 \times10^{15}$ & 2.3$\pm2.0\times10^{-7}$ & 26.8$\pm$0.6 & 8.5$\pm$1.9 \\
 & ($\nu_4 + \nu_5)$ & 7.5  & 12  & 618$\pm$176 & 5.0$\pm0.1 \times10^{16}$ & 7.0$\pm0.8\times10^{-6}$ & 26.0$\pm$0.5 & 8.0$\pm$1.5 \\
 p-NH$_3$ & $\nu_2$ v$_2$=0-1 & 9.5 & 32  & 435$\pm$20 & 1.0$\pm0.5 \times10^{16}$ & 1.0$\pm0.1\times10^{-6}$ & 28.1$\pm$0.4 & 6.7$\pm$1.0  \\
o-NH$_3$ &  $\nu_2$ v$_2$=0-1 & 9.5 & 17  & 493$\pm$24 & 0.9$\pm0.4 \times10^{15}$ & 9.7$\pm0.5\times10^{-7}$ & 27.7$\pm$0.3 & 7.7$\pm$0.9  \\
\enddata
\label{sum2136}
\tablenotetext{}{Note: no abundances relative to H are given for H$_2$O because stellar atmosphere theory is not used in the analysis of \citet{Indriolo2020}.}
\tablenotetext{a}{Indriolo et al. (2020)}
\end{deluxetable*}

\pagebreak

\begin{figure}[h!]
\centering
\begin{tabular}{@{}cccc@{}}
\includegraphics[width=1\textwidth]{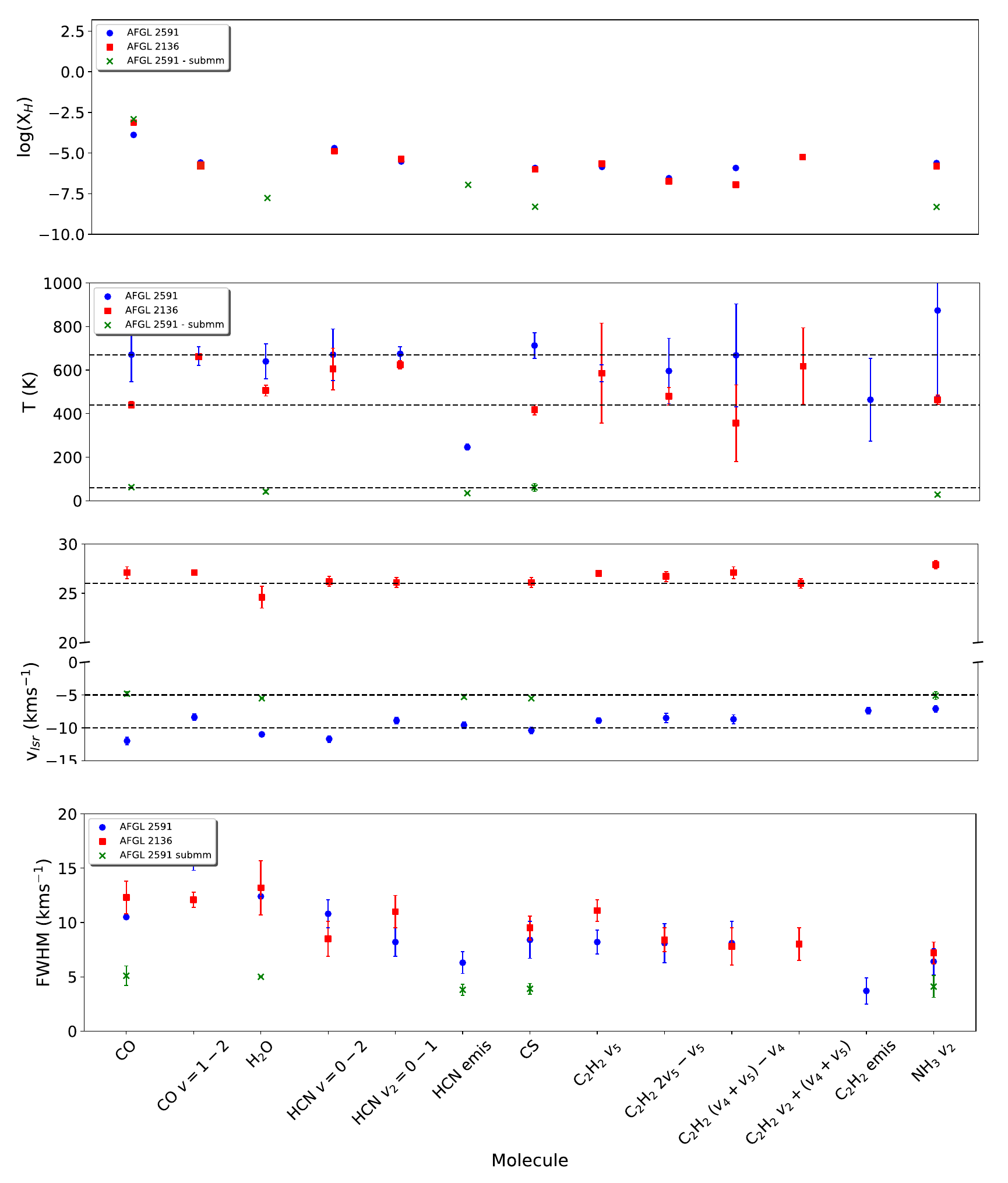} 
\end{tabular}
\caption{Scatter plots comparing the temperature, abundance w.r.t H, velocity and line width of selected molecules observed towards AFGL 2591 (blue circles) and AFGL 2136 (red squares), and also the literature values for AFGL 2591 (green crosses) observed in the sub-mm \citep{Kazmierczak2014}. For the sub-mm abundances, a H column of $\sim 1\times10^{22}$ cm$^{-2}$ was assumed following \citet{Kazmierczak2014}. }
\label{Comp}
\end{figure}

\subsection{Comparison to Sub-millimetre Observations}

Absorption lines at IR wavelengths probe the hot core on very small spatial scales, comparable to the smallest achievable by sub-mm emission line studies (0.02$''$). On scales as large as 10$''$, differences in line widths, peak velocities, temperatures and column densities are observed between sub-mm emission and IR absorption studies. This is summarised in Figure \ref{Comp}, where we compare the results from our MIR line survey of AFGL 2591 with a line survey by \citet{Kazmierczak2014} from 480-1900 GHz (with respective beam sizes 10-40 $''$), with \textit{Herschel}/HIFI. 

Temperatures compare on average with 700 K for IR absorption to 50 K in sub-mm emission. The temperature derived from the emission lines corresponds well with the temperature of the cold $^{13}$CO absorption component (Table \ref{sum2591}). For AFGL 2591 the velocity of the sub-mm emission lines agrees with the systemic velocity and is on average -5.5 kms$^{-1}$, consistent with other studies of this source in the sub-mm \citep{Wang2012, Gieser2019, vanderTak2003}. However from the IR absorption lines the velocity observed is -10 kms$^{-1}$. Furthermore the line widths of the absorption lines are around a factor of 2 higher than those measured in the sub-mm. 

Towards AFGL 2591, 6 lines of HCN were detected in emission and a population diagram results in an estimate of the temperature of 35 K \citep{Kazmierczak2014}. In total, we detect 42 lines of HCN from 2 different transitions in the IR spanning an energy range of 0 to 2500 K in AFGL 2591. The large number of lines combined with the large span in energy range gives a good handle on the physical conditions and a temperature of 670 K is derived. 

Although not many species have been detected towards  AFGL 2136 at sub-mm wavelengths, CS has been detected by \citet{vanderTak2003}. A velocity of 22.8 kms$^{-1}$ and a line width of 3.1 kms$^{-1}$ are derived for sub-mm emission lines. This contrasts with the IR absorption where we derive a v$_{lsr}$ of 26.1 kms$^{-1}$ and a width of 8 kms$^{-1}$. This is consistent with the two velocity components we see in CO presented in section 4.2.1. From a rotation diagram analysis we see that the narrow 22 kms$^{-1}$ component is cold while the broad 26 kms$^{-1}$ component is hot.
 
As mentioned in section 4.2.1, IR CO line profiles of AFGL 2136 reveal two velocity components in the same line of sight (Figure \ref{CO2136}); a cold component with narrow lines centred at 22 kms$^{-1}$ and a hot component with broad lines centred at 26 kms$^{-1}$. \citet{Maud2018} observed this source with the Atacam Large Millimetre Array (ALMA) and detected an unresolved disk-like structure in SiO and dust continuum, at 22 kms$^{-1}$. SiO emission is observed to be spatially extended at 22 kms$^{-1}$, which is the systemic velocity of the cloud as derived from previous sub-mm observations \citep{vanderTak2003}, suggesting that the velocity component we detect in the IR at 26 kms$^{-1}$ is not spatially resolved by \citet{Maud2018}. 

However, recent very high angular resolution (20$\times$15 mas) observations with ALMA by \citet{Maud2019} reveal that sub-mm observations can resolve this velocity component if the spatial resolution is high enough. Velocity maps of the H$_2$O 5$_{5,0} - 6_{4,3}$ $\nu_2=1$ line show that a Keplerian disk is resolved which is centred at a velocity of 26 kms$^{-1}$. Furthermore, since this line lies 3461 K above ground, this gas must be hotter than the temperature estimates given by previous sub-mm studies, which is more consistent with the IR absorption lines. 

A recent study of AFGL 2591, \citep{Gieser2019} analyses chemical complexity in this hot core. The pure rotational CO J=2-1 line profile that is presented in their Figure 7 shows a complex velocity structure, analogous to that which we see in IR absorption. There is a broad, blue-shifted part of the line profile that appears to consist of several velocity components, one of which is centred around -10 kms$^{-1}$. There is also tentative evidence for this same velocity component being present in the emission line profile of SiO, however the other more blue-shifted velocity components are missing. This further indicates that it is possible to resolve in emission this inner part of the hot core that is so evident in IR absorption. 

Overall, IR observations trace hot, dense gas at the centre of the hot core. Sub-mm observations, unless at very high angular resolution (0.02$''$), probe the surrounding envelope and cloud where temperatures are lower.

\section{Discussion}

Summarising the observations presented in Section 4, MIR spectra of AFGL 2136 and 2591 show absorption bands of a number of molecular species, which can be analysed in terms of the velocity and temperature of the absorbing gases. There are a number of differences between IR and sub-mm wavelengths that provide further insight into the structure of these sources. The IR lines are typically seen in absorption and the sub-mm lines in emission, there is a peak velocity offset (Fig \ref{Comp}), and the line width is typically much larger (Fig \ref{Comp}) in the IR than in the sub-mm \citep{vanderTak2003, Kazmierczak2014, Gieser2019}. In addition, the IR lines originate from gas that is typically much warmer (400-600 K) than the sub-mm lines (20-200 K).

\subsection{General Structure of the Sources}

Very high angular resolution observations of AFGL 2136 in the N band with the mid-infrared interferometer (MIDI) instrument on the Very Large Telescope Interferometer (VLTI), H and K band polarisation studies with the Subaru telescope, sparse aperture Keck interferometry and sub-mm dust emission with ALMA reveal the presence of a geometrically thin circumstellar dust disk with an inner and outer radius of ~30 and 120 AU respectively \citep{Murakawa2008, Monnier2009, DeWit2011, Boley2013, Maud2019}. The whole system is embedded in a dusty envelope. The circumstellar disk is optically thick in the MIR continuum and the N-band emission originates from a radius of $\sim$100 AU. Inside the dust sublimation radius ($\leq$30 AU), the disk is thought to be gaseous. Recent ALMA observations at high spatial resolution in a highly excited H$_2$O line (E$_u$=3461.9 K) reveal the presence of a disk in Keplerian rotation around the central $\sim$45 M$_{\odot}$ protostar \citep{Maud2019}. The total disk mass, estimated from the sub-mm continuum, is somewhat sensitive to the adopted temperature but is of the order of 1 M$_{\odot}$ \citep{Maud2019}. 

Both AFGL 2136 and AFGL 2591 have been imaged at 10.7 ${\mu}m$ with the Keck telescope \citep{Monnier2009}. The physical structure of both of these sources is very similar at this wavelength. The MIR continuum is asymmetrically extended to the West around the disk, out to around 250 mas. However, the extended emission is an order of magnitude lower in intensity than the continuum emission from the compact disk. For AFGL 2136, the FWHM size of the MIR continuum is 125$\pm$5$\times$115$\pm$3 mas for the major and minor axes, respectively. This is reasonably consistent with the size of the disk in the ALMA 1.3mm continuum image, 93$\times$71 mas for the major and minor axes, respectively. This 25 \% difference in size, and the somewhat flatter disk seen in the ALMA image, could be due to the fact that at 10 ${\mu}m$ the observer looks less deep into the disk than at sub-mm wavelengths, making it appear more spherical and extended at 10 ${\mu}m$. The size of the MIR continuum in AFGL 2591 is 123$\pm$3$\times$111$\pm$3 mas for the major and minor axes, respectively. Since the MIR continuum is dominated by the disk, we neglect envelope emission contributing to the continuum of our absorption lines. 

At MIR wavelengths, both sources are characterised by extended emission incommensurate with the central protostar. The ALMA observations of AFGL 2136 reveal that the MIR emission coincides with a circumstellar disk in Keplerian rotation. In the following, we presume that this is also the case for AFGL 2591 and will discuss this more in section 5.6.2. As the MIR continuum emission is dominated by the disk, the absorption lines have to originate in the photosphere of this disk.

For AFGL 2136, the observed peak velocity of the IR absorption lines is in good agreement with that of the H$_2$O sub-mm emission line which clearly associates the absorbing gas within the disk (rather than the surrounding envelope or molecular cloud in which the source is embedded). The observed high temperatures also argue for a location in the inner warm regions. The line width of the  CO, HCN, C$_2$H$_2$ and NH$_3$ IR absorption lines are narrower than that of the H$_2$O sub mm emission line. This is also the case for H$_2$O absorption \citep{Indriolo2020}. Absorption may still be associated with the Keplerian disk if the absorption does not trace the full extent of the disk. This will be discussed in more detail in section 5.6.1. 

The fact that the lines are in absorption implies that the disk surface behaves like a photosphere with a temperature which decreases outwards. Hence, rather than heating of the disk surface from the outside by impinging stellar (UV/visible) radiation, the disk is heated from the inside, and radiative diffusion sets up a temperature gradient decreasing outwards in the radial and vertical directions. We attribute the heating to rapid accretion in the mid-plane, which viscously heats the gas to high temperatures. 

At the high densities of the disk, the dust and gas temperature will be closely coupled in this situation. Models for viscous disks imply a radial temperature gradient with an exponent in the range of -0.4 to -0.7 \citep{Dullemond2007}. Following the N-band continuum studies, we set the temperature of the dust disk photosphere to the sublimation temperature (1200 K) at 30 AU, which implies a dust temperature at 120 AU of 450 K. Hence, the disk would indeed be bright in mid-IR continuum and, given the much larger surface area, would outshine the dust photosphere of the protostellar cocoon.

\begin{figure}[h!]
\centering
\begin{tabular}{@{}cccc@{}}
\includegraphics[width=1\textwidth]{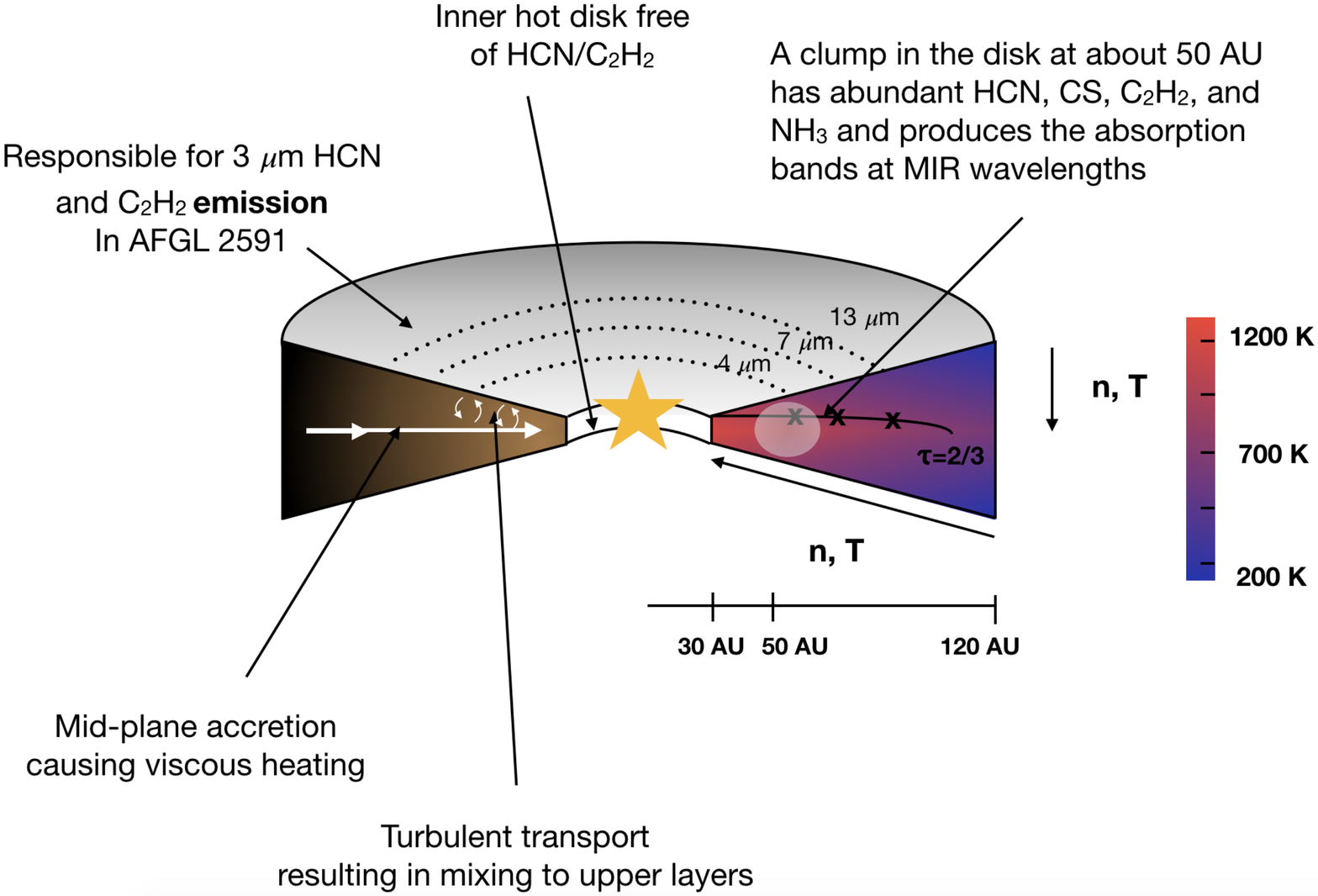}
\end{tabular}
\caption{Slice of the disk showing the structure of the disk and the physical processes taking place. Dotted  lines represent the extent of the disk at the $\tau=2/3$ dust photospheres at 4, 7 and 13 ${\mu}m$. The black solid line indicates the $\tau=2/3$ optical depth profile through the disk with wavelength, where crosses indicate the wavelength corresponding to the dotted lines. The blue/red colour scheme shows the temperature of the disk and is represented by the colourbar on the right. A physical scale is given indicating at what radii the given positions are. Temperature and density increase towards the mid-plane, and also towards the star in the centre. The transparent circle represents the dust clump at 50 AU from the protostar, that produces the absorption lines. The innermost region of the disk is free of HCN and C$_2$H$_2$ gas and dust. Turbulent mixing results in HCN and C$_2$H$_2$ gas being carried up to the cooler, upper layers of the disk where resonant scattering of the disk continuum causes emission lines.
}
\label{slice}
\end{figure}

As sketched in Figure \ref{slice}, the observed spectrum will originate from "characteristic" regions in the disk, with shorter wavelengths coming from the inner regions and longer wavelengths predominantly from further out. For any region in the disk, at any wavelength, the observed flux will originate from a total optical depth ($\tau_L + \tau_c) = 2/3$, where $\tau_L$ and $\tau_c$ are the line and continuum optical depth respectively. The decreasing temperature towards the surface will then lead to an absorption line.

\subsection{Abundance Analysis}

Since we are dealing with an internally heated disk, and therefore an outwardly decreasing temperature gradient in the vertical direction, some further insight into the characteristics of these disks can be obtained by standard stellar atmosphere analysis of absorption lines. We will base this analysis on the radiative transfer equation in the grey approximation, $\kappa\left(\nu\right)=\text{constant}$. It is likely that the dust has coagulated either during the preceding molecular cloud phase or in the disk and will have a grey-like opacity \citep{Ormel2011}. Using the Milne-Eddington solution to the equation of radiative transfer in a grey atmosphere applied to line formation \citep{Mihalas1978}, the observed flux at any wavelength will originate from a total optical depth ($\tau_L + \tau_c) = 2/3$. This assumes an average over viewing angles through the atmosphere. The outward-decreasing temperature gradient will then give rise to an absorption line and, comparing absorption by different species, lines tend to be stronger when the line opacity is larger. The full details of the derivation of the Milne-Eddington solution, along with further information, can be found in the Appendix section A.

We conclude that for a weak line (relative to the continuum), the central depth (or equivalent width) is a measure of the relative column density in the lower level of the molecule. One important conclusion is then that the measured central depths can still be used in a Boltzmann diagram analysis, and the excitation temperatures used from the initial rotation diagram analysis still applies. The column density is relative to the continuum opacity, which itself is a measure of the hydrogen column density (and the dust absorption properties). In this way the line strength is a measure of the abundance with respect to H of the molecular levels, which can be used in a rotation diagram to derive the total abundance of the species. This conversion also involves the temperature gradient in the atmosphere through the coefficients $a$ and $b$ in the source function (see Appendix section A). Since the line opacity is much less than the continuum opacity, the molecules absorb continuum photons therefore coupling the gas and dust.

In the weak line approximation, we can use the curve of growth approach and for the linear portion we arrive at (Mihalas 1978 section 10.3):

\begin{equation}
\label{eqn:4}
\dfrac{W}{2 Y \Delta\nu} = \eta_0 \dfrac{\sqrt{\pi}}{2} 
\end{equation}

with W the equivalent width in frequency space, $\Delta{\nu}$ is the line width in frequency space and $Y$ a parameter which takes care of the gradient in the source function and is defined in the Appendix section A. Here the peak line-to-continuum ratio at line centre, $\eta_0$, is given by

\begin{equation}
\label{eqn:5}
\eta_0 =\frac{\kappa_L (\nu=\nu_0)}{\kappa_c} = \dfrac{A_{ul} \lambda^3}{8 \pi \sqrt{2\pi} \sigma_v} \dfrac{g_u}{g_l} \dfrac{N_l}{\sigma_c N_H} \bigg(1 - \dfrac{g_l N_u}{g_u N_l} \bigg)
\end{equation}

where $A_{ul}$ is the coefficient for spontaneous emission, $g_u$ and $g_l$ are the statistical weights for the upper and lower levels respectively, $\nu_0$ is the line frequency and $\kappa_L$ and $\kappa_c$ are the line and continuum opacities respectively and $N_l$ is the column density in the lower level of the line. $\sigma_v$ is the dispersion in velocity space and $\sigma_c$ the dust opacity per H-nucleus. For the latter we will adopt the value $7\times 10^{-23}$ cm$^2$/H-nucleus appropriate for coagulated interstellar dust \citep{Ormel2011}. This value is not very sensitive to the degree of coagulation. Here the expression for stimulated emission (the term in brackets) has been neglected since it is very close to unity. 

The optical depth of the absorption lines is thus set against the continuum and therefore the level-specific column densities are calculated relative to the hydrogen column density. We measure the equivalent width of the absorption lines and thus calculate level specific abundances which, using equation \ref{eqn:3}, results in a total abundance for a given species (Tables \ref{sum2591} \& \ref{sum2136}). 

We note the dependence on the adopted dust characteristics and hence the quoted absolute abundances carry that uncertainty. For lines that originate from the same spatial zone, derived abundances can still be directly compared. As the disk will show a strong radial temperature gradient in addition to the vertical temperature gradient, we recognise that derived abundances for lines that originate from a wide zone in the disk are ill defined without a detailed model for the disk. Dilution by continuum from regions with low molecular abundances will also result in ill-defined averages of the abundance over the emitting disk surface. Thus, abundance comparisons have to be taken with caution. However, as we will argue below, the observations indicate that much of these molecular absorption lines originate from the same region of the disk. 

In both hot cores, the measured OPRs for C$_2$H$_2$ are less than the statistical equilibrium value of 3 for high temperatures. An ortho to para ratio of 3 is expected for this molecule as the observed (excitation) temperature (600K) is well above the energy difference between these states and, as we will argue in section 5.7, the high observed abundance requires formation of this molecule in hot gas. However we measure 2.1$\pm$0.3 and 1.8$\pm$0.2 for AFGL 2591 and AFGL 2136 respectively. We attribute this slightly lower value for the C$_2$H$_2$ OPR to the finite optical depth of these transitions. The observed strength of a line is only a direct measure of the column density in the weak opacity limit. When the opacity increases, the temperature gradient comes into play. Ortho lines will have 3 times the column density of the para lines. Hence, if line optical depth becomes important relative to that of the dust, our observations of the ortho lines will trace gas slightly higher in the photosphere than the para lines, and hence derived abundances will be smaller than those of the para transitions. The fact that the OPR is not equal to 3 is evidence for temperature gradients. Detailed modelling will be necessary to fully disentangle the effects of gradients.

In the above analysis of the abundances of molecular species, several assumptions are made that are important to emphasise. We assume that the opacity is dominated by dust (weak line limit) and that albedo is small thus $\epsilon=1$. Furthermore we assume that the dust and the gas are well coupled thermally, the level populations are in LTE and dust properties do not vary as a function of distance to the star. We also take a grey atmosphere, namely that the source function is linear, which has a number of assumptions that result in uncertainties in the abundances. These include the degree of dust-to-gas ratio (settling), the extent to which coagulation has proceeded and compaction (e.g., porosity) of the grains. This results in additional uncertainty in the absolute abundances of the molecules however it does not affect the relative abundances of the species, so the actual abundance relative to H will depend on the adopted value of dust opacity, but abundances between species are directly comparable.

Comparing the abundances to the sub-mm observations at high spatial resolution ($>0.5 ''$), we see that the abundances calculated for the IR absorption lines in AFGL 2591 are higher by 2-3 orders of magnitude (Figure \ref{Comp}). This is also true of AFGL 2136 where \citet{vanderTak2003} detect CS from single-dish sub-mm observations and derive an abundance of 4$\times$10$^{-9}$ with respect to H$_2$. This compares to an abundance of 1.2$\times10^{-6}$ with respect to H from the IR absorption lines. For AFGL 2591, the derived abundance for $^{13}$CO corresponds to a $^{12}$CO abundance of 1.6$\pm0.5\times10^{-4}$, assuming a $^{12}$C/$^{13}$C ratio of 60. This agrees well with the typical CO abundance of 1$\times10^{-4}$\citep{Lacy1994}. Taking a $^{16}$O/$^{18}$O ratio of 500 \citep{Asplund2009} we derive a C$^{16}$O ratio of 9.4$\pm5.0\times10^{-4}$ with respect to H for AFGL 2136. This is larger than the typical CO abundance which may reflect that the chosen dust opacity per H-atom in this source was too high. We will return to this issue in section 5.5

\begin{figure}[htb]
\centering
\begin{tabular}{@{}cccc@{}}
\includegraphics[width=0.5\textwidth]{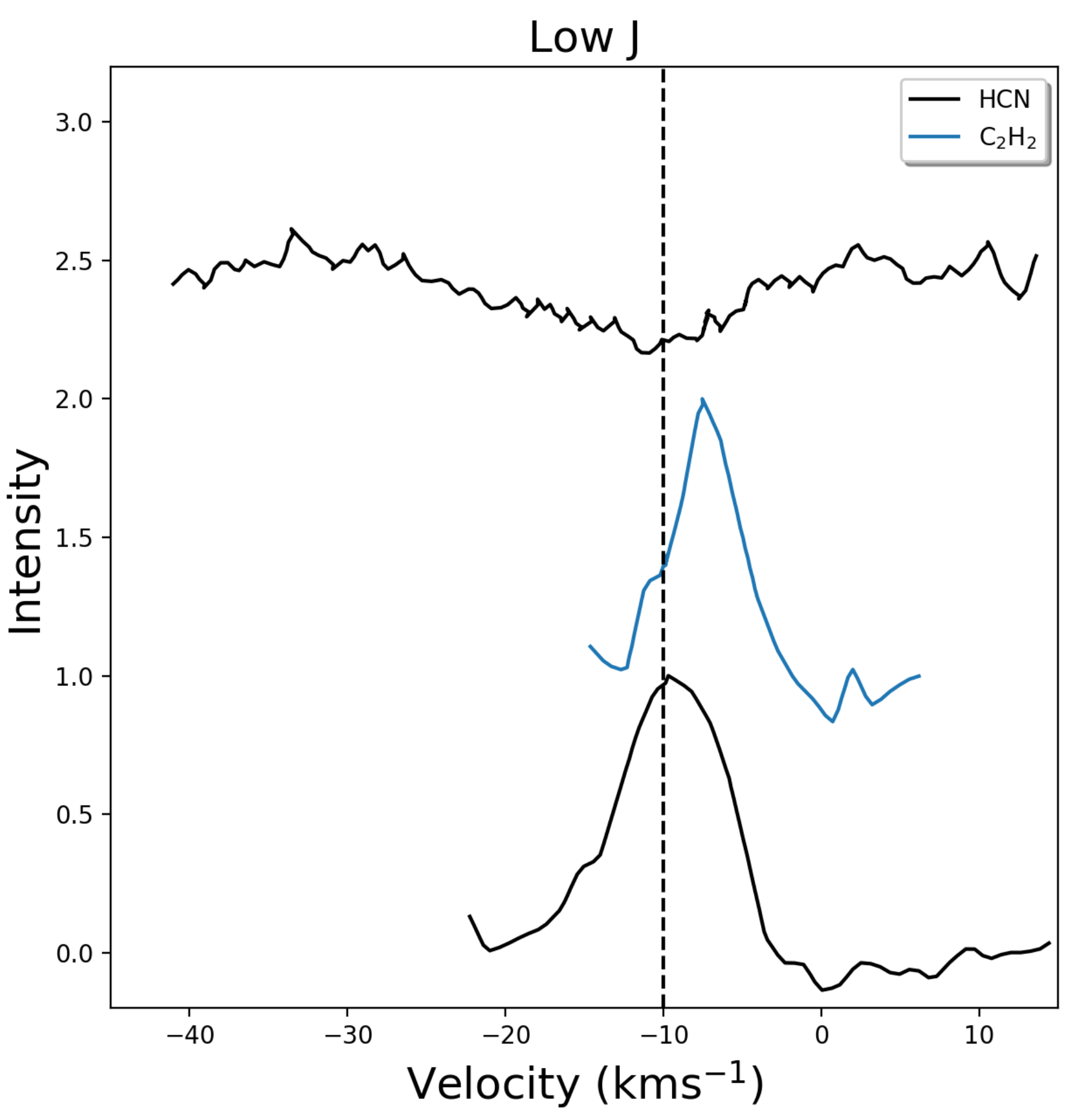}
\includegraphics[width=0.5\textwidth]{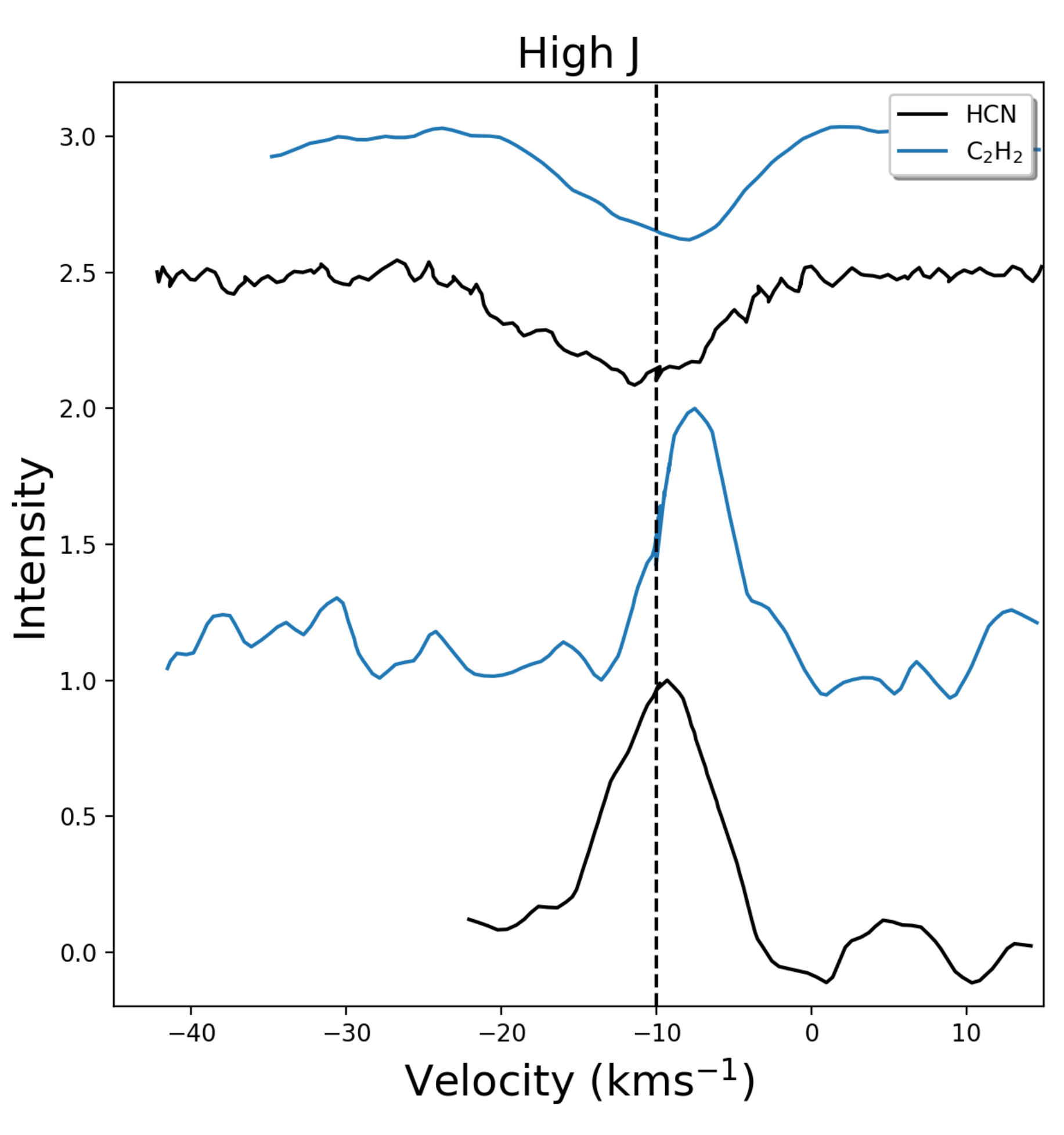} 
\end{tabular}
\caption{Average line profiles comparing HCN and C$_2$H$_2$ emission and absorption line profiles in AFGL 2591 plotted in velocity space. The HCN absorption transition is $\nu_2$ v$_2$=0-2 and the C$_2$H$_2$ absorption band is the $\nu_5$. Transitions are separated into high and low J levels before the average is calculated (high for E$_l$ $<$ 90 K). The black dashed line denotes -10 kms$^{-1}$. }
\label{emisprof}
\end{figure}

\subsection{Emission Lines of HCN and C$_2$H$_2$ in AFGL 2591}

While the 4 to 14 ${\mu}m$ window exclusively shows molecular absorption lines, the 3 ${\mu}m$ region of the spectrum of AFGL 2591 shows emission lines of HCN and C$_2$H$_2$. One major conclusion from these emission lines is that the abundance of HCN and C$_2$H$_2$ are very low in the innermost, warm region that is responsible for the 3 ${\mu}m$ continuum. The absence of broad HCN and C$_2$H$_2$ absorption lines in the 3 ${\mu}m$ region implies that there is an HCN/C$_2$H$_2$ free region in the innermost parts of the disk (Figure \ref{slice}). This should be contrasted with CO which does show absorption lines in the (v=0-2) overtone in the K band (R. Smith 2020, private communication). Hence, the CO abundance must be high throughout the entire disk and produce absorption features, which greatly overwhelm scattered photons from the cooler portions of the disk.

The emitting gas is characterised by low temperatures (250 K) and small line width (3-6 kms$^{-1}$). The line profiles in Figure \ref{emisprof} reveal that the velocities of the emission and absorption lines of HCN are the same, firmly associating this gas with the disk. This is also true of C$_2$H$_2$, suggesting that emission of HCN and C$_2$H$_2$ in the disk originates from the same radial part as the corresponding absorption. As the 3 ${\mu}m$ continuum emission likely arises from a region of the disk further in than the longer wavelength emission, we attribute these emission lines to resonant scattering by C$_2$H$_2$ and HCN molecules high up in the photosphere of the disk at the same radial location that is responsible for the 7 and 13 ${\mu}m$ absorption lines. This would in a natural way explain the relatively low temperatures and the small line widths. For this to occur, at 3 ${\mu}m$ there must be an optically thin gas layer in the upper photosphere of the disk, indicative of dust settling in the disk. Likely, the disk is flaring at large scale heights and the gas is turbulently brought up to high altitudes where it is exposed to near-IR pumping continuum photons from the warm inner disk.

Consider a disk with size $R_{out}$ which emits continuum (black body) radiation at 3 ${\mu}m$ over a small region with surface area $A$. The flux seen by gas, at a distance $R$ from this continuum source to the gas is then

\begin{equation}
\label{eqn:6}
F_R\, =\, \frac{L}{2\pi R^2} = \frac{A \pi B}{2 \pi R^2} = \frac{A B\left(T\right)}{2 R^2}
\quad .
\end{equation}
where we have included an extra factor of 0.5 in the denominator since the disk sees half of the flux. The flux scattered into our direction by gas in an annulus, $dR$, is then,
\begin{equation}
\label{eqn:7}
dF_{s}\, =\, 2\pi R F_R N_s\, \frac{d\sigma}{d\Omega}dR
\quad ,
\end{equation}
where $N_s$ is the column density of scattering molecules along the line of sight from the continuum source to the emitting gas, and $d\sigma/d\Omega$ is the differential scattering cross section, approximated by $\sigma/4\pi$. We find then for the line to continuum ratio that the observer sees,
\begin{equation}
\label{eqn:8}
\frac{F_s-F_c}{F_c}\, =\, \frac{F_L}{F_c} =\, \frac{\tau_{s}}{4\pi}\, \ln\left[R_{out}/R_{in}\right]
\end{equation}
where $\tau_{s}$ is the (scattering) optical depth given by, 

\begin{equation}
\label{eqn:9}
\tau_s = \dfrac{c^3}{8 \pi \sqrt{2\pi} \sigma_v \nu^3} A_{ul} N_u \bigg( \dfrac{g_u N_l}{g_l N_u} - 1 \bigg)
\end{equation}

where $\sigma_v$ is the line standard deviation of the line in velocity space and $N_u$ is the column density in the upper level.  In this derivation, we have assumed that the density is low enough that collisional deactivation and thermalisation of the scattered photon energy is unimportant. We neglect the effect of stimulated emission.

After subtracting the continuum, the required HCN scattering optical depth is 0.03 which results in a total column density of 7.9$\pm0.5\times10^{15}$ cm$^{-2}$. This is not relative to the continuum since the continuum origin is not in the same part of the disk. The temperature of the resonant scattering HCN is only 240$\pm$13 K which is lower than the temperature derived from the HCN absorption lines (670$\pm$118 K). For C$_2$H$_2$ we derive a scattering column density of 6.0$\pm0.9\times10^{15}$ cm$^{-2}$ and a temperature of 464$\pm$190, and again we measure a higher temperature from the absorption lines (600$\pm$50 K). If we assume the abundance of HCN as derived from the absorption lines, this column density would result in a H column of 3$\times10^{22}$ cm$^{-2}$. Therefore the optical depth in the continuum of this layer would be a few times 10$^{-2}$.

\subsection{Abundance Gradients in the Disk}

\begin{figure}[b!]
\centering
\begin{tabular}{@{}cccc@{}}
\includegraphics[width=.5\textwidth]{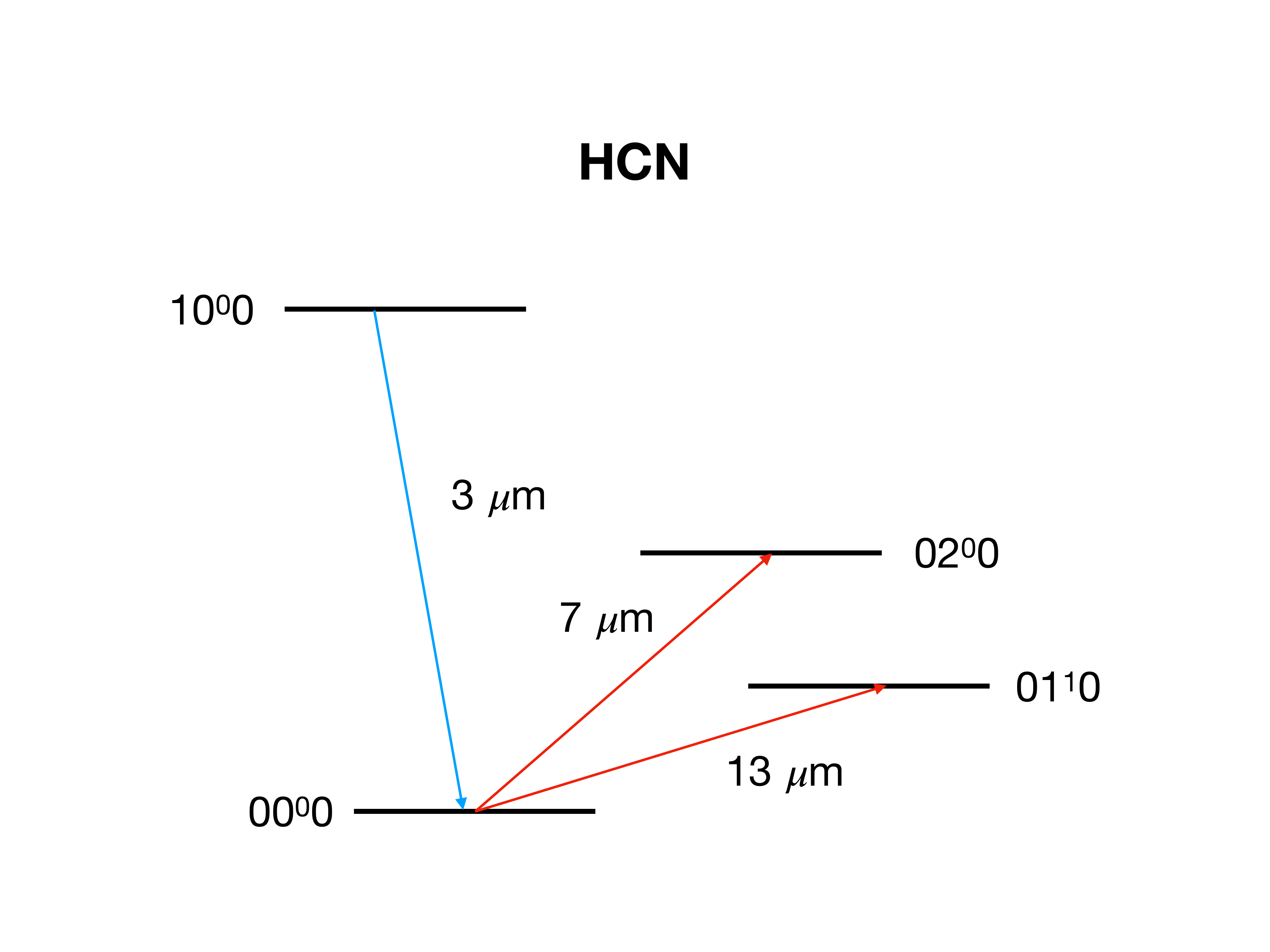} 
\includegraphics[width=.5\textwidth]{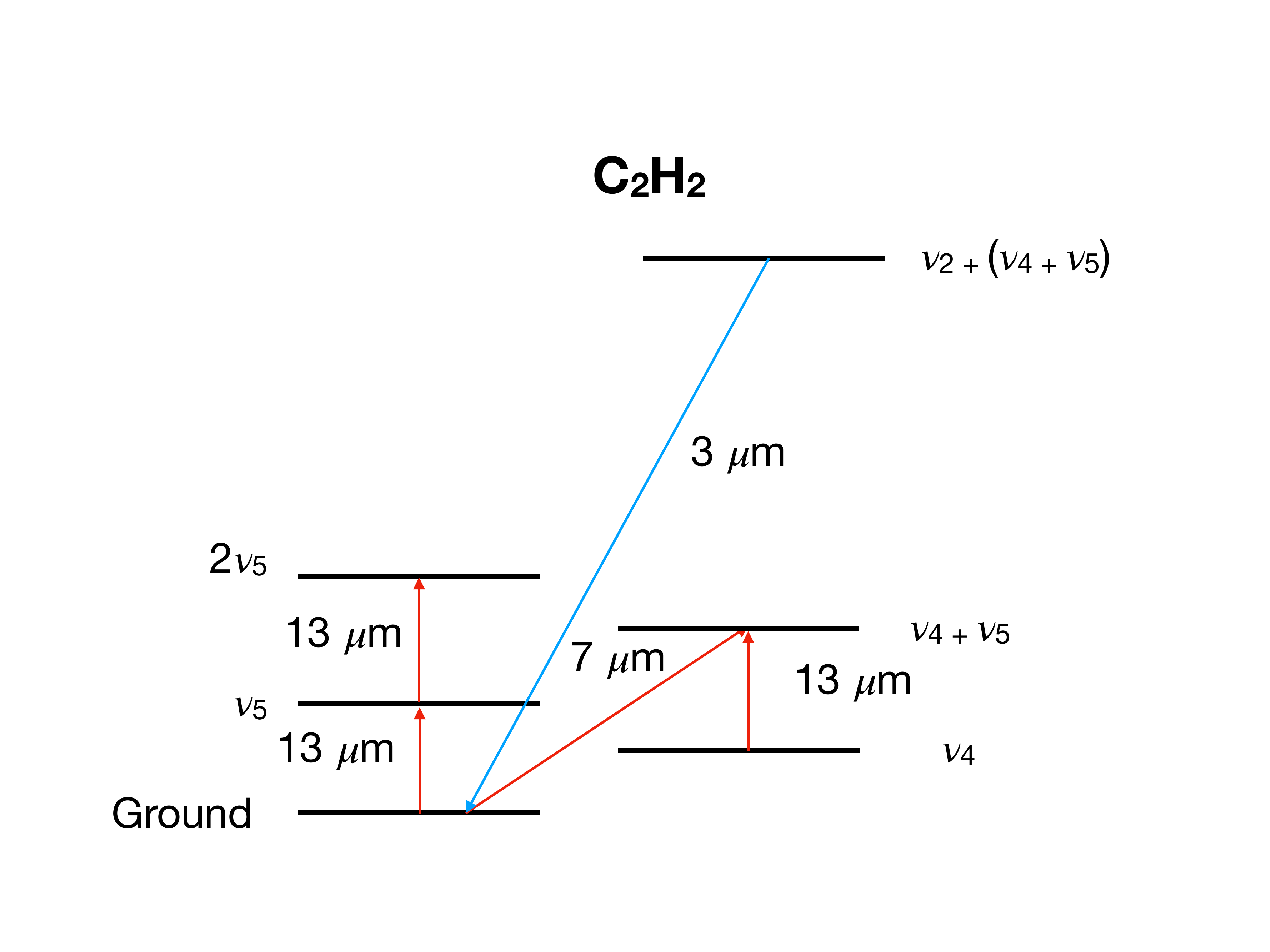} 
\end{tabular}
\caption{Energy level diagrams indicating the observed transitions of HCN (left) and C$_2$H$_2$ (right). The wavelength at which each vibrational transition is observed is indicated. Red arrows indicated transitions seen in absorption and blue arrows transitions seen in emission.}
\label{Elvl}
\end{figure}

Consider the energy level diagrams for HCN and C$_2$H$_2$ shown in Figure \ref{Elvl}. For both species, we measure two transitions originating from the same ground state levels, yet the derived abundances are very different, by up to a factor of 10. Specifically, the abundance of HCN derived from the 13 ${\mu}m$ transition (v$_2$=0-1) is much lower than that derived from the 7 ${\mu}m$ transition (v$_2$=0-2). The same holds for C$_2$H$_2$ in AFGL 2136 where the abundance derived from the ground state to the $\nu_4+\nu_5$ level at 7 ${\mu}m$ is a factor of 4 higher than the abundance derived from the ground state to the $\nu_5$ level. As the 13 ${\mu}m$ continuum originates from further out in the disk than the 7 ${\mu}m$ continuum, this suggests the presence of an abundance gradient with higher abundances in the photosphere of the inner, warmer disk, out to 50 AU (Fig \ref{slice}).  

However, the excitation temperature derived from the rotational diagram analysis is very similar for the transitions probed at 13 ${\mu}m$ and at 7 ${\mu}m$. Most importantly, the observed line profiles are similar (Figs \ref{Lineprof2591}, \ref{Lineprof2136} \& \ref{Comp}). The observed line width will be dominated by Keplerian rotation and suggests absorption by gas at about the same radial location. Thus we locate the absorption in all of these ro-vibrational transitions from HCN and C$_2$H$_2$ to originate from the same gas at about the same depth in the disk photosphere, at about the same radial distance from the protostar. The derived gas temperatures of these transitions are $\sim$600 K which is higher than the dust temperature that is characteristic for 13 ${\mu}m$ continuum, which traces the outer regions of the disk ($\sim$400 K assuming a radial temperature gradient $\propto$ T$^{-0.7}$) This implies a location of the absorbing gas in the warmer inner regions. At 13 ${\mu}m$ however, due to the lack of absorbing species, the outer disk will contribute to the continuum, and only very little to the absorption feature. This dilution by continuum emission will lead to an underestimate of the abundance of the absorbing species (Fig \ref{slice}). Detailed models will be required to assess the importance of this effect. 

In their pioneering study of C$_2$H$_2$ absorption lines in the spectrum of Orion IRC2, \citet{Evans1991} noted a similar effect with a 3 times higher column density derived from the $\nu_4+\nu_5$ (7 ${\mu}m$) transition  than from the $\nu_5$ (13 ${\mu}m$) transition. They located the absorbing material in the protostellar photosphere and attributed this difference to a higher continuum (dust) opacity at 13 ${\mu}m$ than at 7 ${\mu}m$. This would locate the C$_2$H$_2$ gas absorbing at 13 ${\mu}m$ higher up in the photosphere where the abundance could be lower. This means that extinction can only explain around a factor of 2 difference and not the factor of up to 10 we observe. Furthermore it does not explain the similar line profile and temperatures observed at 7 and 13 ${\mu}m$. \citet{BL2012} have seen the same effect for C$_2$H$_2$ absorption in the spectrum of NGC 7538 IRS9. They attributed this to a filling in of absorption by resonant scattering which would be more important for the fundamental ($\nu_5$) band than for the combination band ($\nu_4+\nu_5)$ (Fig \ref{Elvl}). However, for material along the line of sight towards the continuum source, resonant scattering does not play a role as this line scattering process emits into 2$\pi$ steradians. Moreover, even in the depth of the absorption band, the photon emission is dominated by continuum processes rather than line transitions because the continuum opacity is much larger than the line opacity.

The rotation diagrams of HCN and C$_2$H$_2$ at 7 ${\mu}m$ reveal non-linearity. Line profiles of low and high J of these species are consistent with each other which implies that these lines originate from the same physical component (Figs \ref{Lineprof2591} \& \ref{Lineprof2136}). We attribute this to the presence of a temperature gradient in the disk with colder gas higher up in the disk photosphere absorbing more predominantly in the lowest rotational transitions. The fact that this is only observed in these two species, and not, for example, in CS, suggests that these species have a more extended vertical distribution in the disk measuring a relatively steep vertical temperature gradient along the disk photospheres. This could reflect that CS absorption is more concentrated close to $\tau_c=2/3$ than HCN and C$_2$H$_2$. A full chemical and radiative transfer modelling effort would be required to ascertain the exact structure of the disk photospheres, however this is beyond the scope of this paper.

In AFGL 2591, all species are at the same temperature. It is therefore plausible that absorption by all species originates in the same gas. For AFGL 2136 however, a scatter of rotational temperatures of the molecules, indicative of the scatter seen in ISO-SWS studies \citep{Boonman2003, BoonmanVD2003, Keane2001, Lahuis2000}, reveals complex temperature variations with wavelength. This is a reflection of the interplay between the vertical and radial temperature gradients which will require radiative transfer modelling to disentangle. As a result of the grey atmosphere in the disk, all wavelengths will probe to the same depth in the disk. However, due to a combination of the disk being larger at longer wavelengths, and a decrease in the density with increasing radius from the central star, the photosphere layer structure is such that longer wavelengths probe to deeper layers of the mid-plane where the temperatures are higher (Figure \ref{slice})

In contrast to C$_2$H$_2$ and HCN, the spectra do not show absorption lines associated with the N-H bending mode of NH$_3$ in the EXES data range. As the umbrella mode (at 10 ${\mu}m$) is 5 times stronger than the N-H bending mode, the inferred abundance derived from the 10 ${\mu}m$ would not give rise to discernible absorption features at 6 ${\mu}m$. However, we can exclude much higher NH$_3$ abundances in the region responsible for the 6 ${\mu}m$ continuum emission. So, unlike for C$_2$H$_2$ and HCN, NH$_3$ does not show a strong increase in abundance in the radial (inwards) direction.

In the discussion of 3 \& 6 ${\mu}m$ H$_2$O absorption lines, \citet{Indriolo2020} already emphasised that absorption lines originate from a total optical depth of 2/3 which for strong lines is higher up in the disk photosphere than for weaker lines.  Our complete 5-8 ${\mu}m$ spectra of these two sources contains hundreds of H$_2$O lines. This will allow a much more detailed analysis of the origin and distribution of H$_2$O in the disk photosphere and we will come back to this in Barr et al. 2020 (in prep).

\subsection{Vibrationally Excited Absorption}

As presented in section 4.1.1 and 4.21, the vibrational temperatures for $^{12}$CO are 623$\pm$292 K and 490$\pm$39 K for AFGL 2591 and AFGL 2136, respectively. These are in agreement with the rotational temperatures for $^{13}$CO and C$^{18}$O which suggests that CO is in vibrational equilibrium. The vibrational temperatures for the $\nu_5$ band of C$_2$H$_2$ are 670$\pm$95 K and 494$\pm$139 K for AFGL 2591 and AFGL 2136, respectively. Again these are in agreement with the rotational temperatures which suggests that this band is in vibrational equilibrium. \citet{Knez2009} noted the presence of vibrationally excited transitions (C$_2$H$_2$ $2\nu_5-\nu_5$) in their spectrum of NGC 7538 IRS1 and concluded that this must imply very dense gas. 

Collisional coupling of the rotational population to the temperature of the gas requires densities in excess of $\sim$ 10$^{10}$ cm$^{-3}$. This limit is consistent with densities in protoplanetary disks, therefore the presence of such high density gas would further imply the presence of a dust disk that dominates the continuum emission for this source. The scale height of the disk is given by $H=C_s/\Omega_k$ with $C_s$ the sound speed ($\sim$ 2 kms$^{-1}$ for 600 K) and $\Omega_k=v_k/r$ with $v_k$ the Keplerian velocity and $r$ the distance to the star. $v_k$ is 30 kms$^{-1}$ at a distance of 50 AU from a 50 M$_{\odot}$ star. This scale height would correspond to a dust optical depth of unity or $\sim$ 7$\times10^{22}$ H-nuclei/cm$^2$. This results in a density of $\sim$ 1$\times10^9$ cm$^{-3}$. Hence, the radiation field plays an important role in the population of the molecular gas. As gas in the photosphere of the disk receives only radiation from 2$\pi$ steradians, the molecular excitation temperature is given by \citep{Tielens2005},

\begin{equation}
 T_x/T_R= 1 + (kT_x/h\nu) \ln[2] 
\end{equation}

in the absence of collisions. $T_x$ is the excitation temperature and $T_R$ is the radiation temperature. As a result, the excitation temperature will be about 25 \% lower than the radiation field temperature. For a reduced black body, collisional excitation dominates over radiative pumping if: 

\begin{equation}
\dfrac{W}{exp(h\nu/kT_R)} \lesssim \dfrac{n}{n_{crti}}
\end{equation}

where $W$ is the dilution factor, $n$ is the density and $n_{crit}$ is the critical density \citep{Tielens2005}. For $W=0.5$, this occurs for $n > 1.7\times10^{10}$ cm$^{-3}$ for the species considered here. As the actual density is likely only 10$^9$ cm$^{-3}$, the radiation field couples the level populations of these species to the continuum (dust) temperature. This implies that we observe weak scattering lines in our spectra.

In both sources, the $^{12}$CO v=0-1 transitions go close to zero, with the low J lines being saturated. Therefore these are strong scattering lines and the $^{12}$CO v=0-1 scattering line opacity is strong compared to the continuum opacity. This implies that collisional de-excitation is unimportant and hence the density is much less than the critical density for the v=0-1 ro-vibrational transitions of CO (10$^{10}$ cm$^{-3}$). 

For AFGL 2136, we derived a C$^{16}$O abundance that was higher than the typical value of 1$\times10^{-4}$ based on the C$^{18}$O abundance. Since the $^{12}$C$^{16}$O absorption lines are saturated in this source, we can make an estimate on what the dust opacity per H-atom should be for these values to agree. If we set the line-to-continuum ratio equal to 1, using equation \ref{eqn:5} we calculate a dust opacity per H-atom of 5.3$\times10^{-22}$ cm$^2$/H-nucleus. Taking this value instead of that quoted in section 5.2 would reduce the calculated absolute abundances in AFGL 2136 by an order of magnitude.

\subsection{The Schematic Structure of AFGL 2136 and AFGL 2591}

\subsubsection{AFGL 2136}

\begin{figure}[h!]
\centering
\begin{tabular}{@{}cccc@{}}
\includegraphics[width=0.8\textwidth]{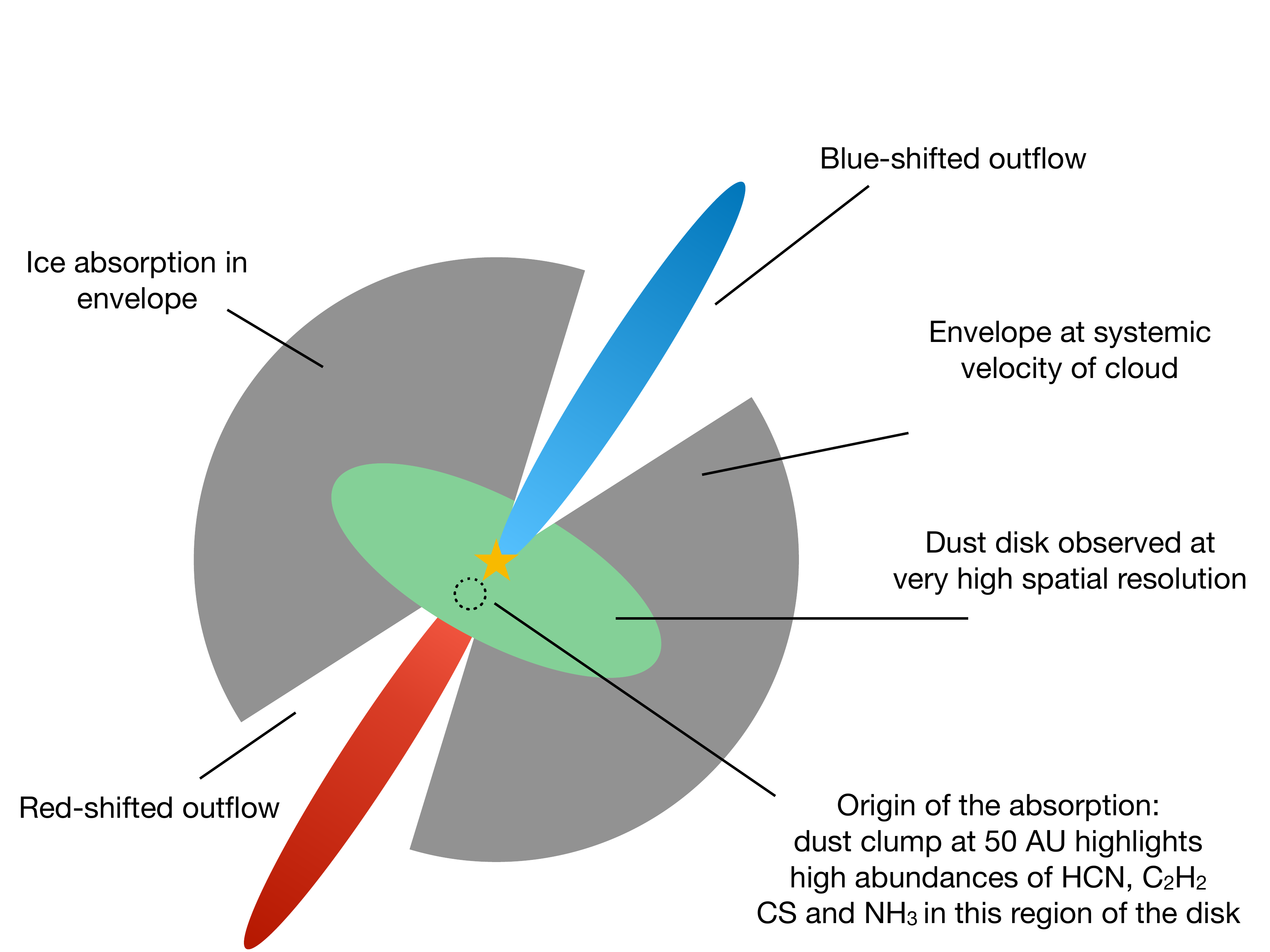}
\end{tabular}
\caption{Cartoon schematic illustrating the different physical components in the large scale environs of the hot cores discussed. Drawing is not to scale.}
\label{cartoon}
\end{figure}

In order to place our results in context with respect to the specifics of each source, we outline a proposed schematic in Figure \ref{cartoon}. For AFGL 2136, as mentioned in section 4.3, at a spatial resolution of 0.2$''$, an unresolved disk-like structure is observed in SiO emission at the systemic velocity, 22 kms$^{-1}$, which traces a disk wind out to 250 AU \citep{Maud2018, Maud2019}. At higher spatial resolution however, of 0.02$''$, SiO kinematics are observed to follow the H$_2$O Keplerian disk \citep{Maud2019} on scales $<$120 AU. 

In the H$_2$O velocity map and position-velocity diagrams that are presented by \citet{Maud2019}, this disk is centred at around 26 kms$^{-1}$ and is seen almost edge-on. The IR absorption that we detect at 26 kms$^{-1}$ is associated with gas in the centre of the Keplerian disk. Therefore we propose that our IR absorption observations trace the minor axis of the disk seen in H$_2$O emission. This is further supported by the high temperatures that we measure of around 400-600 K (Table \ref{sum2136}) which suggest that gas seen in the IR comes from warm gas close to the protostar. 

H$_2$O emission is present across the entire extent of the dust disk. The line width of the 232 GHz H$_2$O line is broader than the line widths of the absorption lines, which would suggest that the absorption actually probes regions further out in the disk. This may be reconciled however by considering the 1.3 mm continuum image in Figure 1b of \citet{Maud2019} where clumps are observed to occur in the disk along the minor axis. These clumps are observed as peaks in the 1.3 mm continuum image. \citet{Indriolo2020} propose that, assuming the continuum structure is the same in the MIR as it is in the sub-mm, locating the H$_2$O absorption in these clumps would explain the difference in line width between the disk absorption and emission. One of these clumps coincides with the velocity of the absorption lines (26 kms$^{-1}$) of the species we discuss here and occurs about 50 AU from the protostar. It should be noted that there is no obvious sign of the 33 kms$^{-1}$ clump, which is present in H$_2$O absorption and 1.3mm continuum, for any of the other molecules except perhaps CO, as seen from the v=2-0 band at 2 ${\mu}m$ \citep{Goto2019} and marginally in a handful of the lines in Figure \ref{2136CO_prof}.

Since the sub-mm observations show the H$_2$O $5_{5,0}-6_{4,3}$ $\nu_2 = 1$ line is in emission \citep{Maud2019}, the disk will be optically thin at sub-mm wavelengths. For interstellar dust properties at 230 GHz, N$_H$ = N(H) + 2N(H$_2$) must be less than 2$\times$10$^{26}$ cm$^{-2}$ to be in the optically thin regime \citep{Draine2003}. This line provides support for the presence of strong temperature gradients in the disk. It originates from a level $\sim$ 3500K above the ground and its emission peaks at a radius of about 50 AU. The very high energy required to excite the H$_2$O emission line requires that it is produced in a very hot environment. This suggests that the line emission originates in the mid-plane of the disk where viscous heating originates, while the absorption lines that we study at the optically thick, mid-IR wavelengths originate from about the same radial location but then in the much cooler, disk photosphere.

The line width of the 22 kms$^{-1}$ velocity component of CO absorption is narrow (2 kms$^{-1}$; Fig \ref{CO2136}). Combined with the low temperature of only 27 K and difference in peak velocity, this suggests that this contribution comes from the cloud or surrounding envelope but not the disk.

\subsubsection{AFGL 2591}

In the case of AFGL 2591 the picture is less clear. We observe 5 different velocity components of $^{13}$CO in absorption. This points to a complex environment in this hot core, which has been previously illustrated by \citet{vanderTak1999}. These authors suggest that the different CO components are picked up along the cavity walls excavated by the outflow which points towards the observer, implying that any disk in this source would be seen face-on. The main velocity component in the IR at -10 kms$^{-1}$ is seen in all of the species presented in section 4.1, not just CO. There have been several studies that point towards the presence of a disk in AFGL 2591, despite no disk being spatially resolved \citep{Wang2012, vanderTak2006, Preibisch2003, VM2005}. Sub-mm observations at 0.5$''$ spatial resolution suggest the presence of a disk wind \citep{Wang2012}. Line profiles best match sub-Keplerian motion along with a contribution from expansion which \citet{Wang2012} attribute to a disk structure with a wind blowing across the surface layer. \citet{vanderTak2006} find evidence for a face-on disk in H$_2$O and dust emission on scales of $<$ 800 AU. In this study, the emission lines are still centred at -5.5 km$^{-1}$ with a line width of up to 5.4 kms$^{-1}$. The beam size of the observations was 2-0.9$''$, and combined with the line profiles shows that the -10 kms$^{-1}$ component is not resolved at these scales.  

We propose that we are viewing AFGL 2591 down to the base of the blue-shifted outflow close to the protostar, the same as \citet{Indriolo2015} suggest for H$_2$O absorption towards AFGL 2591 and \citet{Barr2018} suggest for CS. Due to the similarity in the IR absorption properties of both hot cores, and the tentative evidence for a disk in AFGL 2591, although no disk has been clearly imaged in this source, we propose a similar scenario as for AFGL 2136. The -10 kms$^{-1}$ velocity component seen in IR absorption traces a clump in a circumstellar disk around a high mass protostar(s). It is fortuitous that this clump would also arise from close to the zero point velocity of the disk, as in AFGL 2136, and not at the extreme velocities in the disk, placing the absorption around 50 AU from the protostar. This disk is heated from the inside by viscous heating due to accretion in the mid-plane. \citet{VM2005} see tentative evidence for a compact ionised accretion flow onto the protostar for AFGL 2591, although results may also be consistent with dust emission from a disk. We propose that the other velocity components seen in CO are also due to dust clumps in the disk, as in AFGL 2136. These are present only in CO which is omnipresent throughout the disk.

In a recent sub-mm study of rotational transitions of COMs by \citet{Gieser2019}, the temperature and density structure of AFGL 2591 is modelled along with chemical complexity. The authors find that, at a spatial resolution of 0.5$''$, the source is not resolved below 700 AU. They conclude that no disk structure is observed. \citet{Gieser2019} also show, based on modelling of the temperature distribution in AFGL 2591, that a temperature of around 700 K is reached at a radius of 56 AU from the central star, in agreement with predictions of the source size based on the CS column density and critical density \citep{Barr2018}. This is consistent with inner disk extent estimates from chemical modelling of Herbig disks \citep{Agundez2018, Walsh2015}. For the many other simple and complex species that are detected by \citet{Gieser2019}, low temperatures are derived compared to the results in Table \ref{sum2591}, with a maximum of 212 K for HC$_3$N. Furthermore, the systemic velocity of -5.5 kms$^{-1}$ is observed for all species. This would place the emission from these species in the surrounding envelope of the hot core, likely at the location of the sublimation temperature of the ices.

In the line profiles in Figure \ref{Lineprof2591}, NH$_3$ absorption is seen to occur at a peak velocity of -7 kms$^{-1}$, slightly more red-shifted compared to the other absorption lines. This is still blue-shifted with respect to the systemic velocity, associating the NH$_3$ absorption with the disk. This is supported by the high temperature that we derive for NH$_3$ absorption, of 870 K. The line width is also observed to be narrower than the other absorption lines. Furthermore, from Figure \ref{Lineprof2591}, we see that HCN and C$_2$H$_2$ absorption are slightly shifted in velocity space with respect to each other. These line profiles point towards compositional gradients in a clumpy distribution in the disk. It is difficult to ascertain the exact kinematical nature of the disk without an image of it from the sub-mm, therefore very high spatial resolution ($\lesssim$ 0.05) observations of AFGL 2591 will be necessary for this.

The hot core phase in the evolution of massive stars ends when ionising radiation breaks out of the cocoon around the protostar itself and creates a hypercompact {H\sc{ii}} region. At that point, EUV and FUV photons will start to illuminate the photosphere of the disk and the temperature gradient can then be reversed. At that point, the absorption lines will disappear and gas in the upper regions of the disk will radiate in emission lines. We note that several sources in the neighbourhood of AFGL 2591 are connected with ionised gas \citep{Johnston2013} and hence have already entered in this subsequent phase. We surmise that low mass protostars in the class 0 phase will also show molecular absorption lines in their mid-IR disk spectrum. Finally, we note that ultra-luminous infrared galaxies (ULIRGS) show molecular absorption lines in their mid-IR spectrum \citep{Spoon2013} also indicative of an outwardly decreasing temperature gradient in an optically thick, grey atmosphere.

\subsection{Chemistry in the warm and dense disk photosphere}

The chemistry of the inner regions of TTauri and Herbig disks has been modelled extensively, including HCN, C$_2$H$_2$, NH$_3$ CH$_4$, CS and H$_2$O \citep{Markwick2002, WW2009, Walsh2010, Agundez2008, Agundez2018}, with high abundances of these species predicted to arise due to high temperature gas phase chemistry \citep{Agundez2018, Walsh2015}. For Herbig disks, the high abundance region is expected to be larger and extend to further out regions of the disk than TTauri disks due to the higher temperatures involved, although this is contrary to observations \citep{Pontoppidan2010, Fedele2011}. Detailed modelling studies of the chemistry of massive disks have not been carried out, yet. These would need to be undertaken to determine the exact nature of the chemistry in these disks which may be quite different due to the much higher masses and luminosities of the central object, and also the different heating mechanism which results in an outwardly decreasing temperature gradient in the vertical direction. However, if the same trend of increasing abundance with mass of the disk applies, we can expect that massive disks will show even larger abundances to further out distances in the disk from the protostar, compared with Herbig and TTauri disks.

High abundances of C$_2$H$_2$ and HCN are characteristic of chemistry in warm, dense gas \citep{Agundez2008, Bast2013}. When C and N are broken out of their main reservoir, CO and N$_2$, high temperatures will allow the reaction of C with H$_2$ to CH to dominate the chemistry and a rich chemistry ensues with high abundances of C$_2$H$_2$ and HCN, as well as CH$_4$ and NH$_3$. The chemical models of TTAuri and Herbig disks rely on UV and X-ray photons to break the C and N out of their very stable main parent species. That is relevant for these objects where the stellar photons freely impinge on the photosphere of the disk. However, as the presence of absorption lines attests, disks around these massive stars do not have a photon-dominated region (PDR) surface. The destruction of N$_2$ and CO must therefore rely on cosmic ray ionisation. Cosmic ray ionisation of H$_2$ will lead to protonated H$_2$ and this proton can be handed over to CO and N$_2$. However, dissociative recombination of these protonated species will merely reform the parent species without breaking the strong bonds involved. The formation of C and N atoms requires then cosmic ray ionisation of helium. 

We can then derive a minimum timescale for the disk by requiring that each cosmic ray ionised helium atom leads to either C$_2$H$_2$ or HCN. This results in 
\begin{equation}
\label{eqn:10}
\tau_{chem}\, =\, \frac{X\left(C2H2\right)+X\left(HCN\right)}{X\left(He\right)\zeta_{CR} }
\quad ,
\end{equation}
where $X\left(i\right)$ is the abundance of species $i$ and $\zeta_{CR}$ is the cosmic ray ionisation rate for which we take the nominal value of $3\times 10^{-17}$ s$^{-1}$ commonly adopted for molecular clouds. Adopting a He abundance of 0.06 \citep{Peimbert2007} we derive a timescale of 7.3$\pm0.5\times10^{5}$ yr and 6.2$\pm0.5\times10^{5}$ yr for AFGL 2591 and AFGL 2136 respectively. These timescales are consistent with these two objects being similar in evolutionary stage \citep{BoonmanVD2003}. With the recently modelled values for the cosmic ray ionisation rate towards hot cores being up to 10 times greater than the value we have adopted here \citep{BG2020, Padovani2015, Padovani2016}, we place an upper limit on these calculated chemical timescales. This timescale compares to an age of 2$\times10^4$ yr based on chemical modelling of AFGL 2591 on scales of $>$ 0.4 $''$ \citep{Doty2002, Gieser2019}.

\citet{Indriolo2013, Indriolo2020} detect H$_2$O absorption in the K band in AFGL 2136 which means that this innermost region still has H$_2$O present. We expect that H$_2$O and CO will be omnipresent throughout the disk photosphere, tracing the entire extent of the disk seen in H$_2$O emission at 232 GHz, and do not show abundance gradients. H$_2$O gas-phase formation at high temperature has a low energy barrier of 1736 K \citep{Mcelroy2013}. Enhanced abundances of H$_2$O are predicted to occur for temperatures greater than 250 K \citep{Charnley1997, Doty2002}. 

The abundance of HCN and C$_2$H$_2$, however, will be strongly influenced by chemistry. Both HCN and C$_2$H$_2$ require high temperature gas for their formation as the first step in the chemical pathway to either of these species (the conversion of C to hydrocarbons through reaction C$+$H$_2$ -$>$ CH $+$ H) has a high activation barrier of 12,000 K \citep{Bast2013}. Thus, due to the lower temperature in the outer regions of the disk, chemical enhancement is confined to the inner region only. In the inner dense warm gas regions, however, three body reactions will dissociate H$_2$ and atomic H will break OH out of H$_2$O. The OH radical will eventually burn these hydrocarbons to CO \citep{Kress2008}. Hence, we expect that HCN and C$_2$H$_2$ will occupy a small region in the disk limited on the inside by this burning process and on the outside by kinetics in the low temperature gas (Fig \ref{slice}). 

However, since the size of the dust disk gets larger with wavelength, the continuum source will be larger at 13 ${\mu}m$ than at 7 ${\mu}m$ (Fig \ref{slice}). The longer wavelength continuum will be dominated by lower temperature regions further out in the disk. The consequent abundance gradient will result in a filling in of absorption lines at these wavelengths by dust emission from the photosphere in the outer regions, which has low abundances of absorbing molecules. This results in a dilution of the observed flux at 13 ${\mu}m$, affecting the relative strength of 7 and 13 ${\mu}m$ absorption bands (cf., section 5.4), hence we infer a strong radial gradient in the HCN and C$_2$H$_2$ abundance.

\citet{Agundez2008, Agundez2018} and \citet{Walsh2015} find that the chemistry of HCN and C$_2$H$_2$ in the inner region of disks is similar, therefore it is expected that, since these species exhibit similar chemical behaviour at high temperature, it is likely that they also exhibit similar physical behaviour. Indeed, the derived gas temperatures of these bands are similar ($\sim$600 K), both species show behaviour in the rotation diagrams characteristic of a temperature gradient, and both species exhibit an abundance gradient with wavelength. Furthermore, L-band data of AFGL 2591 show that HCN and C$_2$H$_2$ are detected in emission and NH$_3$ is not detected.

Upper limits of NH$_3$ emission from the inner disks of TTauri disks suggests that NH$_3$ is efficiently destroyed in the inner disk and that the main N-carriers are N$_2$ followed by HCN \citep{Pontoppidan2019}. Disk models, on the other hand, with full nitrogen chemical networks predict gas-phase NH$_3$ abundances of 10$^{-5}$ or 10$^{-6}$ with respect to H for the majority of N originating in NH$_3$ or N$_2$ respectively \citep{SB2014}. We have very accurately determined the NH$_3$ abundance in the disks of both AFGL 2591 and AFGL 2136. High abundances (10$^{-6}$) suggest that NH$_3$ is one of the main N-carriers, along with HCN and most likely N$_2$, in the inner region.

\subsection{Molecular absorption lines in the IR spectra of massive protostars}

This study focuses on AFGL 2136 and AFGL 2591, however several other hot cores have been observed before at high spectral resolution at MIR wavelengths using a few specifically chosen settings. These studies reveal MIR transitions in absorption towards hot cores \citep{Mitchell1989, Mitchell1990, Evans1991, Knez2009, BL2012, Rangwala2018, Dungee2018, Indriolo2013, Indriolo2015, Indriolo2020}. Hot cores have also been observed at low spectral resolution in the MIR with ISO \citep{Lahuis2000, Keane2001, Boonman2003, BoonmanVD2003}.

All theses studies conclude that absorption lines in the IR originate in a region in very close proximity to the illuminating protostar. In the past, molecular absorption lines in the mid-IR spectra of high mass protostars have been interpreted in terms of either a disk, or small foreground blobs along the line of sight towards the star. However, if the continuum emission is associated with a circumstellar disk, as in the case for AFGL 2136, these blobs must by necessity cover a substantial fraction of the disk surface. Moreover, as the rotational temperature is very high, this gas must be located close to a heating source. It is then likely that this gas is really associated with the surface layers of an internally heated disk. This also explains, in a natural way, the observed peak velocity and width of these absorption lines. 

In view of the presence of absorption lines, a temperature gradient of the form we proposed in section 5.1 must also apply to these other hot cores. Viscous heating from the mid-plane up results in a vertically decreasing temperature gradient. These results are in contrast to TTauri stars and Herbig AeBe stars whose MIR spectra are dominated by strong molecular emission lines \citep{CN2011} indicative of an inwardly decreasing temperature gradient in the vertical direction. The high abundances measured indicate the presence of a hot gas-phase chemistry, which has been discussed in section 5.7. This chemistry implies a timescale for these hot cores between 7$\times10^4$ yr and $7\times10^5$ years. We speculate that during the deeply embedded phase of low mass protostars, the Hot Corino gas \citep{Ceccarelli2008} is characterised by mid-IR absorption lines with a disk with an outwardly decreasing temperature gradient in the vertical direction. 

Several massive YSOs show hot CO v=2-0 emission at NIR wavelengths which is best described as arising from a circumstellar disk. This emission arises from within the dust sublimation radius where the disk is gaseous \citep{Ilee2013}. One of the hot cores in this survey, AFGL 4176, shows CO v=2-0 bandhead emission, as well as H$_2$O and CO absorption at 2.5 ${\mu}m$ and 4.6 ${\mu}m$ respectively (A. Karska et al. 2020, in preparation). The absorption is indicative of an outwardly decreasing temperature gradient in the vertical direction, therefore this may be a general characteristic possessed by disks around massive stars, which would suggest that these disks are accreting.

\section{Conclusions}

We have presented here results of the first ever full spectral survey of the 4.5-13 ${\mu}m$ region at high spectral resolution towards a hot core, which utilises EXES, TEXES and iSHELL. MIR ro-vibrational  lines are observed to be in absorption in CO, CS, HCN, C$_2$H$_2$ and NH$_3$. However HCN and C$_2$H$_2$ in the L-band are seen in emission towards AFGL 2591. A rotational diagram analysis reveals that the absorption lines probe high temperature gas of the order of 600 K. Line profiles reveal a difference in line width and peak velocity when compared to the surrounding envelope which is traced by sub-mm emission lines. IR absorption lines are broader and offset by several kms$^{-1}$ with respect to sub-mm lines seen by single-dish telescopes.

A disk around AFGL 2136 has been inferred at MIR wavelengths and has recently been resolved as a Keplerian disk with ALMA. There is also evidence for a potential disk around AFGL 2591. We propose that the IR absorption lines trace this circumstellar disk and due to the presence of absorption lines, we propose that this disk is characterised by an outwardly decreasing temperature gradient in the vertical direction, with viscous heating due to accretion originating in the mid-plane. Temperatures and line profiles are consistent with clumps in these disks at a distance of 50 AU from the protostar.

We have derived abundances for all species detected using stellar atmosphere theory to account for the temperature gradient in the disk. We derive high abundances for all species, of the order 10$^{-6}$ with respect to H. In view of the lower temperature of the emission lines in the L-band, we conclude that these lines trace gas higher up in the photosphere of the disk.

Differences of up to an order of magnitude in the abundances of transitions that trace the same ground state level are measured for HCN and C$_2$H$_2$ in both hot cores. The abundance derived from the transition at 7 ${\mu}m$ is higher than that at 13 ${\mu}m$, however the temperature and line profiles of the two transitions are in agreement. We conclude that the transitions originate from the same gas, whilst we attribute the abundance difference to filling in of absorption lines by dust continuum emission from the outer part of the disk where the abundance of these species is low. The extent of the outer part of the disk increases with wavelength, thus diluting the gas at 13 ${\mu}m$ more than at 7 ${\mu}m$. Thus abundance gradients exist in the disk with the high abundance material constrained to the inner 50 AU of the disk.

The location of the absorption coming from 50 AU from the protostar is supported by chemical models of TTauri and Herbig Ae/Be disks, where high temperatures and abundances are derived for all species observed. Comparison to these models is limited however due to the different physical structure of the disks. In the case of these massive disks, we suggest that C and N are broken out of their main reservoirs by cosmic ray ionisation. The presence of HCN and C$_2$H$_2$ emission at 3 ${\mu}m$ implies that these species may be destroyed by OH radicals in the innermost regions of the disk resulting in a depleted region of HCN and C$_2$H$_2$ gas. Turbulent mixing then transports HCN and C$_2$H$_2$ gas to the upper, cooler layers of the disk photosphere where it is illuminated by the warm inner disk and resonant scattering produces emission lines. High abundances of NH$_3$ implies that this is one of the main N-bearing species at radii of around 50 AU from the central object, along with HCN and likely N$_2$. 

Finally, MIR absorption lines have been observed along sight lines towards several other hot cores. All of these studies observe high abundances of detected species. These observations are consistent with the scenario that we propose in this paper for AFGL 2591 and AFGL 2136. Therefore MIR absorption lines indicating high rotational temperatures towards hot cores may be an indication of massive disks at the centre of these sources.

\acknowledgments

Based [in part] on observations made with the NASA/DLR Stratospheric Observatory for Infrared Astronomy (SOFIA). SOFIA is jointly operated by the Universities Space Research Association, Inc. (USRA), under NASA contract NAS2-97001, and the Deutsches SOFIA Institut (DSI) under DLR contract 50 OK 0901 to the University of Stuttgart. A.G.G.M.T thanks the Spinoza premie of the NWO. D.A.N gratefully acknowledges the support of an USRA SOFIA grant, SOF05-0041. The authors also thank Rachel Smith and Agata Karska for their helpful contributions to this paper, as well as the anonymous referee for the constructive suggestions for improving the manuscript.

\appendix

\section{Derivation of Milne-Eddington Solution}

We will adopt a linear source function on the continuum optical depth scale, $\tau_c$:

\begin{equation}
\label{eqn:11}
S = a + b \tau_c = a + p_{\nu} \tau_{\nu}
\end{equation}

where $p_{\nu} = b/(1+\eta_{\nu})$ transforms the continuum optical depth scale to the total line plus continuum optical depth scale, $\tau_{\nu}$. Here $\eta_{\nu}$ is the line-to-continuum opacity ratio at frequency $\nu$. Ignoring scattering in the continuum, the Milne-Eddington approximation leads to the residual flux of an absorption line \citep{Mihalas1978}:

\begin{equation}
\label{eqn:12}
R_{\nu} = \dfrac{F_{\nu}}{F_c} =  \dfrac{2(p_{\nu} + \sqrt{3\lambda_{\nu}} a)}{(1+\sqrt{\lambda_{\nu}})(b+\sqrt{3}a)}
\end{equation}

Examining this solution we see that, in the case of a pure absorption line, the residual flux will not be equal to 1, even for a strong line. For a pure absorption line, $\epsilon=1$ and $\lambda_{\nu}=1$, where the thermalisation length is: 

\begin{equation}
\lambda_{\nu} = \frac{1+\epsilon \eta_{\nu}}{1+\eta_{\nu}}
\end{equation}

where $\epsilon$ ranges from 0 to 1 and is a measure of the thermalisation of the line, such that the fraction of scattered and absorbed photons during line formation is $(1-\epsilon)$ and $\epsilon$, respectively (see section B for further definition).

For a weak line the residual flux is given by,

\begin{equation}
\label{eqn:14}
R_\nu = \frac{\sqrt{3} a+ b/(1+\eta_\nu)}{\sqrt{3} a+ b}
\end{equation}

This leads to a central depth of 

\begin{equation}
\label{eqn:15}
A_0 = 1 - R_{\nu} = \frac{b (1 - (1+\eta_\nu)^{-1})}{b + \sqrt{3} a}
\end{equation}

For weak lines ($\eta_\nu <<1$), we can use the Taylor expansion to arrive at,

\begin{equation}
\label{eqn:16}
A_0 \simeq \frac{\eta_\nu}{1+\sqrt{3} a/b}
\end{equation}

Thus the strength of the line scales with $\eta_{\nu}$. For a strong line ($\eta_{\nu}\rightarrow \infty$), the central depth of the line is given by,

\begin{equation}
\label{eqn:13}
A_0 = \frac{b}{b+\sqrt{3} a}
\end{equation}

Thus, if the column density is large enough, lines will go down to the same value, $A_0$, independent of the line column density. In this case, the continuum optical depth will be small and hence the stellar atmosphere approximation will break down as the line opacity will be much larger than the continuum opacity.

The expression in eqn \ref{eqn:16} is the same for the weak line approximation in a pure scattering line \citep{Collins2003}. This is because for weak lines the opacity is dominated by the continuum and a scattered photon is lost to the continuum opacity process (thermalisation). So, line scattered photons do not stray from the place where they initially interacted and we will still measure the flux at a depth of $\tau=\tau_c + \tau_l$ =2/3. Strong scattering lines, however, will go to 0 residual flux as photons at any depth will wander around until they are lost.

If we consider equation \ref{eqn:4} in section 5.2, the source function gradient is given by $Y=(1+3a/2b)^{-1}$ \citep{Mihalas1978}. For $a$ we will adopt the Planck function at the temperature derived from the rotation diagram and $b=3/8 X_0 a$, where $X_0$ is given by eqn \ref{eqn:18} below. For a grey atmosphere, we can evaluate the source function gradient to arrive at (Mihalas 1978, section 10.2),

\begin{equation}
\label{eqn:17}
Y = \dfrac{X_0}{X_0 + 4}
\end{equation}

with

\begin{equation}
\label{eqn:18}
 X_0 = \dfrac{h\nu_0/kT_0}{1 - e^{-h\nu_0/kT_0}}
\end{equation}

with $T_0$ the temperature at the depth where the continuum originates and $\nu_0$ the frequency of the lines. At the frequency of the absorption lines ($\sim5 \times 10^{13}$ Hz) and the excitation temperature derived from the rotation diagrams, $X_0$ is in the range 3-19.

\section{Derivation of Stellar Atmosphere Parameters}

Let us consider the equation of radiative transfer:

\begin{equation}
\label{eqn:19}
\mu \dfrac{dI_{\nu}(\mu)}{dx} = (\kappa^c_{\nu} + \kappa^L_{\nu} + \sigma)I_{\nu} - \epsilon^c_{\nu} - \epsilon^L_{\nu} - \sigma J_{\nu}
\end{equation}

where $\kappa_i$ are the absorption coefficients and $\epsilon_i$ are the emission coefficients. $\sigma$ is the scattering contribution to the total absorption coefficient. Assuming that we have LTE in the continuum ($\epsilon^c_{\nu}/\kappa^c_{\nu} = B_{\nu}(T) )$ and that scattering is negligible in the continuum $(\sigma << \kappa^c_{\nu})$, we come to:

\begin{equation}
\label{eqn:20}
\mu \dfrac{dI_{\nu}}{dx} = (\kappa^c_{\nu} + \kappa^L_{\nu}) \big[ I_{\nu} - \dfrac{\kappa^c_{\nu}}{\kappa^c_{\nu} + \kappa^L_{\nu}} B_{\nu} - \dfrac{\kappa^L_{\nu}}{\kappa^c_{\nu} + \kappa^L_{\nu}} S^L_{\nu} \big]
\end{equation}

Now let us define $\eta_{\nu} = \dfrac{\kappa^L_{\nu}}{\kappa^c_{\nu}}$ and $\lambda_{\nu} = \dfrac{1+\epsilon \eta_{\nu}}{1 + \eta_{\nu}}$. Making a change of variables from $x$ to $\tau_{\nu}$:

\begin{equation}
\label{eqn:21}
d\tau_{\nu} = (\kappa^c_{\nu} + \kappa^L_{\nu})dx = k^c_{\nu}(1+ \eta_{\nu})dx
\end{equation}

for the equation of radiative transfer we arrive at the Milne-Eddington approximation:

\begin{equation}
\label{eqn:22}
\mu \dfrac{dI_{\nu}}{d\tau_{\nu}} = I_{\nu} - \lambda_{\nu}B_{\nu} - (1-\lambda_{\nu}) \int_{0}^{\infty} \phi_{\nu} J_{\nu} d{\nu}
\end{equation}

The definition of the thermalisation of the line $\epsilon$ used in equations \ref{eqn:13} and \ref{eqn:14} is:

\begin{equation}
\label{eqn:23}
\epsilon = \dfrac{\kappa_c}{\kappa_c + \sigma}
\end{equation}

where $\kappa_c$ is the continuum opacity and $\sigma$ the term from equation \ref{eqn:19}. Note that this is not the same as the emission coefficient in the equation of radiative transfer, $\epsilon_{\nu}^c$ (Eqn \ref{eqn:19}), but rather this $\epsilon$ without a subscript is that from section 5.2. When scattering is important, $\sigma \rightarrow \infty$, thus $\epsilon$ becomes 0. When scattering is negligible, $\sigma=0$ therefore $\epsilon=1$.

\LTcapwidth=\textwidth

\tabcolsep=0.05cm
\begin{longtable}{cccccccccc}  
\caption{Line Parameters for AFGL 2591. E$_l$ is the energy of the lower level of the transition, g$_l$ is the statistical weight of the lower level, A$_{ij}$ is the Einstein A coefficient of the transition, v$_{lsr}$ is the centroid velocity of the line, ${\Delta}v$ is the FWHM of the line, $\tau_0$ is the optical depth at line centre and $N_l/N_H$ is the abundance in the lower level of the transition. Line data were taken from the HITRAN database \citep{Gordon2017}.}\\
\hline\hline
Species & Transition & $\lambda$ (${\mu}m$) & E$_l$ (K)  & g$_l$ & A$_{ij}$ (s$^{-1}$)  & $v_{lsr}$ (kms$^{-1}$) & ${\Delta}v$ (kms$^{-1}$) & ${\tau}_0$ & $N_l/N_H$ ($\times10^{-7}$) \\
\hline
\endfirsthead
\caption{continued.}\\
\hline\hline
Species & Transition & $\lambda$ (${\mu}m$) & E$_l$ (K)  & g$_l$ & A$_{ij}$ (s$^{-1}$)  & $v_{lsr}$ (kms$^{-1}$) & ${\Delta}v$ (kms$^{-1}$) & ${\tau}_0$ & $N_l/N_H$ ($\times10^{-7}$) \\
\hline
\endhead
\hline
\label{gl2591}
\endfoot
$^{13}$CO v=0-1 & P(1) & 4.7792 & 5.2 & 6 & 32.4 & -9.4 $\pm0.2$ & 1.5 & 0.50 $\pm0.04$ & 72.4 $\pm7.3$\\
&P(2) & 4.7877 & 15.8 & 10 & 21.5 & -9.3 $\pm0.1$ & 1.5 & 0.57 $\pm0.02$ & 77.5 $\pm3.6$ \\
& R(2) & 4.7463 & 15.8 & 10 & 14.2 & -9.7 $\pm0.3$ & 1.5 & 0.68 $\pm0.07$ & 67.0 $\pm8.0$ \\
&R(3) & 4.7383 & 31.7 & 14 & 14.8 & -9.3 $\pm0.1$ & 1.5 & 0.70 $\pm0.02$ & 68.6 $\pm2.7$ \\
& P(3) & 4.7963 & 31.7 & 14 & 19.3 & -9.3 $\pm0.2$ & 1.5 & 0.61 $\pm0.04$ & 79.0 $\pm6.2$ \\
& P(4) & 4.8050 & 52.8 & 18 & 18.2 & -9.3 $\pm0.2$ & 1.5 & 0.57 $\pm0.03$ & 72.4 $\pm5.2$ \\
&R(5) & 4.7227 & 79.3 & 22 & 15.5 & -8.3 $\pm0.2$ & 1.5 & 0.44 $\pm0.06$ & 38.6 $\pm5.7$ \\
& R(6) & 4.7150 & 111.0 & 26 & 15.8 & -9.1 $\pm0.3$ & 1.5 & 0.31 $\pm0.06$ & 29.3 $\pm5.9$ \\
& R(9) & 4.6927 & 237.9 & 38 & 16.4 & -12.1 $\pm0.7$ & 10.5 & 0.14 $\pm0.01$ & 56.5 $\pm7.5$ \\
& P(9) & 4.8501 & 237.9 & 38 & 16.4 & -13.2 $\pm0.8$ & 10.5 & 0.10 $\pm0.01$ & 42.7$\pm6.3$ \\
& R(10) & 4.6853 & 290.8 & 42 & 16.5 & -10.5 $\pm0.6$ & 10.5 & 0.10 $\pm0.01$ & 41.4 $\pm6.3$ \\
& R(11) & 4.6782 & 349.0 & 46 & 16.7 & -12.0 $\pm0.5$ & 10.5 & 0.11 $\pm0.01$ & 39.0 $\pm5.0$ \\
&P(11) & 4.8689 & 349.0 & 46 & 16.1 & -13.0 $\pm0.6$ & 10.5 &  0.05 $\pm0.01$ & 23.9 $\pm3.8$ \\
& R(12) & 4.6711 & 412.3 & 50 & 16.8 & -12.0 $\pm0.6$ & 10.5 & 0.09 $\pm0.01$ & 38.9 $\pm3.8$ \\
& R(13) & 4.6641 & 481.1 & 54 & 16.9 & -12.0 $\pm0.4$ & 10.5 & 0.09 $\pm0.01$ & 37.7 $\pm3.8$ \\
& R(16) & 4.6437 & 718.8 & 66 & 17.3 & -10.8 $\pm0.5$ & 10.5 & 0.10 $\pm0.01$ & 38.9 $\pm 3.8$ \\
\hline
$^{12}$CO v=1-2	&	R(6)	&	4.6675	&	3199.5	&	13	&	33.2	&	-9.9	$\pm$	0.6	&	7.6	$\pm$	0.7	&	0.107	$\pm$	0.007	&	42.7	$\pm$	3.8	\\
&	R(7)	&	4.6598	&	3237.9	&	15	&	33.7	&	-8.3	$\pm$	0.7	&	7.7	$\pm$	0.8	&	0.112	$\pm$	0.008	&	41.4	$\pm$	6.3	\\
&	R(8)	&	4.6523	&	3281.8	&	17	&	34.1	&	-9.2	$\pm$	0.4	&	7	$\pm$	0.4	&	0.128	$\pm$	0.006	&	38.9	$\pm$	6.3	\\
&	R(9)	&	4.6448	&	3331	&	19	&	34.4	&	-8.6	$\pm$	0.4	&	6.8	$\pm$	0.4	&	0.116	$\pm$	0.005	&	40.2	$\pm$	2.5	\\
&	R(17)	&	4.5887	&	3922.5	&	35	&	36.6	&	-8.6	$\pm$	0.3	&	6.2	$\pm$	0.4	&	0.105	$\pm$	0.005	&	27.6	$\pm$	2.5	\\
&	R(18)	&	4.8948	&	4020.9	&	37	&	31.7	&	-7.2	$\pm$	0.3	&	4.8	$\pm$	0.4	&	0.086	$\pm$	0.004	&	27.6	$\pm$	2.5	\\
&	R(19)	&	4.9054	&	4124.8	&	39	&	31.4	&	-9.9	$\pm$	0.8	&	5.2	$\pm$	1.6	&	0.105	$\pm$	0.009	&	30.1	$\pm$	5.0	\\
&	R(20)	&	4.5694	&	4234.4	&	41	&	37.2	&	-6.3	$\pm$	0.4	&	5.8	$\pm$	0.5	&	0.088	$\pm$	0.005	&	25.1	$\pm$	3.8	\\
 \hline
CS v=0-1	&	P(8)	&	7.9443	&	84.6	&	17	&	8.1	&	-10.9	$\pm$	0.5	&	9.7	$\pm$	1.5	&	0.055	$\pm$	0.006	&	12.6	$\pm$	2.5	\\
&	R(3)	&	7.8212	&	14.1	&	7	&	7.2	&	-11.9	$\pm$	0.5	&	8.3	$\pm$	1.5	&	0.031	$\pm$	0.004	&	5.0	$\pm$	1.3	\\
&	R(4)	&	7.8116	&	23.5	&	9	&	7.4	&	-9.4	$\pm$	0.5	&	6.2	$\pm$	1.3	&	0.032	$\pm$	0.004	&	5.0	$\pm$	1.3	\\
&	R(5)	&	7.802	&	35.3	&	11	&	7.6	&	-10.1	$\pm$	0.5	&	7.8	$\pm$	1.3	&	0.039	$\pm$	0.004	&	6.3	$\pm$	1.3	\\
&	R(7)	&	7.7833	&	65.8	&	15	&	7.8	&	-10.4	$\pm$	0.2	&	6.4	$\pm$	0.6	&	0.056	$\pm$	0.003	&	8.8	$\pm$	1.3	\\
&	R(9)	&	7.7649	&	105.8	&	19	&	8	&	-9.5	$\pm$	0.6	&	11.2	$\pm$	1.8	&	0.051	$\pm$	0.005	&	11.3	$\pm$	2.5	\\
&	R(10)	&	7.7559	&	129.2	&	21	&	8	&	-11.1	$\pm$	0.3	&	10.4	$\pm$	0.9	&	0.061	$\pm$	0.004	&	12.6	$\pm$	1.3	\\
&	R(11)	&	7.7469	&	155.1	&	23	&	8.1	&	-12.1	$\pm$	0.3	&	9.9	$\pm$	0.8	&	0.06	$\pm$	0.003	&	12.6	$\pm$	1.3	\\
&	R(18)	&	7.6868	&	401.8	&	37	&	8.5	&	-12	$\pm$	0.4	&	11	$\pm$	1.2	&	0.063	$\pm$	0.004	&	13.8	$\pm$	2.5	\\
&	R(22)	&	7.6545	&	594.5	&	45	&	8.7	&	-12.4	$\pm$	0.9	&	11.3	$\pm$	3.7	&	0.081	$\pm$	0.015	&	17.6	$\pm$	6.3	\\
&	R(23)	&	7.6467	&	648.5	&	47	&	8.8	&	-10.3	$\pm$	0.7	&	8.4	$\pm$	1.7	&	0.064	$\pm$	0.009	&	11.3	$\pm$	3.8	\\
&	R(24)	&	7.6389	&	704.7	&	49	&	8.8	&	-5.2	$\pm$	0.8	&	15.2	$\pm$	2.2	&	0.06	$\pm$	0.006	&	16.3	$\pm$	5.0	\\
&	R(26)	&	7.6237	&	824.5	&	53	&	8.9	&	-8.4	$\pm$	0.6	&	7.9	$\pm$	2.1	&	0.048	$\pm$	0.007	&	7.5	$\pm$	2.5	\\
&	R(27)	&	7.6162	&	887.8	&	55	&	9	&	-11.2	$\pm$	0.4	&	4.9	$\pm$	1	&	0.076	$\pm$	0.008	&	10.0	$\pm$	2.5	\\
&	R(28)	&	7.6088	&	953.4	&	57	&	9	&	-10	$\pm$	0.6	&	6.8	$\pm$	1.5	&	0.055	$\pm$	0.008	&	8.8	$\pm$	2.5	\\
&	R(29)	&	7.6015	&	1021.5	&	59	&	9.1	&	-11	$\pm$	0.4	&	6.7	$\pm$	1.2	&	0.078	$\pm$	0.008	&	12.6	$\pm$	2.5	\\
&	]R(31)	&	7.5871	&	1164.5	&	63	&	9.2	&	-7.9	$\pm$	0.9	&	11.5	$\pm$	4.2	&	0.024	$\pm$	0.005	&	5.0	$\pm$	2.5	\\
&	R(33)	&	7.5731	&	1316.8	&	67	&	9.2	&	-9.1	$\pm$	0.6	&	5.2	$\pm$	1.5	&	0.051	$\pm$	0.009	&	7.5	$\pm$	2.5	\\
\hline
HCN $\nu_2$ v$_2$=0-1	&	R(11e)	&	13.3796	&	280.6	&	138	&	1.2	&	-11.7	$\pm$	0.4	&	8.5	$\pm$	1.1	&	0.098	$\pm$	0.009	&	26.4	$\pm$	3.8	\\
&	R(12e)	&	13.3271	&	331.7	&	150	&	1.2	&	-11.7	$\pm$	0.7	&	9.4	$\pm$	1.8	&	0.078	$\pm$	0.011	&	23.9	$\pm$	6.3	\\
&	R(13e)	&	13.275	&	387	&	162	&	1.3	&	-11.1	$\pm$	0.3	&	8.7	$\pm$	0.8	&	0.102	$\pm$	0.007	&	28.9	$\pm$	3.8	\\
&	R(14e)	&	13.2233	&	446.4	&	174	&	1.3	&	-9.5	$\pm$	0.3	&	10.1	$\pm$	0.8	&	0.096	$\pm$	0.005	&	30.1	$\pm$	3.8	\\
&	R(16e)	&	13.1213	&	578.2	&	198	&	1.3	&	-8.4	$\pm$	0.6	&	9.1	$\pm$	1.4	&	0.102	$\pm$	0.012	&	30.1	$\pm$	6.3	\\
&	R(17e)	&	13.0709	&	650.5	&	210	&	1.3	&	-7.7	$\pm$	0.3	&	6.5	$\pm$	0.8	&	0.07	$\pm$	0.006	&	16.3	$\pm$	2.5	\\
&	R(18e)	&	13.0209	&	727	&	222	&	1.3	&	-8.7	$\pm$	0.2	&	8.8	$\pm$	0.6	&	0.09	$\pm$	0.004	&	26.4	$\pm$	2.5	\\
&	R(19e)	&	12.9713	&	807.8	&	234	&	1.3	&	-7.7	$\pm$	0.4	&	7.7	$\pm$	0.9	&	0.077	$\pm$	0.007	&	20.1	$\pm$	3.8	\\
&	R(20e)	&	12.9221	&	892.7	&	246	&	1.3	&	-9.4	$\pm$	0.2	&	8.5	$\pm$	0.7	&	0.062	$\pm$	0.004	&	17.6	$\pm$	2.5	\\
&	R(21e)	&	12.8734	&	981.9	&	258	&	1.4	&	-9.4	$\pm$	0.4	&	8.3	$\pm$	1	&	0.058	$\pm$	0.005	&	16.3	$\pm$	2.5	\\
&	R(23e)	&	12.7771	&	1172.9	&	282	&	1.4	&	-8.3	$\pm$	0.3	&	8	$\pm$	0.7	&	0.05	$\pm$	0.003	&	13.8	$\pm$	1.3	\\
&	R(24e)	&	12.7295	&	1274.8	&	294	&	1.4	&	-9.5	$\pm$	0.5	&	7.3	$\pm$	1.3	&	0.039	$\pm$	0.005	&	10.0	$\pm$	2.5	\\
&	R(25e)	&	12.6823	&	1381	&	306	&	1.4	&	-8.5	$\pm$	0.3	&	7	$\pm$	0.7	&	0.05	$\pm$	0.004	&	12.6	$\pm$	1.3	\\
&	R(27e)	&	12.5891	&	1605.8	&	330	&	1.4	&	-8.2	$\pm$	0.3	&	6	$\pm$	0.9	&	0.034	$\pm$	0.003	&	7.5	$\pm$	1.3	\\
&	R(29e)	&	12.4974	&	1847.6	&	354	&	1.5	&	-6.1	$\pm$	0.7	&	9.1	$\pm$	1.7	&	0.028	$\pm$	0.004	&	8.8	$\pm$	2.5	\\
&	R(30e)	&	12.4521	&	1974.8	&	366	&	1.5	&	-8	$\pm$	0.8	&	5.8	$\pm$	2.1	&	0.026	$\pm$	0.007	&	6.3	$\pm$	2.5	\\
&	R(31e)	&	12.4072	&	2106.2	&	378	&	1.5	&	-10.3	$\pm$	1.0	&	7.5	$\pm$	2.6	&	0.021	$\pm$	0.005	&	5.0	$\pm$	2.5	\\
&	R(32e)	&	12.3627	&	2241.7	&	390	&	1.5	&	-6.4	$\pm$	1.0	&	8.6	$\pm$	2.7	&	0.013	$\pm$	0.003	&	3.8	$\pm$	1.3	\\
&	R(33e)	&	12.3185	&	2381.6	&	402	&	1.5	&	-6.3	$\pm$	0.8	&	8.4	$\pm$	2	&	0.013	$\pm$	0.002	&	3.8	$\pm$	1.3	\\
&	R(34e)	&	12.2747	&	2525.5	&	414	&	1.5	&	-9.4	$\pm$	0.6	&	8.5	$\pm$	1.5	&	0.015	$\pm$	0.002	&	5.0	$\pm$	1.3	\\
\hline
HCN $\nu_2$ v$_2$=0-2	&	P(14)	&	7.2923	&	446.4	&	174	&	1.1	&	-13	$\pm$	0.4	&	12.7	$\pm$	1.2	&	0.083	$\pm$	0.005	&	208.5	$\pm$	22.6	\\
&	P(13)	&	7.2775	&	387	&	162	&	1.1	&	-11.2	$\pm$	0.4	&	9.3	$\pm$	1.2	&	0.063	$\pm$	0.005	&	126.0	$\pm$	16.3	\\
&	P(12)	&	7.2627	&	331.7	&	150	&	1.1	&	-11.2	$\pm$	0.3	&	10.5	$\pm$	0.7	&	0.078	$\pm$	0.003	&	169.6	$\pm$	12.6	\\
&	P(10)	&	7.2332	&	233.9	&	126	&	1.1	&	-10.9	$\pm$	0.6	&	9.8	$\pm$	1.5	&	0.076	$\pm$	0.008	&	154.5	$\pm$	27.6	\\
&	P(9)	&	7.2184	&	191.4	&	114	&	1.1	&	-10.9	$\pm$	0.3	&	12.2	$\pm$	1	&	0.074	$\pm$	0.004	&	189.7	$\pm$	17.6	\\
&	P(8)	&	7.2037	&	153.1	&	102	&	1.1	&	-10.9	$\pm$	0.4	&	12.9	$\pm$	1.9	&	0.078	$\pm$	0.007	&	178.4	$\pm$	27.6	\\
&	P(6)	&	7.1741	&	89.2	&	78	&	1.2	&	-11.8	$\pm$	0.4	&	7.9	$\pm$	1	&	0.092	$\pm$	0.007	&	170.8	$\pm$	21.4	\\
&	P(5)	&	7.1593	&	63.8	&	66	&	1.2	&	-11.1	$\pm$	0.3	&	9.1	$\pm$	0.9	&	0.059	$\pm$	0.004	&	126.9	$\pm$	13.8	\\
&	P(4)	&	7.1445	&	42.5	&	54	&	1.2	&	-12.4	$\pm$	0.4	&	13	$\pm$	0.2	&	0.069	$\pm$	0.005	&	197.2	$\pm$	16.3	\\
&	P(3)	&	7.1297	&	25.5	&	42	&	1.3	&	-11	$\pm$	0.6	&	13	$\pm$	1.5	&	0.05	$\pm$	0.004	&	152.0	$\pm$	21.4	\\
&	P(2)	&	7.1148	&	12.7	&	30	&	1.4	&	-10.3	$\pm$	0.8	&	10.6	$\pm$	2	&	0.046	$\pm$	0.006	&	124.3	$\pm$	27.6	\\
&	R(0)	&	7.0702	&	0	&	6	&	0.7	&	-22.2	$\pm$	0.5	&	5.5	$\pm$	1.2	&	0.083	$\pm$	0.01	&	59.0	$\pm$	10.0	\\
&	R(2)	&	7.0404	&	12.7	&	30	&	0.9	&	-11.3	$\pm$	0.9	&	10.1	$\pm$	2.8	&	0.078	$\pm$	0.014	&	139.4	$\pm$	41.4	\\
&	R(4)	&	7.0106	&	42.5	&	54	&	1	&	-11.7	$\pm$	0.4	&	11	$\pm$	0.9	&	0.102	$\pm$	0.006	&	207.2	$\pm$	1.6	\\
&	R(5)	&	6.9956	&	63.8	&	66	&	1	&	-11.1	$\pm$	0.7	&	13	$\pm$	2.3	&	0.06	$\pm$	0.018	&	147.0	$\pm$	46.5	\\
&	R(8)	&	6.9509	&	153.1	&	102	&	1.1	&	-9.8	$\pm$	0.4	&	5.5	$\pm$	0.4	&	0.095	$\pm$	0.009	&	129.4	$\pm$	15.1	\\
&	R(9)	&	6.936	&	191.4	&	114	&	1.1	&	-12.7	$\pm$	0.4	&	13	$\pm$	0.5	&	0.08	$\pm$	0.009	&	204.7	$\pm$	25.1	\\
&	R(10)	&	6.9211	&	233.9	&	126	&	1.1	&	-12.9	$\pm$	0.5	&	8.8	$\pm$	1.3	&	0.079	$\pm$	0.007	&	148.2	$\pm$	22.6	\\
&	R(12)	&	6.8914	&	331.7	&	150	&	1.1	&	-13.6	$\pm$	0.4	&	11.7	$\pm$	1.2	&	0.096	$\pm$	0.007	&	222.3	$\pm$	25.1	\\
&	R(13)	&	6.8766	&	387	&	162	&	1.1	&	-13.8	$\pm$	0.8	&	10.8	$\pm$	2.6	&	0.067	$\pm$	0.01	&	149.5	$\pm$	36.4	\\
&	R(14)	&	6.8619	&	446.4	&	174	&	1.1	&	-12.6	$\pm$	0.7	&	9.2	$\pm$	1.8	&	0.08	$\pm$	0.011	&	162.0	$\pm$	35.2	\\
&	R(15)	&	6.8472	&	510.3	&	186	&	1.1	&	-11.1	$\pm$	0.6	&	13	$\pm$	0.7	&	0.083	$\pm$	0.007	&	218.5	$\pm$	26.4	\\
\hline
HCN $\nu_1$ v$=0-1$ & P(18) & 3.0721 & 727.0 & 222 & 40.9 & -9.2$\pm$0.9 & 4.6$\pm$2.4 & 0.008$\pm$0.003 & -- \\
 & P(17) & 3.0690 & 650.5 & 210 & 41.0 & -9.9$\pm$0.8 & 4.6$\pm$0.8 & 0.012$\pm$0.003 & -- \\
 & P(16) & 3.0659 & 578.2 & 198 & 41.2 & -9.8$\pm$0.6 & 4.6$\pm$0.5 & 0.014$\pm$0.005 & -- \\
 & P(14) & 3.0598 & 446.4 & 174 & 41.5 & -9.6$\pm$0.2 & 5.9$\pm$0.6 & 0.022$\pm$0.002 & -- \\
 & P(13) & 3.0568 & 387.0 & 162 & 41.7 & -9.0$\pm$0.5 & 5.5$\pm$1.2 & 0.022$\pm$0.003  & -- \\
 & P(12) & 3.0538 & 331.7 & 150 & 41.9 & -9.4$\pm$0.3 & 5.9$\pm$0.7 & 0.029$\pm$0.002 & -- \\
 & P(11) & 3.0508 & 280.6 & 138 & 42.2 & -10.4$\pm$0.5 & 7.3$\pm$1.3 & 0.025$\pm$0.003 & -- \\
 & P(10) & 3.0479 & 233.9 & 126 & 42.5 & -9.6$\pm$0.2 & 5.3$\pm$0.6 & 0.026$\pm$0.002 & -- \\
 & P(8) & 3.0421 & 153.1 & 102 & 43.1 & -9.3$\pm$0.3 & 7.9$\pm$0.8 & 0.034$\pm$0.003 &  -- \\
 & P(6) & 3.0363 & 89.2 & 78 & 44.3 & -10.3$\pm$0.6 & 9.9$\pm$1.5 & 0.029$\pm$0.003 & -- \\
 & P(4) & 3.0307 & 42.5 & 54 & 46.5 & -9.8$\pm$0.4 & 8.7$\pm$1.0 & 0.029$\pm$0.002 & -- \\
 & P(3) & 3.0280 & 25.5 & 42 & 48.9 & -10.1$\pm$0.2 & 5.3$\pm$0.5 & 0.028$\pm$0.002 &  -- \\
 & P(2) & 3.0252 & 12.7 & 30 & 54.4 & -9.1$\pm$0.6 & 6.1$\pm$1.6 & 0.016$\pm$0.003 & -- \\
 & R(0) & 3.0171 & 0.0 & 6 & 27.3 & -10.2$\pm$0.7 & 4.6$\pm$1.5 & 0.013$\pm$0.003 & -- \\
 & R(3) & 3.0092 & 25.5 & 42 & 36.5 & -8.8$\pm$0.2 & 6.5$\pm$0.6 & 0.035$\pm$0.002 & -- \\
 & R(5) & 3.0041 & 63.8 & 66 & 38.0 & -8.4$\pm$0.2 & 5.0$\pm$0.5 & 0.045$\pm$0.003 & -- \\
 & R(6) & 3.0016 & 89.2 & 78 & 38.5 & -9.8$\pm$0.2 & 6.3$\pm$0.5 & 0.024$\pm$0.001 & -- \\
 & R(8) & 2.9966 & 153.1 & 102 & 39.1 & -10.6$\pm$0.6 & 8.5$\pm$1.6 & 0.038 $\pm$0.005 & -- \\
 & R(11) & 2.9893 & 280.6 & 138 & 39.7 & -9.5$\pm$0.4 & 6.7$\pm$1.1 & 0.015$\pm$0.002 & -- \\
 & R(13) & 2.9845 & 387.0 & 162 & 40.0 & -10.6$\pm$0.8 & 11.1$\pm$0.5 & 0.022$\pm$0.004 & -- \\
\hline
p-C$_2$H$_2$ $\nu_5$	&	R(10e)	&	13.245	&	186.2	&	21	&	3.5	&	-9.8	$\pm$	0.3	&	8.4	$\pm$	0.9	&	0.079	$\pm$	0.006	&	7.9	$\pm$	1.0	\\
&	R(12e)	&	13.1632	&	264.1	&	25	&	3.5	&	-10	$\pm$	0.4	&	9.6	$\pm$	1.1	&	0.073	$\pm$	0.006	&	8.8	$\pm$	1.3	\\
&	R(14e)	&	13.0825	&	355.5	&	29	&	3.6	&	-8.4	$\pm$	0.3	&	7.9	$\pm$	0.8	&	0.072	$\pm$	0.006	&	8.3	$\pm$	1.0	\\
&	R(16e)	&	13.0028	&	460.4	&	33	&	3.6	&	-7.7	$\pm$	0.3	&	5.1	$\pm$	0.8	&	0.06	$\pm$	0.006	&	4.9	$\pm$	0.8	\\
&	R(20e)	&	12.8466	&	710.8	&	41	&	3.8	&	-10	$\pm$	0.5	&	7.3	$\pm$	1.3	&	0.038	$\pm$	0.005	&	4.6	$\pm$	0.8	\\
&	R(22e)	&	12.77	&	856.2	&	45	&	3.9	&	-8.1	$\pm$	0.3	&	6.8	$\pm$	0.8	&	0.047	$\pm$	0.003	&	4.1	$\pm$	0.5	\\
&	R(24e)	&	12.6944	&	1015.2	&	49	&	3.9	&	-7.6	$\pm$	0.4	&	7.4	$\pm$	1.2	&	0.039	$\pm$	0.004	&	4.1	$\pm$	0.6	\\
&	R(26e)	&	12.6198	&	1187.6	&	53	&	4	&	-8.3	$\pm$	0.5	&	9.4	$\pm$	1.9	&	0.035	$\pm$	0.005	&	4.3	$\pm$	1.0	\\
&	R(28e)	&	12.5461	&	1373.4	&	57	&	4.1	&	-8.1	$\pm$	0.5	&	9	$\pm$	1.8	&	0.029	$\pm$	0.004	&	2.6	$\pm$	0.8	\\
\hline
o-C$_2$H$_2$ $\nu_5$	&	R(9e)	&	13.2862	&	152.3	&	57	&	3.5	&	-9.4	$\pm$	0.9	&	6	$\pm$	2.4	&	0.106	$\pm$	0.027	&	7.5	$\pm$	2.5	\\
&	R(11e)	&	13.2039	&	223.4	&	69	&	3.5	&	-9.9	$\pm$	0.8	&	7.9	$\pm$	3.1	&	0.083	$\pm$	0.02	&	7.5	$\pm$	2.5	\\
&	R(13e)	&	13.1227	&	308	&	81	&	3.6	&	-9.8	$\pm$	0.3	&	9.6	$\pm$	0.7	&	0.132	$\pm$	0.007	&	13.8	$\pm$	1.3	\\
&	R(15e)	&	13.0425	&	406.3	&	93	&	3.6	&	-10	$\pm$	0.2	&	9.6	$\pm$	0.5	&	0.133	$\pm$	0.005	&	13.8	$\pm$	1.0	\\
&	R(17e)	&	12.9634	&	517.9	&	105	&	3.7	&	-9.9	$\pm$	0.3	&	8.8	$\pm$	0.8	&	0.108	$\pm$	0.007	&	10.0	$\pm$	1.1	\\
&	R(19e)	&	12.8853	&	643	&	117	&	3.7	&	-9.3	$\pm$	0.3	&	9	$\pm$	0.7	&	0.102	$\pm$	0.006	&	10.0	$\pm$	1.1	\\
&	R(21e)	&	12.8082	&	781.7	&	129	&	3.8	&	-9.4	$\pm$	0.2	&	10.5	$\pm$	0.6	&	0.085	$\pm$	0.004	&	10.0	$\pm$	0.8	\\
&	R(23e)	&	12.7321	&	934	&	141	&	3.9	&	-9.1	$\pm$	0.2	&	8	$\pm$	0.5	&	0.072	$\pm$	0.003	&	6.3	$\pm$	0.5	\\
&	R(25e)	&	12.657	&	1099.6	&	153	&	4	&	-8.8	$\pm$	0.3	&	7.4	$\pm$	0.7	&	0.061	$\pm$	0.004	&	5.0	$\pm$	0.6	\\
&	R(27e)	&	12.5828	&	1278.8	&	165	&	4.1	&	-8.7	$\pm$	0.8	&	11.9	$\pm$	2.1	&	0.057	$\pm$	0.007	&	7.5	$\pm$	0.1	\\
&	R(29e)	&	12.5097	&	1471.4	&	177	&	4.2	&	-7.2	$\pm$	0.5	&	8.2	$\pm$	1.3	&	0.038	$\pm$	0.004	&	3.8	$\pm$ 0.8	\\
&	R(31e)	&	12.4374	&	1677.5	&	189	&	4.3	&	-8.1	$\pm$	0.5	&	8.5	$\pm$	1.2	&	0.033	$\pm$	0.003	&	3.8	$\pm$	0.6	\\
&	R(33e)	&	12.3661	&	1896.9	&	201	&	4.4	&	-7.8	$\pm$	0.5	&	7.9	$\pm$	1.3	&	0.025	$\pm$	0.003	&	2.5	$\pm$	0.5	\\
&	R(35e)	&	12.2958	&	2129.8	&	213	&	4.5	&	-9.5	$\pm$	0.6	&	8.9	$\pm$	1.5	&	0.017	$\pm$	0.002	&	1.3	$\pm$	0.4	\\
&	R(37e)	&	12.2263	&	2376.1	&	225	&	4.6	&	-7	$\pm$	0.8	&	8.1	$\pm$	2.1	&	0.018	$\pm$	0.003	&	1.3	$\pm$	0.6	\\
\hline
o-C$_2$H$_2$ $2\nu_5 - \nu_5$		&	R(10e)	&	13.2341	&	1235.5	&	63	&	3.6	&	-9	$\pm$	0.6	&	8.5	$\pm$	1.8	&	0.061	$\pm$	0.009	&	4.6	$\pm$	1.1	\\
&	R(11f)	&	13.2028	&	1273.6	&	69	&	4.1	&	-9.6	$\pm$	0.7	&	7.4	$\pm$	1.9	&	0.047	$\pm$	0.008	&	4.4	$\pm$	1.0	\\
&	R(12e)	&	13.1479	&	1313.4	&	75	&	3.3	&	-8	$\pm$	1.2	&	6	$\pm$	2.5	&	0.026	$\pm$	0.019	&	2.9	$\pm$	1.3	\\
&	R(14e)	&	13.0619	&	1404.7	&	87	&	3.1	&	-9	$\pm$	0.4	&	9	$\pm$	1	&	0.029	$\pm$	0.002	&	4.8	$\pm$	0.5	\\
&	R(15f)	&	13.0411	&	1457.1	&	93	&	4.1	&	-9.2	$\pm$	0.4	&	9.6	$\pm$	1.1	&	0.045	$\pm$	0.004	&	4.4	$\pm$	0.6	\\
&	R(16e)	&	12.9761	&	1509.6	&	99	&	2.9	&	-7.5	$\pm$	0.5	&	9.9	$\pm$	1.3	&	0.026	$\pm$	0.003	&	4.6	$\pm$	0.8	\\
&	R(17f)	&	12.9619	&	1569.2	&	105	&	4.1	&	-8.1	$\pm$	0.4	&	7.7	$\pm$	1.1	&	0.033	$\pm$	0.003	&	2.8	$\pm$	0.5	\\
&	R(19f)	&	12.8838	&	1694.9	&	117	&	4.2	&	-8.3	$\pm$	0.5	&	10.2	$\pm$	2	&	0.031	$\pm$	0.004	&	2.6	$\pm$	0.8	\\
&	R(21f)	&	12.8066	&	1834	&	129	&	4.3	&	-8.3	$\pm$	0.3	&	7.4	$\pm$	1	&	0.029	$\pm$	0.003	&	2.8	$\pm$	0.4	\\
&	R(23f)	&	12.7305	&	1986.9	&	141	&	4.4	&	-7	$\pm$	0.4	&	8.4	$\pm$	1.1	&	0.022	$\pm$	0.002	&	2.3	$\pm$	0.4	\\
&	R(25f)	&	12.6555	&	2153.1	&	153	&	4.5	&	-8	$\pm$	0.6	&	8.3	$\pm$	1.6	&	0.029	$\pm$	0.004	&	3.0	$\pm$	0.6	\\
&	R(27f)	&	12.5814	&	2333	&	165	&	4.6	&	-9.9	$\pm$	1.3	&	8.3	$\pm$	3.4	&	0.028	$\pm$	0.008	&	2.5	$\pm$	1.3	\\
\hline
p-C$_2$H$_2$ $(\nu_4 + \nu_5)^2 - \nu_4^1$	&	R(11f)	&	13.1697	&	1104.4	&	23	&	2.2	&	-6.3	$\pm$	0.6	&	6.7	$\pm$	1.8	&	0.027	$\pm$	0.005	&	3.8	$\pm$	1.3	\\	
&	R(14e)	&	13.0289	&	1235.3	&	29	&	1.8	&	-12.8	$\pm$	1.1	&	5.8	$\pm$	2.7	&	0.019	$\pm$	0.006	&	2.5	$\pm$	1.3	\\
&	R(16e)	&	12.9437	&	1340.1	&	33	&	1.7	&	-10.6	$\pm$	1	&	9.7	$\pm$	2.8	&	0.017	$\pm$	0.004	&	3.8	$\pm$	1.3	\\
&	R(17f)	&	12.9311	&	1399.8	&	35	&	2.3	&	-7.4	$\pm$	0.6	&	6.3	$\pm$	1.6	&	0.02	$\pm$	0.004	&	3.1	$\pm$	0.9	\\
&	R(23f)	&	12.7022	&	1817.3	&	47	&	2.6	&	-9.7	$\pm$	0.7	&	8	$\pm$	2.1	&	0.017	$\pm$	0.003	&	3.3	$\pm$	0.9	\\
\hline
o-C$_2$H$_2$ $(\nu_4 + \nu_5)^2 - \nu_4^1$	&	R(10f) & 13.2104	&	1067.1	&	63	&	2.2	&	-9.2	$\pm$	0.9	&	10.4	$\pm$	2.5	&	0.056	$\pm$	0.01	&	10.0	$\pm$	3.8	\\
&	R(11e)	&	13.157	&	1103.5	&	69	&	1.9	&	-8.2	$\pm$	0.5	&	10.4	$\pm$	1.3	&	0.043	$\pm$	0.004	&	10.0	$\pm$	1.3	\\
&	R(12f)	&	13.1292	&	1145.3	&	75	&	2.2	&	-9	$\pm$	0.8	&	11.9	$\pm$	2.6	&	0.052	$\pm$	0.008	&	7.5	$\pm$	2.5	\\
&	R(13e)	&	13.0715	&	1188	&	81	&	1.8	&	-6.7	$\pm$	0.4	&	6	$\pm$	1.5	&	0.029	$\pm$	0.004	&	5.0	$\pm$	1.3	\\
&	R(17e)	&	12.9012	&	1397.5	&	105	&	1.7	&	-8.5	$\pm$	0.4	&	6	$\pm$	0.5	&	0.04	$\pm$	0.009	&	7.5	$\pm$	1.3	\\
&	R(19e)	&	12.8162	&	1522.6	&	117	&	1.6	&	-7.3	$\pm$	0.3	&	9.9	$\pm$	1	&	0.031	$\pm$	0.002	&	7.7	$\pm$	1.1	\\
&	R(20f)	&	12.8155	&	1593.4	&	123	&	2.4	&	-8.3	$\pm$	0.5	&	7.2	$\pm$	1.7	&	0.039	$\pm$	0.006	&	5.0	$\pm$	1.3	\\
\hline
C$_2$H$_2$ $\nu_2 + (\nu_4+\nu_5)$ & P(17e) & 3.0857 & 517.9 & 105 & 26.1 & -7.1$\pm$0.9 & 7.8$\pm$2.2 & 0.014$\pm$0.003 & -- \\
 & P(15e) & 3.0810 & 406.3 & 93 & 25.9 & -7.6$\pm$0.3 & 2.5$\pm$0.8 & 0.019$\pm$0.003 & -- \\
 & P(13e) & 3.0763 & 308.1 & 81 & 25.7 & -7.2$\pm$0.7 & 5.4$\pm$1.7 & 0.012$\pm$0.003 & -- \\
 & P(11e) & 3.0717 & 223.5 & 69 & 25.6 & -7.8$\pm$0.4 & 5.0$\pm$1.1 & 0.016$\pm$0.002 & -- \\
 & P(5e) & 3.0581 & 50.8 & 33 & 26.4 & -7.1$\pm$0.4 & 4.2$\pm$1.0 & 0.012$\pm$0.002 & -- \\
 & R(1e) & 3.0427 & 3.4 & 9 & 18.9 & -7.4$\pm$0.4 & 2.5$\pm$1.0 & 0.011$\pm$0.003 & -- \\
 & R(11e) & 3.0216 & 223.5 & 69 & 23.2 & -7.9$\pm$0.3 & 3.7$\pm$1.2 & 0.019$\pm$0.003 & -- \\
 & R(13e) & 3.0176 & 308.1 & 81 & 23.5 & -6.9$\pm$0.4 & 2.8$\pm$1.0 & 0.019$\pm$0.004 & -- \\
\hline
p-NH$_3$ $\nu_2$ v$_2$=0-1	&	sP(8,2) E' E'	& 12.3805	&	1002.5	&	102	&	4.6	&	-6.9	$\pm$	0.7	&	7.8	$\pm$	1.7	&	0.024	$\pm$	0.004	&	2.6	$\pm$	0.9	\\
&	s(8,1) E'' E''	& 12.3782	&	1018	&	102	&	4.8	&	-6.7	$\pm$	0.9	&	6.2	$\pm$	2.4	&	0.015	$\pm$	0.004	&	1.6	$\pm$	0.8	\\
&	aP(6,4) E' E''	& 12.3107	&	514.3	&	78	&	3.1	&	-8.4	$\pm$	0.5	&	6.3	$\pm$	1.2	&	0.02	$\pm$	0.003	&	4.1	$\pm$	0.8	\\
&	aP(6,1) E'' E'	& 12.2491	&	593.5	&	78	&	5.4	&	-6	$\pm$	0.3	&	4.5	$\pm$	0.5	&	0.039	$\pm$	0.006	&	3.3	$\pm$	0.5	\\
\hline
o-NH$_3$ $\nu_2$	v$_2$=0-1 &	aP(8,3) A1'' A2'	&	12.8487	&	977.2	&	204	&	4	&	-6.4	$\pm$	0.4	&	6.1	$\pm$	1.1	&	0.028	$\pm$	0.003	&	2.8	$\pm$	0.6	\\
&	aP(8,0) A1' A2''	&	12.8112	&	1023.9	&	204	&	4.7	&	-6.7	$\pm$	0.2	&	9.5	$\pm$	0.7	&	0.032	$\pm$	0.002	&	3.9	$\pm$	0.4	\\
&	aP(6,3) A1'' A2'	&	12.2814	&	551.3	&	156	&	4.2	&	-9.4	$\pm$	0.4	&	8.5	$\pm$	1.1	&	0.033	$\pm$	0.003	&	5.7	$\pm$	0.9	\\
&	aP(6,0) A1' A2''	&	12.2451	&	598.7	&	156	&	5.5	&	-7.3	$\pm$	0.5	&	7.2	$\pm$	1.4	&	0.044	$\pm$	0.006	&	4.4	$\pm$	1.1	\\
\end{longtable}

\tabcolsep=0.09cm
\begin{longtable}{cccccccccc}  
\caption{Line Parameters for AFGL 2136. E$_l$ is the energy of the lower level of the transition, g$_l$ is the statistical weight of the lower level, A$_{ij}$ is the Einstein A coefficient of the transition, v$_{lsr}$ is the centroid velocity of the line, ${\Delta}v$ is the FWHM of the line, $\tau_0$ is the optical depth at line centre and $N_l/N_H$ is the abundance in the lower level of the transition. Line data were taken from the HITRAN database \citep{Gordon2017}.}\\
\hline\hline
Species & Transition & $\lambda$ (${\mu}m$) & E$_l$ (K)  & g$_l$ & A$_{ij}$ (s$^{-1}$)  & $v_{lsr}$ (kms$^{-1}$) & ${\Delta}v$ (kms$^{-1}$) & ${\tau}_0$ & $N_l/N_H$ ($\times10^{-7}$) \\
\hline
\endfirsthead
\caption{continued.}\\
\hline\hline
Species & Transition & $\lambda$ (${\mu}m$) & E$_l$ (K)  & g$_l$ & A$_{ij}$ (s$^{-1}$)  & $v_{lsr}$ (kms$^{-1}$) & ${\Delta}v$ (kms$^{-1}$) & ${\tau}_0$ & $N_l/N_H$ ($\times10^{-7}$) \\
\hline
\endhead
\hline
\label{gl2136}
\endfoot
C$^{18}$O v=0-1	&	P(2)	&	4.7967	&	15.7	&	5	&	21.3	&	22.6	$\pm$	0.1	&	2	$\pm$	0.2	&	0.218	$\pm$	0.01	&	37.7	$\pm$	2.5	\\
&	P(1)	&	4.7882	&	5.2	&	3	&	32.2	&	22.1	$\pm$	0.1	&	2.4	$\pm$	0.2	&	0.167	$\pm$	0.007	&	25.1	$\pm$	2.5	\\
&	R(0)	&	4.7716	&	0	&	1	&	10.8	&	22.1	$\pm$	0.1	&	2.8	$\pm$	0.3	&	0.265	$\pm$	0.015	&	16.3	$\pm$	1.3	\\
&	R(3)	&	4.7473	&	31.5	&	7	&	14.7	&	21.9	$\pm$	0.2	&	3	$\pm$	0.5	&	0.198	$\pm$	0.017	&	23.9	$\pm$	3.8	\\
&	R(4)	&	4.7395	&	52.7	&	9	&	15.1	&	22.5	$\pm$	0.2	&	1.6	$\pm$	0.5	&	0.178	$\pm$	0.019	&	17.6	$\pm$	2.5	\\
&	R(5)	&	4.7317	&	79	&	11	&	15.4	&	22.2	$\pm$	0.3	&	2.5	$\pm$	0.8	&	0.151	$\pm$	0.022	&	17.6	$\pm$	3.8	\\
&	P(17)	&	4.9369	&	805.5	&	35	&	15.1	&	27.2	$\pm$	0.3	&	11.2	$\pm$	1	&	0.054	$\pm$	0.003	&	18.8	$\pm$	2.5	\\
&	P(15)	&	4.9168	&	631.9	&	31	&	15.4	&	27.1	$\pm$	0.2	&	11.3	$\pm$	0.7	&	0.061	$\pm$	0.002	&	21.4	$\pm$	1.3	\\
&	P(12)	&	4.8874	&	410.7	&	25	&	15.8	&	26.8	$\pm$	0.7	&	12.8	$\pm$	2.3	&	0.075	$\pm$	0.009	&	30.1	$\pm$	6.3	\\
&	R(6)	&	4.724	&	110.5	&	13	&	15.7	&	26.3	$\pm$	0.4	&	9.8	$\pm$	1.2	&	0.11	$\pm$	0.009	&	30.1	$\pm$	3.8	\\
&	R(11)	&	4.6872	&	347.6	&	23	&	16.5	&	25.9	$\pm$	0.3	&	13	$\pm$	1	&	0.09	$\pm$	0.005	&	32.7	$\pm$	2.5	\\
&	R(12)	&	4.6801	&	410.7	&	25	&	16.7	&	25.8	$\pm$	0.7	&	14.3	$\pm$	0.7	&	0.117	$\pm$	0.028	&	47.7	$\pm$	11.3	\\
&	R(14)	&	4.6662	&	552.9	&	29	&	16.9	&	27.3	$\pm$	0.2	&	13.1	$\pm$	0.8	&	0.08	$\pm$	0.003	&	28.9	$\pm$	2.5	\\
&	R(15)	&	4.6594	&	631.9	&	31	&	17	&	26.3	$\pm$	0.1	&	10.9	$\pm$	0.7	&	0.058	$\pm$	0.002	&	15.1	$\pm$	1.3	\\
&	R(16)	&	4.6527	&	716.1	&	33	&	17.1	&	27.8	$\pm$	0.4	&	11.4	$\pm$	1.1	&	0.06	$\pm$	0.004	&	20.1	$\pm$	2.5	\\
&	R(17)	&	4.6461	&	805.5	&	35	&	17.2	&	28.1	$\pm$	0.3	&	11.8	$\pm$	1.1	&	0.056	$\pm$	0.003	&	18.8	$\pm$	2.5	\\
&	R(18)	&	4.6396	&	900.2	&	37	&	17.3	&	30	$\pm$	0.6	&	14.3	$\pm$	0.8	&	0.046	$\pm$	0.007	&	18.8	$\pm$	2.5	\\
&	R(20)	&	4.6268	&	1105.3	&	41	&	17.5	&	27.8	$\pm$	0.4	&	10.7	$\pm$	1.3	&	0.038	$\pm$	0.003	&	11.3	$\pm$	1.3	\\
&	R(21)	&	4.6206	&	1215.6	&	43	&	17.6	&	27.5	$\pm$	0.2	&	10.2	$\pm$	0.7	&	0.034	$\pm$	0.001	&	10.0	$\pm$	1.3	\\
&	R(22)	&	4.6145	&	1331.2	&	45	&	17.7	&	28	$\pm$	0.3	&	11.1	$\pm$	1	&	0.025	$\pm$	0.001	&	7.5	$\pm$	1.3	\\
\hline
$^{12}$CO v=1-2	&	P(19)	&	4.9054	&	4124.8	&	39	&	31.4	&	26.2	$\pm$	0.2	&	13.4	$\pm$	0.6	&	0.165	$\pm$	0.005	&	36.4	$\pm$	2.5	\\
&	P(18)	&	4.8948	&	4020.9	&	37	&	31.7	&	27	$\pm$	0.3	&	12.6	$\pm$	0.7	&	0.163	$\pm$	0.006	&	25.2	$\pm$	2.5	\\
&	P(17)	&	4.8843	&	3922.5	&	35	&	31.9	&	27.1	$\pm$	0.1	&	11.1	$\pm$	0.4	&	0.171	$\pm$	0.004	&	31.4	$\pm$	1.3	\\
&	P(16)	&	4.8739	&	3829.4	&	33	&	32.2	&	27.6	$\pm$	0.2	&	11.8	$\pm$	0.7	&	0.17	$\pm$	0.006	&	33.9	$\pm$	2.5	\\
&	P(15)	&	4.8637	&	3741.9	&	31	&	32.5	&	27.6	$\pm$	0.2	&	12.8	$\pm$	0.7	&	0.181	$\pm$	0.006	&	38.9	$\pm$	2.5	\\
&	P(8)	&	4.7954	&	3281.8	&	17	&	34.9	&	27.2	$\pm$	0.2	&	11.7	$\pm$	0.5	&	0.213	$\pm$	0.006	&	41.1	$\pm$	2.5	\\
&	P(7)	&	4.7861	&	3237.9	&	15	&	35.5	&	27.3	$\pm$	0.3	&	11.7	$\pm$	1	&	0.227	$\pm$	0.012	&	44.0	$\pm$	3.8	\\
&	P(5)	&	4.7678	&	3166.7	&	11	&	37	&	28	$\pm$	0.3	&	12	$\pm$	0.9	&	0.189	$\pm$	0.01	&	42.7	$\pm$	3.8	\\
&	P(4)	&	4.7589	&	3139.2	&	9	&	38.3	&	27.5	$\pm$	0.4	&	10.6	$\pm$	1	&	0.18	$\pm$	0.011	&	35.2	$\pm$	3.8	\\
&	P(3)	&	4.75	&	3117.3	&	7	&	40.5	&	28	$\pm$	0.6	&	14.6	$\pm$	1.7	&	0.133	$\pm$	0.01	&	36.4	$\pm$	5.0	\\
&	P(2)	&	4.7413	&	3100.9	&	5	&	45.2	&	27.8	$\pm$	0.3	&	6.6	$\pm$	0.7	&	0.129	$\pm$	0.009	&	18.8	$\pm$	2.5	\\
&	P(1)	&	4.7327	&	3089.8	&	3	&	68.2	&	27.1	$\pm$	0.2	&	9.7	$\pm$	0.5	&	0.063	$\pm$	0.002	&	16.3	$\pm$	1.3	\\
&	R(4)	&	4.6832	&	3139.2	&	9	&	32	&	27.8	$\pm$	0.3	&	12.3	$\pm$	0.9	&	0.195	$\pm$	0.009	&	35.2	$\pm$	2.5	\\
&	R(5)	&	4.6753	&	3166.7	&	11	&	32.7	&	27.1	$\pm$	0.1	&	13	$\pm$	0.4	&	0.222	$\pm$	0.004	&	40.2	$\pm$	1.3	\\
&	R(6)	&	4.6675	&	3199.5	&	13	&	33.2	&	26.7	$\pm$	0.2	&	14	$\pm$	0.5	&	0.234	$\pm$	0.006	&	46.5	$\pm$	2.5	\\
&	R(7)	&	4.6598	&	3237.9	&	15	&	33.7	&	28.4	$\pm$	0.3	&	12.8	$\pm$	0.9	&	0.223	$\pm$	0.01	&	40.2	$\pm$	3.8	\\
&	R(8)	&	4.6523	&	3281.8	&	17	&	34.1	&	27.2	$\pm$	0.2	&	12.1	$\pm$	0.5	&	0.227	$\pm$	0.006	&	40.2	$\pm$	2.5	\\
&	R(12)	&	4.623	&	3511.9	&	25	&	35.3	&	27.4	$\pm$	0.4	&	11.7	$\pm$	0.9	&	0.246	$\pm$	0.013	&	41.5	$\pm$	3.8	\\
&	R(12)	&	4.616	&	3583	&	27	&	35.6	&	27.5	$\pm$	0.2	&	12.6	$\pm$	0.5	&	0.215	$\pm$	0.006	&	40.2	$\pm$	1.3	\\
&	R(16)	&	4.5954	&	3829.4	&	33	&	36.3	&	28.3	$\pm$	0.4	&	10.7	$\pm$	1.2	&	0.191	$\pm$	0.015	&	32.7	$\pm$	3.8	\\
&	R(17)	&	4.5887	&	3922.5	&	35	&	36.6	&	28.1	$\pm$	0.2	&	13	$\pm$	0.5	&	0.192	$\pm$	0.005	&	37.7	$\pm$	1.3	\\
&	P(24)	&	4.9603	&	4726.1	&	49	&	30.3	&	26.5	$\pm$	0.2	&	9.3	$\pm$	0.4	&	0.103	$\pm$	0.003	&	17.6	$\pm$	1.3	\\
&	P(25)	&	4.9716	&	4862.7	&	51	&	30.1	&	27	$\pm$	0.3	&	11.4	$\pm$	0.8	&	0.08	$\pm$	0.004	&	16.3	$\pm$	1.3	\\
&	P(26)	&	4.9831	&	5004.6	&	53	&	29.8	&	27	$\pm$	0.4	&	6.7	$\pm$	0.9	&	0.07	$\pm$	0.007	&	8.8	$\pm$	1.3	\\
&	P(27)	&	4.9947	&	5151.9	&	55	&	29.6	&	26.9	$\pm$	0.2	&	8.5	$\pm$	0.6	&	0.064	$\pm$	0.003	&	10.0	$\pm$	1.3	\\
&	P(28)	&	5.0065	&	5304.8	&	57	&	29.4	&	26.2	$\pm$	0.3	&	11.1	$\pm$	0.9	&	0.065	$\pm$	0.004	&	12.6	$\pm$	1.3	\\
\hline
CS v=0-1	&	R(2)	&	7.8309	&	7.1	&	5	&	6.9	&	25.7	$\pm$	0.4	&	8	$\pm$	1	&	0.037	$\pm$	0.003	&	5.0	$\pm$	1.0	\\
&	R(3)	&	7.8212	&	14.1	&	7	&	7.2	&	25.1	$\pm$	0.5	&	8.7	$\pm$	1.2	&	0.042	$\pm$	0.004	&	6.3	$\pm$	1.1	\\
&	R(4)	&	7.8116	&	23.5	&	9	&	7.4	&	25.2	$\pm$	0.4	&	6.2	$\pm$	1.1	&	0.044	$\pm$	0.005	&	5.0	$\pm$	1.3	\\
&	R(5)	&	7.802	&	35.3	&	11	&	7.6	&	25.4	$\pm$	0.3	&	9.2	$\pm$	0.8	&	0.057	$\pm$	0.003	&	8.8	$\pm$	1.0	\\
&	R(7)	&	7.7833	&	65.8	&	15	&	7.8	&	25.4	$\pm$	0.2	&	8.2	$\pm$	0.5	&	0.089	$\pm$	0.003	&	12.6	$\pm$	6.3	\\
&	R(8)	&	7.774	&	84.6	&	17	&	7.9	&	26.3	$\pm$	0.1	&	7.7	$\pm$	0.4	&	0.1	$\pm$	0.003	&	13.8	$\pm$	3.8	\\
&	R(10)	&	7.7559	&	129.2	&	21	&	8	&	26.5	$\pm$	0.2	&	10.3	$\pm$	0.5	&	0.112	$\pm$	0.003	&	18.8	$\pm$	6.3	\\
&	R(11)	&	7.7469	&	155.1	&	23	&	8.1	&	26	$\pm$	0.3	&	10.3	$\pm$	0.7	&	0.098	$\pm$	0.004	&	16.3	$\pm$	11.3	\\
&	R(23)	&	7.6467	&	648.5	&	47	&	8.8	&	25	$\pm$	0.4	&	2	$\pm$	1.1	&	0.059	$\pm$	0.008	&	5.0	$\pm$	0.4	\\
&	R(25)	&	7.6313	&	763.5	&	51	&	8.9	&	31.3	$\pm$	1	&	8.6	$\pm$	2.7	&	0.054	$\pm$	0.011	&	7.5	$\pm$	1.0	\\
&	R(26)	&	7.6237	&	824.5	&	53	&	8.9	&	27.3	$\pm$	0.6	&	5.9	$\pm$	1.6	&	0.053	$\pm$	0.008	&	6.3	$\pm$	0.3	\\
&	R(27)	&	7.6162	&	887.8	&	55	&	9	&	24.9	$\pm$	0.5	&	6.5	$\pm$	1.3	&	0.048	$\pm$	0.005	&	6.3	$\pm$	0.3	\\
&	R(31)	&	7.5871	&	1164.5	&	63	&	9.2	&	28.3	$\pm$	0.5	&	4.9	$\pm$	1.2	&	0.045	$\pm$	0.006	&	5.0	$\pm$	0.3	\\
\hline
HCN $\nu_2$ v$_2$=0-1	&	R(7e)	&	13.5941	&	119	&	90	&	1.2	&	26.4	$\pm$	0.2	&	12.3	$\pm$	0.7	&	0.155	$\pm$	0.006	&	50.2	$\pm$	3.8	\\
&	R(10e)	&	13.4326	&	233.9	&	126	&	1.2	&	26.3	$\pm$	0.2	&	9.6	$\pm$	0.6	&	0.151	$\pm$	0.006	&	42.7	$\pm$	2.5	\\
&	R(11e)	&	13.3796	&	280.6	&	138	&	1.2	&	26.5	$\pm$	0.2	&	10.1	$\pm$	0.6	&	0.156	$\pm$	0.007	&	46.5	$\pm$	3.8	\\
&	R(12e)	&	13.3271	&	331.7	&	150	&	1.2	&	25.3	$\pm$	0.4	&	11.5	$\pm$	1.5	&	0.117	$\pm$	0.009	&	40.2	$\pm$	6.3	\\
&	R(14e)	&	13.2233	&	446.4	&	174	&	1.3	&	27.8	$\pm$	0.2	&	11.5	$\pm$	0.6	&	0.145	$\pm$	0.005	&	49.0	$\pm$	2.5	\\
&	R(16e)	&	13.1213	&	578.2	&	198	&	1.3	&	26.7	$\pm$	0.3	&	12.4	$\pm$	0.9	&	0.166	$\pm$	0.008	&	61.5	$\pm$	5.0	\\
&	R(17e)	&	13.0709	&	650.5	&	210	&	1.3	&	27.7	$\pm$	0.3	&	11.8	$\pm$	0.8	&	0.106	$\pm$	0.005	&	37.7	$\pm$	2.5	\\
&	R(18e)	&	13.0209	&	727	&	222	&	1.3	&	27.6	$\pm$	0.2	&	10.3	$\pm$	0.5	&	0.121	$\pm$	0.004	&	38.9	$\pm$	2.5	\\
&	R(19e)	&	12.9713	&	807.8	&	234	&	1.3	&	28.2	$\pm$	0.4	&	13	$\pm$	1.2	&	0.085	$\pm$	0.005	&	33.9	$\pm$	3.8	\\
&	R(20e)	&	12.9221	&	892.7	&	246	&	1.3	&	28	$\pm$	0.2	&	11.2	$\pm$	0.6	&	0.079	$\pm$	0.003	&	27.6	$\pm$	1.3	\\
&	R(21e)	&	12.8734	&	981.9	&	258	&	1.4	&	27.2	$\pm$	0.2	&	12.2	$\pm$	0.7	&	0.083	$\pm$	0.003	&	31.4	$\pm$	2.5	\\
&	R(23e)	&	12.7771	&	1172.9	&	282	&	1.4	&	28	$\pm$	0.2	&	11.6	$\pm$	0.5	&	0.069	$\pm$	0.002	&	25.1	$\pm$	1.3	\\
&	R(24e)	&	12.7295	&	1274.8	&	294	&	1.4	&	26.5	$\pm$	0.3	&	9.2	$\pm$	0.9	&	0.048	$\pm$	0.004	&	15.1	$\pm$	1.3	\\
&	R(25e)	&	12.6823	&	1381	&	306	&	1.4	&	26.7	$\pm$	0.2	&	9.7	$\pm$	0.5	&	0.073	$\pm$	0.003	&	22.6	$\pm$	1.3	\\
&	R(26e)	&	12.6355	&	1491.2	&	318	&	1.4	&	26.5	$\pm$	1	&	10.8	$\pm$	2.7	&	0.036	$\pm$	0.007	&	12.6	$\pm$	3.8	\\
\hline
HCN $\nu_2$ v$_2$=0-2	&	P(16)	&	7.3219	&	578.2	&	198	&	1.1	&	27.2	$\pm$	0.4	&	8.2	$\pm$	1	&	0.073	$\pm$	0.006	&	126.9	$\pm$	18.8	\\
&	P(14)	&	7.2923	&	446.4	&	174	&	1.1	&	28.5	$\pm$	0.3	&	10	$\pm$	0.9	&	0.087	$\pm$	0.005	&	170.8	$\pm$	18.8	\\
&	P(13)	&	7.2775	&	387	&	162	&	1.1	&	25.3	$\pm$	0.7	&	7.6	$\pm$	0	&	0.07	$\pm$	0.013	&	118.1	$\pm$	20.1	\\
&	P(12)	&	7.2627	&	331.7	&	150	&	1.1	&	26.5	$\pm$	0.3	&	10.8	$\pm$	0.9	&	0.097	$\pm$	0.005	&	207.2	$\pm$	21.4	\\
&	P(11)	&	7.248	&	280.6	&	138	&	1.1	&	26.3	$\pm$	0.5	&	9.9	$\pm$	1.9	&	0.062	$\pm$	0.007	&	118.1	$\pm$	26.4	\\
&	P(10	&	7.2332	&	233.9	&	126	&	1.1	&	26.2	$\pm$	0.6	&	9.4	$\pm$	1.6	&	0.071	$\pm$	0.008	&	133.1	$\pm$	26.4	\\
&	P(9)	&	7.2184	&	191.4	&	114	&	1.1	&	25.5	$\pm$	0.9	&	7.6	$\pm$	2.8	&	0.068	$\pm$	0.035	&	113.0	$\pm$	71.6	\\
&	P(8)	&	7.2037	&	153.1	&	102	&	1.1	&	26.7	$\pm$	0.4	&	8.3	$\pm$	1	&	0.077	$\pm$	0.006	&	136.9	$\pm$	20.1	\\
&	P(6)	&	7.1741	&	89.2	&	78	&	1.2	&	25.2	$\pm$	0.5	&	7.9	$\pm$	1.1	&	0.064	$\pm$	0.006	&	113.0	$\pm$	20.1	\\
&	P(5)	&	7.1593	&	63.8	&	66	&	1.2	&	27.7	$\pm$	0.4	&	8.8	$\pm$	0.9	&	0.087	$\pm$	0.006	&	169.6	$\pm$	21.4	\\
&	P(4)	&	7.1445	&	42.5	&	54	&	1.2	&	27.3	$\pm$	0.5	&	10.7	$\pm$	1.4	&	0.069	$\pm$	0.006	&	159.5	$\pm$	26.4	\\
&	P(3)	&	7.1297	&	25.5	&	42	&	1.3	&	26.3	$\pm$	0.4	&	7.6	$\pm$	1.1	&	0.08	$\pm$	0.008	&	149.5	$\pm$	26.4	\\
&	P(2)	&	7.1148	&	12.7	&	30	&	1.4	&	26	$\pm$	0.5	&	9.3	$\pm$	1.3	&	0.054	$\pm$	0.005	&	129.4	$\pm$	21.4	\\
&	R(0)	&	7.0702	&	0	&	6	&	0.7	&	23.2	$\pm$	0.5	&	7.8	$\pm$	1.4	&	0.094	$\pm$	0.011	&	76.6	$\pm$	16.3	\\
&	R(2)	&	7.0404	&	12.7	&	30	&	0.9	&	25.8	$\pm$	0.6	&	10.9	$\pm$	2	&	0.086	$\pm$	0.01	&	150.7	$\pm$	32.7	\\
&	R(3)	&	7.0255	&	25.5	&	42	&	1	&	25.4	$\pm$	0.3	&	10.6	$\pm$	0.9	&	0.082	$\pm$	0.005	&	149.5	$\pm$	15.1	\\
&	R(5)	&	6.9956	&	63.8	&	66	&	1	&	26.7	$\pm$	0.5	&	10.9	$\pm$	3.3	&	0.063	$\pm$	0.013	&	98.0	$\pm$	36.4	\\
&	R(8)	&	6.9509	&	153.1	&	102	&	1.1	&	27.4	$\pm$	0.4	&	7.7	$\pm$	1.5	&	0.099	$\pm$	0.011	&	153.2	$\pm$	35.2	\\
&	R(9)	&	6.936	&	191.4	&	114	&	1.1	&	24.3	$\pm$	0.4	&	9	$\pm$	1	&	0.09	$\pm$	0.006	&	162.0	$\pm$	21.4	\\
&	R(10)	&	6.9211	&	233.9	&	126	&	1.1	&	28.2	$\pm$	0.4	&	7.6	$\pm$	1.3	&	0.082	$\pm$	0.008	&	133.1	$\pm$	26.4	\\
&	R(12)	&	6.8914	&	331.7	&	150	&	1.1	&	24.9	$\pm$	0.7	&	9.6	$\pm$	1.9	&	0.07	$\pm$	0.009	&	135.6	$\pm$	33.9	\\
&	R(13)	&	6.8766	&	387	&	162	&	1.1	&	24.5	$\pm$	0.6	&	9.4	$\pm$	2	&	0.089	$\pm$	0.012	&	167.0	$\pm$	42.7	\\
&	R(14)	&	6.8619	&	446.4	&	174	&	1.1	&	24.9	$\pm$	0.6	&	7.6	$\pm$	1.6	&	0.103	$\pm$	0.014	&	173.3	$\pm$	45.2	\\
&	R(15)	&	6.8472	&	510.3	&	186	&	1.1	&	26.4	$\pm$	0.7	&	7.7	$\pm$	3.5	&	0.078	$\pm$	0.015	&	123.1	$\pm$	61.5	\\
&	R(16)	&	6.8326	&	578.2	&	198	&	1.1	&	27.4	$\pm$	0.6	&	7.6	$\pm$	1.8	&	0.063	$\pm$	0.009	&	108.0	$\pm$	30.1	\\
&	R(17)	&	6.818	&	650.5	&	210	&	1.1	&	24.3	$\pm$	1	&	8.2	$\pm$	2.2	&	0.065	$\pm$	0.014	&	116.8	$\pm$	41.4	\\
\hline
p-C$_2$H$_2$ $\nu_5$	&	R(6e)	&	13.4118	&	71.1	&	13	&	3.4	&	25.4	$\pm$	0.5	&	9.4	$\pm$	1.5	&	0.131	$\pm$	0.014	&	12.6	$\pm$	2.5	\\
&	R(10e)	&	13.245	&	186.2	&	21	&	3.5	&	27	$\pm$	0.2	&	11.4	$\pm$	0.8	&	0.093	$\pm$	0.004	&	11.3	$\pm$	1.3	\\
&	R(12e)	&	13.1632	&	264.1	&	25	&	3.5	&	27	$\pm$	0.2	&	11.6	$\pm$	0.7	&	0.093	$\pm$	0.004	&	11.3	$\pm$	1.3	\\
&	R(14e)	&	13.0825	&	355.5	&	29	&	3.6	&	28.7	$\pm$	0.4	&	11.4	$\pm$	1.3	&	0.104	$\pm$	0.008	&	12.6	$\pm$	1.3	\\
&	R(16e)	&	13.0028	&	460.4	&	33	&	3.6	&	27.1	$\pm$	0.2	&	10	$\pm$	0.6	&	0.071	$\pm$	0.003	&	10.0	$\pm$	1.3	\\
&	R(24e)	&	12.6944	&	1015.2	&	49	&	3.9	&	26.4	$\pm$	0.5	&	12.1	$\pm$	1.5	&	0.052	$\pm$	0.005	&	6.3	$\pm$	1.3	\\
\hline
o-C$_2$H$_2$ $\nu_5$	&	Q(9e)	&	13.7069	&	152.4	&	57	&	6.1	&	25.8	$\pm$	0.3	&	12.7	$\pm$	1.1	&	0.221	$\pm$	0.011	&	15.1	$\pm$	1.3	\\
&	Q(21e)	&	13.6758	&	781.8	&	129	&	6.1	&	26	$\pm$	0.4	&	13.3	$\pm$	1.7	&	0.147	$\pm$	0.012	&	11.3	$\pm$	1.3	\\
&	R(11e)	&	13.2039	&	223.5	&	69	&	3.5	&	25	$\pm$	0.9	&	9.6	$\pm$	2.8	&	0.196	$\pm$	0.037	&	20.1	$\pm$	6.3	\\
&	R(13e)	&	13.1227	&	308.1	&	81	&	3.6	&	28.1	$\pm$	0.2	&	11.2	$\pm$	0.8	&	0.173	$\pm$	0.008	&	20.1	$\pm$	1.3	\\
&	R(15e)	&	13.0425	&	406.3	&	93	&	3.6	&	26.9	$\pm$	0.2	&	11.9	$\pm$	0.6	&	0.162	$\pm$	0.006	&	20.1	$\pm$	1.3	\\
&	R(17e)	&	12.9634	&	517.9	&	105	&	3.7	&	26.9	$\pm$	0.1	&	12.3	$\pm$	0.5	&	0.14	$\pm$	0.003	&	17.6	$\pm$	1.3	\\
&	R(19e)	&	12.8853	&	643.1	&	117	&	3.7	&	27.2	$\pm$	0.1	&	11.4	$\pm$	0.4	&	0.131	$\pm$	0.003	&	15.1	$\pm$	1.3	\\
&	R(21e)	&	12.8082	&	781.8	&	129	&	3.8	&	27.3	$\pm$	0.2	&	10.4	$\pm$	0.4	&	0.108	$\pm$	0.003	&	12.6	$\pm$	1.3	\\
&	R(23e)	&	12.7321	&	934	&	141	&	3.9	&	28.1	$\pm$	0.1	&	10.9	$\pm$	0.5	&	0.08	$\pm$	0.002	& 10.0	$\pm$	1.3	\\
&	R(25e)	&	12.657	&	1099.7	&	153	&	4	&	27.4	$\pm$	0.1	&	10.4	$\pm$	0.3	&	0.092	$\pm$	0.002	&	10.0	$\pm$	1.3	\\
\hline
o-C$_2$H$_2$ $(2\nu_5 - \nu_5)$		&	R(6e)	&	13.4074	&	1120.4	&	39	&	4.1	&	26.3	$\pm$	0.5	&	9.2	$\pm$	2.2	&	0.035	$\pm$	0.005	&	2.6	$\pm$	0.8	\\
&	R(10e)	&	13.2341	&	1235.5	&	63	&	3.6	&	27.5	$\pm$	0.5	&	9.9	$\pm$	0.1	&	0.045	$\pm$	0.007	&	6.7	$\pm$	0.9	\\
&	R(14e)	&	13.0619	&	1404.7	&	87	&	3.1	&	25.5	$\pm$	0.8	&	9.9	$\pm$	0.9	&	0.019	$\pm$	0.006	&	3.0	$\pm$	1.0	\\
&	R(15f)	&	13.0411	&	1457.1	&	93	&	4.1	&	25.6	$\pm$	0.3	&	4.6	$\pm$	0.8	&	0.033 $\pm$ 0.004	&	3.3	$\pm$	0.4	\\
&	R(16e)	&	12.9761	&	1509.6	&	99	&	2.9	&	27.6	$\pm$	0.4	&	9.9	$\pm$	0.5	&	0.03	$\pm$	0.003	&	5.1	$\pm$	0.6	\\
&	R(17f)	&	12.9619	&	1569.2	&	105	&	4.1	&	28.5	$\pm$	0.5	&	10	$\pm$	1.4	&	0.026 $\pm$ 0.003		&	2.8	$\pm$	0.5	\\
&	R(19f)	&	12.8838	&	1694.9	&	117	&	4.2	&	24.9	$\pm$	0.5	&	4.8	$\pm$	1.3	&	0.022 $\pm$ 0.004	&	2.1	$\pm$	0.4	\\
&	R(21f)	&	12.8066	&	1834	&	129	&	4.3	&	26.2	$\pm$	0.6	&	7.8	$\pm$	1.8	&	0.024 $\pm$ 0.004		&	2.8 $\pm$	0.6	\\
&	R(23f)	&	12.7305	&	1986.9	&	141	&	4.4	&	27.2	$\pm$	0.3	&	8.4	$\pm$	1.1	&	0.024 $\pm$ 0.002		&	1.6	$\pm$	0.3	\\
&	R(25f)	&	12.6555	&	2153.1	&	153	&	4.5	&	27	$\pm$	0.4	&	8.3	$\pm$	1.1	&	0.019 $\pm$ 0.002		&	1.4	$\pm$	0.3	\\
\hline
o-C$_2$H$_2$ $(\nu_4 + \nu_5)^2 - \nu_4^1$	&	R(6f)	&	13.3762	&	951.6	&	39	&	2.2	&	26.4	$\pm$	0.2	&	7.9	$\pm$	0.7	&	0.06	$\pm$	0.004	&	5.5	$\pm$	0.5	\\
&	R(9e)	&	13.2428	&	1032.4	&	57	&	2	&	27.3	$\pm$	0.7	&	9.7	$\pm$	1.8	&	0.035	$\pm$	0.005	&	6.3	$\pm$	1.3	\\
&	R(10f)	&	13.2104	&	1067.1	&	63	&	2.2	&	28.5	$\pm$	0.7	&	8.1	$\pm$	2	&	0.033	$\pm$	0.006	&	4.5	$\pm$	0.9	\\
&	R(11e)	&	13.157	&	1103.5	&	69	&	1.9	&	28	$\pm$	0.6	&	8.9	$\pm$	2.2	&	0.022	$\pm$	0.003	&	0.5	$\pm$	1.0	\\
&	R(14f)	&	13.0492	&	1236.9	&	87	&	2.2	&	25.9	$\pm$	0.6	&	9.9	$\pm$	1.6	&	0.025	$\pm$	0.003	&	4.4	$\pm$	0.6	\\
&	R(16f)	&	12.9702	&	1342.1	&	99	&	2.3	&	27.9	$\pm$	0.6	&	4.3	$\pm$	1.7	&	0.031	$\pm$	0.007	&	2.8	$\pm$	0.8	\\
&	R(18f)	&	12.8923	&	1461	&	111	&	2.3	&	28.6	$\pm$	0.6	&	5.8	$\pm$	1.6	&	0.033	$\pm$	0.006	&	2.6	$\pm$	0.8	\\
\hline
o-C$_2$H$_2$ $(\nu_4 + \nu_5)^0 - \nu_4^1$	&	R(10f)	&	13.2493	&	1067.1	&	63	&	1.4	&	27.8	$\pm$	0.8	&	4.7	$\pm$	2.2	&	0.029	$\pm$	0.008	&	5.0	$\pm$	2.5	\\
&	R(16f)	&	13.007	&	1342.1	&	99	&	1.6	&	26.5	$\pm$	0.4	&	9.4	$\pm$	1.1	&	0.028	$\pm$	0.002	&	7.9	$\pm$	1.0	\\
&	R(18f)	&	12.9284	&	1461	&	111	&	1.6	&	28.6	$\pm$	0.5	&	6.9	$\pm$	1.3	&	0.018	$\pm$	0.003	&	4.3	$\pm$	09	\\
&	R(20f)	&	12.8509	&	1593.4	&	123	&	1.6	&	28.1	$\pm$	0.4	&	9.5	$\pm$	1.4	&	0.018	$\pm$	0.002	&	5.5	$\pm$	0.8	\\
&	R(22f)	&	12.7745	&	1739.3	&	135	&	1.6	&	26.9	$\pm$	0.8	&	12.4	$\pm$	3	&	0.012	$\pm$	0.002	&	3.8	$\pm$	1.3	\\
&	R(24f)	&	12.6991	&	1898.8	&	147	&	1.6	&	25.6	$\pm$	0.7	&	8.2	$\pm$	1.9	&	0.015	$\pm$	0.003	&	4.5	$\pm$	1.0	\\
\hline
o-C$_2$H$_2$ $(\nu_4 + \nu_5)$	&	P(17e)	&	7.7579	&	517.9	&	105	&	2.1	&	25.1	$\pm$	0.1	&	6.1	$\pm$	0.4	&	0.121	$\pm$	0.004	&	42.7	$\pm$	2.5	\\
&	P(15e)	&	7.7308	&	406.3	&	93	&	2.1	&	25.3	$\pm$	0.4	&	7.4	$\pm$	1	&	0.054	$\pm$	0.004	&	40.2	$\pm$	6.3	\\
&	P(11e)	&	7.677	&	223.5	&	69	&	2.2	&	27.3	$\pm$	0.3	&	10.4	$\pm$	2.6	&	0.122	$\pm$	0.026	&	50.2	$\pm$	15.1	\\
&	P(5e)	&	7.5966	&	50.8	&	33	&	2.4	&	24.8	$\pm$	0.3	&	11.2	$\pm$	1	&	0.121	$\pm$	0.007	&	130.6	$\pm$	12.6	\\
&	P(3e)	&	7.5698	&	20.3	&	21	&	2.6	&	24.8	$\pm$	0.3	&	11.8	$\pm$	1.2	&	0.084	$\pm$	0.006	&	104.2	$\pm$	12.6	\\
&	R(7e)	&	7.423	&	94.8	&	45	&	2.1	&	24.8	$\pm$	0.3	&	9.7	$\pm$	1	&	0.113	$\pm$	0.007	&	96.7	$\pm$	11.3	\\
&	R(11e)	&	7.37	&	223.5	&	69	&	2.2	&	25.7	$\pm$	1.4	&	6.7	$\pm$	3.4	&	0.064	$\pm$	0.02	&	41.4	$\pm$	21.4	\\
&	R(13e)	&	7.3436	&	308.1	&	81	&	2.2	&	28.2	$\pm$	0.2	&	5.6	$\pm$	0.7	&	0.082	$\pm$	0.005	&	50.2	$\pm$	6.3	\\
&	R(15e)	&	7.3174	&	406.3	&	93	&	2.2	&	29.4	$\pm$	0.5	&	7.1	$\pm$	1.3	&	0.075	$\pm$	0.008	&	52.8	$\pm$	10.0	\\
&	R(17e)	&	7.2913	&	517.9	&	105	&	2.2	&	26.5	$\pm$	0.5	&	7.7	$\pm$	1.5	&	0.067	$\pm$	0.008	&	51.5	$\pm$	10.0	\\
&	R(19e)	&	7.2654	&	643.1	&	117	&	2.2	&	23.8	$\pm$	0.6	&	10.4	$\pm$	1.5	&	0.051	$\pm$	0.005	&	47.3	$\pm$	8.8	\\
&	R(21e)	&	7.2398	&	781.8	&	129	&	2.2	&	26.4	$\pm$	0.7	&	5.6	$\pm$	1.9	&	0.049	$\pm$	0.009	&	31.4	$\pm$	10.0	\\
\hline
p-NH$_3$ $\nu_2$ v$_2$=0-1	&	aP(7,4) E' E''	&	12.5906	&	713.5	&	90	&	3.4	&	29	$\pm$	0.5	&	4.2	$\pm$	1.4	&	0.025	$\pm$	0.005	&	2.4	$\pm$	0.6	\\
&	sP(8,5) E'' E''	&	12.3956	&	892.5	&	102	&	3.0	&	29.5	$\pm$	0.3	&	4.6	$\pm$	0.9	&	0.021	$\pm$	0.003	&	2.0	$\pm$	0.1	\\
&	sP(8,4) E' E'	&	12.3894	&	939.8	&	102	&	3.7	&	28.5	$\pm$	0.4	&	6.9	$\pm$	1.1	&	0.024	$\pm$	0.002	&	3.3	$\pm$	0.4	\\
&	sP(8,1) E'' E''	&	12.3782	&	1018	&	102	&	4.8	&	28.4	$\pm$	0.4	&	5.1	$\pm$	1.1	&	0.023	$\pm$	0.003	&	2.3	$\pm$	0.3	\\
&	aP(6,5) E'' E'	&	12.35	&	466.7	&	78	&	1.7	&	28.8	$\pm$	0.3	&	6.3	$\pm$	0.9	&	0.03	$\pm$	0.003	&	6.9	$\pm$	0.9	\\
&	aP(6,4) E''E''	&	12.3107	&	514.3	&	78	&	3.1	&	28.3	$\pm$	0.2	&	6.2	$\pm$	0.4	&	0.043	$\pm$	0.002	&	5.5	$\pm$	0.4	\\
&	aP(6,2) E' 'E'	&	12.261	&	577.7	&	78	&	4.9	&	27.1	$\pm$	0.5	&	7.8	$\pm$	1.3	&	0.06	$\pm$	0.007	&	5.4	$\pm$	1.0	\\
&	aP(6,1) E'' E'	&	12.2491	&	593.5	&	78	&	5.4	&	29.7	$\pm$	0.2	&	6.6	$\pm$	0.5	&	0.048	$\pm$	0.002	&	4.5	$\pm$	0.5	\\
&	sP(7,5) E'' E''	&	12.094	&	665.2	&	90	&	2.7	&	28.3	$\pm$	0.4	&	5.6	$\pm$	1.0	&	0.026	$\pm$	0.003	&	4.0	$\pm$	0.6	\\
&	sP(7,4) E' E'	&	12.089	&	712.7	&	90	&	3.6	&	28.4	$\pm$	0.4	&	7.3	$\pm$	1.0	&	0.032	$\pm$	0.003	&	4.5	$\pm$	0.5	\\
&	sP(7,1) E'' E''	&	12.0797	&	791.4	&	90	&	5.2	&	29.3	$\pm$	0.5	&	7.9	$\pm$	1.5	&	0.026	$\pm$	0.003	&	3.1	$\pm$	0.5	\\
&	aP(5,4) E' E''	&	12.0387	&	343.4	&	66	&	2.2	&	28.2	$\pm$	0.3	&	5.8	$\pm$	0.8	&	0.03	$\pm$	0.003	&	6.0	$\pm$	0.8	\\
&	aP(5,1) E'' E'	&	11.9786	&	422.8	&	66	&	5.8	&	28.1	$\pm$	0.2	&	7.7	$\pm$	0.5	&	0.054	$\pm$	0.003	&	4.9	$\pm$	0.4	\\
&	sP(6,5) E'' E''	&	11.8057	&	465.6	&	78	&	1.8	&	26.9	$\pm$	0.6	&	5.8	$\pm$	1.6	&	0.024	$\pm$	0.004	&	5.1	$\pm$	1.3	\\
&	sP(6,4) E' E'	&	11.8017	&	513.3	&	78	&	3.2	&	28.0	$\pm$	0.5	&	6.6	$\pm$	1.4	&	0.032	$\pm$	0.004	&	4.3	$\pm$	0.9	\\
&	sP(6,2) E' E'	&	11.7958	&	576.8	&	78	&	5.2	&	27.8	$\pm$	0.4	&	5.7	$\pm$	1.0	&	0.045	$\pm$	0.005	&	4.5	$\pm$	0.6	\\
&	sP(6,1) E'' E''	&	11.7942	&	592.6	&	78	&	5.7	&	29.1	$\pm$	0.3	&	5.6	$\pm$	0.9	&	0.046	$\pm$	0.005	&	3.0	$\pm$	0.5	\\
&	aP(4,1) E'' E'	&	11.7158	&	280.3	&	54	&	6.3	&	28.0	$\pm$	0.1	&	7.6	$\pm$	0.4	&	0.065	$\pm$	0.002	&	5.8	$\pm$	0.3	\\
&	sP(5,4) E' E	&	11.5271	&	342.2	&	66	&	2.3	&	30.6	$\pm$	0.1	&	6.9	$\pm$	0.5	&	0.031	$\pm$	0.001	&	6.4	$\pm$	0.5	\\
&	sP(5,2) E' E'	&	11.5224	&	406	&	66	&	5.3	&	27.4	$\pm$	0.3	&	6.1	$\pm$	0.8	&	0.039	$\pm$	0.004	&	4.9	$\pm$	0.5	\\
&	aP(3,2) E' E'	&	11.4714	&	150.1	&	42	&	4.2	&	32.5	$\pm$	0.3	&	4.3	$\pm$	0.7	&	0.035	$\pm$	0.003	&	4.1	$\pm$	0.5	\\
&	aP(3,1) E'' E'	&	11.4604	&	166.1	&	42	&	6.7	&	28.4	$\pm$	0.2	&	7.6	$\pm$	0.6	&	0.043	$\pm$	0.002	&	4.3	$\pm$	0.4	\\
&	sP(4,2) E' E' 	&	11.2613	&	263.4	&	54	&	5.2	&	26.0	$\pm$	0.3	&	6.0	$\pm$	0.8	&	0.038	$\pm$	0.003	&	4.6	$\pm$	0.5	\\
&	sP(4,1) E'' E''	&	11.2603	&	279.3	&	54	&	6.5	&	25.0	$\pm$	0.3	&	6.3	$\pm$	0.7	&	0.04	$\pm$	0.003	&	2.9	$\pm$	0.4	\\
&	aP(2,1) E'' E'	&	11.2122	&	80.4	&	30	&	6.7	&	24.4	$\pm$	0.2	&	7.7	$\pm$	0.6	&	0.041	$\pm$	0.002	&	5.4	$\pm$	0.4	\\
&	aQ(8,7) E'' E'	&	10.7789	&	766.8	&	102	&	10.4	&	28.9	$\pm$	0.5	&	7.7	$\pm$	1.3	&	0.031	$\pm$	0.004	&	1.6	$\pm$	0.3	\\
&	sQ(2,1) E'' E''	&	10.7732	&	79.3	&	30	&	6.9	&	26.4	$\pm$	0.5	&	6.5	$\pm$	1.4	&	0.032	$\pm$	0.005	&	3.8	$\pm$	0.9	\\
&	aQ(5,5) E'' E'	&	10.7671	&	295.3	&	66	&	12.7	&	25.2	$\pm$	0.5	&	6.6	$\pm$	1.4	&	0.068	$\pm$	0.009	&	3.4	$\pm$	0.5	\\
&	aQ(4,4) E' E''	&	10.7539	&	200.5	&	54	&	12.2	&	27.6	$\pm$	0.2	&	9.0	$\pm$	0.7	&	0.079	$\pm$	0.004	&	4.4	$\pm$	0.3	\\
&	aQ(6,5) E'' E'	&	10.7491	&	466.7	&	78	&	9.1	&	26.3	$\pm$	0.6	&	10.0	$\pm$	1.8	&	0.04	$\pm$	0.005	&	3.6	$\pm$	0.6	\\
&	aQ(5,4) E' E''	&	10.7391	&	343.4	&	66	&	8.1	&	27.1	$\pm$	0.4	&	6.0	$\pm$	1.2	&	0.043	$\pm$	0.006	&	3.6	$\pm$	0.5	\\
&	aQ(1,1) E'' E'	&	10.7339	&	23.2	&	18	&	7.6	&	25.5	$\pm$	0.4	&	8.9	$\pm$	1.2	&	0.047	$\pm$	0.004	&	4.3	$\pm$	0.5	\\
\hline
o-NH$_3$ $\nu_2$ v$_2$=0-1	&	aP(7,3) A1'' A2'	&	12.5607	&	750.4	&	180	&	4.2	&	29.0	$\pm$	0.4	&	6.6	$\pm$	1.2	&	0.062	$\pm$	0.007	&	5.1	$\pm$	1.0	\\
&	sP(8,3) A2'' A2''	&	12.3843	&	976.3	&	204	&	4.2	&	30.2	$\pm$	0.3	&	6.1	$\pm$	0.8	&	0.047	$\pm$	0.004	&	4.5	$\pm$	0.5	\\
&	aP(6,3) A1'' A2'	&	12.2814	&	551.3	&	156	&	4.2	&	27.6	$\pm$	0.2	&	8.2	$\pm$	0.5	&	0.067	$\pm$	0.003	&	7.7	$\pm$	0.6	\\
&	aP(6,0) A1' A2''	&	12.2451	&	598.7	&	156	&	5.5	&	28.6	$\pm$	0.2	&	7.8	$\pm$	0.4	&	0.074	$\pm$	0.003	&	6.9	$\pm$	0.4	\\
&	sP(7,6) A2' A2'	&	12.0997	&	606.7	&	180	&	1.4	&	28.6	$\pm$	0.2	&	5.2	$\pm$	0.6	&	0.03	$\pm$	0.002	&	7.9	$\pm$	0.8	\\
&	sP(7,3) A2'' A2''	&	12.0848	&	749.5	&	180	&	4.4	&	27.8	$\pm$	0.3	&	6.7	$\pm$	0.7	&	0.047	$\pm$	0.003	&	4.8	$\pm$	0.5	\\
&	sP(7,0) A2' A2'	&	12.0791	&	796.6	&	180	&	5.3	&	28.6	$\pm$	0.3	&	5.8	$\pm$	1	&	0.045	$\pm$	0.005	&	2.6	$\pm$	0.1	\\
&	aP(5,3) A1'' A2'	&	12.0101	&	380.5	&	132	&	3.9	&	27.8	$\pm$	0.3	&	9.0	$\pm$	0.8	&	0.069	$\pm$	0.004	&	10.9	$\pm$	1.0	\\
&	sP(6,3) A2'' A2''	&	11.7983	&	550.5	&	156	&	4.4	&	28.2	$\pm$	0.2	&	6.3	$\pm$	0.4	&	0.061	$\pm$	0.003	&	6.7	$\pm$	0.4 \\
&	aP(4,0) A1' A2''	&	11.7121	&	285.7	&	108	&	6.6	&	27.5	$\pm$	0.1	&	8.3	$\pm$	0.4	&	0.101	$\pm$	0.003	&	9.0	$\pm$	0.5	\\
&	sP(5,3) A2'' A2''	&	11.5245	&	379.5	&	132	&	4.1	&	31.7	$\pm$	0.2	&	8.0	$\pm$	0.5	&	0.052	$\pm$	0.002	&	7.7	$\pm$	0.5	\\
&	aP(4,3) A2'' A1''	&	11.2628	&	236.7	&	108	&	3.0	&	25.3	$\pm$	0.2	&	8.0	$\pm$	0.7	&	0.049	$\pm$	0.003	&	11.1	$\pm$	1.0	\\
&	aP(2,0) A1' A2''	&	11.2088	&	85.8	&	60	&	8.9	&	24.5	$\pm$	0.1	&	8.4	$\pm$	0.4	&	0.065	$\pm$	0.002	&	7.0	$\pm$	0.4	\\
&	aQ(3,3) A1'' A2?' &	10.7439	&	123.5	&	84	&	11.4	&	24.8	$\pm$	0.3	&	9.9	$\pm$	1	&	0.086	$\pm$	0.006	&	5.3	$\pm$	0.5	\\
&	aQ(8,6) A1' A2''	&	10.7397	&	835.6	&	204	&	7.6	&	26.5	$\pm$	1	&	9.8	$\pm$	3.1	&	0.025	$\pm$	0.006	&	3.6	$\pm$	1.1	\\
&	aQ(4,3) A1'' A2' &	10.7322	&	237.7	&	108	&	6.8	&	27.6	$\pm$	0.3	&	10.2	$\pm$	0.8	&	0.067	$\pm$	0.004	&	6.8	$\pm$	0.6	\\
&	aQ(5,3) A1'' A2' &	10.7182	&	380.5	&	132	&	4.5	&	26.4	$\pm$	0.4	&	10.6	$\pm$	1.4	&	0.05	$\pm$	0.004	&	8.0	$\pm$	1.1	\\
\end{longtable}

\end{document}